\documentclass[12pt]{article}
\usepackage[english]{babel}
\usepackage[latin1]{inputenc}
\usepackage{amsfonts,amssymb,amsmath, epsfig}
\usepackage{color,graphicx,graphics,psfrag}
\usepackage{amsmath,amstext,amssymb,amsfonts, amscd}
\usepackage[title]{appendix}
\usepackage{caption}
\usepackage{subcaption}

\usepackage{hyperref}          

\def\Box{\leavevmode\vbox{\hrule
     \hbox{\vrule\kern4pt\vbox{\kern4pt}%
           \vrule}\hrule}}
\def\blackbox{\leavevmode\vrule height 5pt width 4pt depth 0pt\relax}
\def\endproof{\null\hfill {$\blackbox$}\bigskip}

\def\paragraph#1{{\bf #1\ }}

\newtheorem{lemma}{Lemma}[section]  

\newtheorem{theorem}[lemma]{Theorem}

\newtheorem{definition}[lemma]{Definition}

\newtheorem{proposition}[lemma]{Proposition}

\newtheorem{remark}{Remark}[section]
\title{Hydrodynamics of the Kuramoto-Vicsek model of rotating self-propelled particles} 
\author{Pierre Degond$^{1,2}$, Giacomo Dimarco$^{1,2}$, Thi Bich Ngoc Mac$^{1,2}$} 
\date{} 
\begin{document}

\maketitle

\begin{center}
1-Université de Toulouse; UPS, INSA, UT1, UTM ;\\ 
Institut de Mathématiques de Toulouse ; \\
F-31062 Toulouse, France. \\
$\mbox{}$ \\
2-CNRS; Institut de Mathématiques de Toulouse UMR 5219 ;\\ 
F-31062 Toulouse, France.\\
email: pierre.degond, giacomo.dimarco, thi-bich-ngoc.mac@math.univ-toulouse.fr \\
$\mbox{}$ 
\end{center}

\vspace{0.5 cm}
\begin{abstract}
We consider an Individual-Based Model for self-rotating particles interacting through local alignment and investigate its macroscopic limit. Self-propelled particles moving in the plane try to synchronize their rotation motion like in the Kuramoto model. But the partners which the particle synchronize with are recruited locally, like in the Vicsek model. We study the mean-field kinetic and hydrodynamic limits of this system within two different scalings. In the small angular velocity regime, the resulting model is a slight modification of the 'Self-Organized Hydrodynamic' model \cite{Degond_etal_MAA13}. In the large angular velocity case, differences with previous models are more striking. A preliminary study of the linearized stability is proposed.  
\end{abstract}

\medskip
\noindent
{\bf Acknowledgments:} This work has been supported by the French 'Agence Nationale pour la Recherche (ANR)' in the frame of the contract 'MOTIMO' (ANR-11-MONU-009-01).

\medskip
\noindent
{\bf Key words: } Alignment, Fokker-Planck equation, macroscopic limit, von Mises-Fisher distribution, order parameter, Generalized Collision Invariants, dispersion relation

\medskip
\noindent
{\bf AMS Subject classification: } 35Q80, 35L60, 82C22, 82C70, 92D50
\vskip 0.4cm

%%%%%%%%%%%%%%%%%%%%%%%%%%%%%%%%%%%%%%%%%%%%%%%%%%%%%%%%%%%%%%%%%%%%%%%%%%%%%%%%%%%%%%%%%%%%%%%%
%%%%%%%%%%%%%%%%%%%%%%%%%%%%%%%%%%%%%%%%%%%%%%%%%%%%%%%%%%%%%%%%%%%%%%%%%%%%%%%%%%%%%%%%%%%%%%%%
\setcounter{equation}{0}
\section{Introduction}
\label{sec:intro}

This paper is concerned with the study of large system of rotating self-propelled particles subject to collective `social' interactions. Specifically, we consider particles evolving in the plane under the following influences: (i) self-propulsion, (ii) proper rotation, (iii) `social interaction' resulting in velocity alignment with their neighbors' average velocity and (iv) random velocity fluctuations in the form of Brownian motions in  the velocity direction. Proper rotation means that the self-propelled particle trajectories, in the absence of any other influence (i.e. without (iii) or (iv)) are circles of constant centers and radii. Moreover, the centers and radii of different particles can be different. The goal of the present work is to establish a set of hydrodynamic equations for the density and mean-velocity of these particles. Such hydrodynamic equations will be valid at large time and space scales compared with the typical interaction time and distance between the particles. 

Systems of self-propelled particles interacting through local alignment have received considerable interest since the early work of Vicsek and coauthors \cite{Vicsek_etal_PRL95}. This is because despite its simplicity, this paradigm is able to reproduce many of the collective patterns observed in nature. It also exhibits complex behaviors such as phase transitions which have motivated a huge literature (see e.g. \cite{Aldana_etal_PRL07, Chate_etal_PRE08, Degond_etal_preprint13, Gretoire_Chate_PRL04, Vicsek_etal_PRL95}). We refer to \cite{Vicsek_Zafeiris_PhysRep12} for a recent review on the subject. But in the vast majority of previous works, the influence of proper rotation (see item (ii) above) has been ignored. 

Furthermore, a majority of works on such systems use Individual-Based Models (IBM) which consist in following the evolution of each particle (or individual, or agent) in time (see e.g. in \cite{Carrillo_etal_SIMA10, Chate_etal_PRE08, Chuang_etal_PhysicaD07, Cucker_Smale_IEEETransAutCont07, Ha_Liu_CMS09, Mogilner_etal_JMB03, Motsch_Tadmor_JSP11}). These models aim at describing systems of swarming biological agents such as animals living in groups \cite{Aoki_BullJapSocSciFish92, Couzin_etal_JTB02, Gautrais_etal_PlosCB12} or  bacterial colonies \cite{Czirok_etal_PRE96}, among others. Alignment interaction has also been shown to result from volume exclusion interaction in the case of elongated self-propelled particles \cite{Baskaran_Marchetti_PRL10, Peruani_etal_PRE06}. 

When the number of agents is large, it is legitimate to consider mean-field kinetic models \cite{Bellomo_Soler_M3AS12, Bertin_etal_JPA09, Bolley_etal_M3AS11, Fornasier_etal_PhysicaD11, Ha_Tadmor_KRM08}, where the state of the system is described by the probability distribution of a single particle. It is even possible to reduce the description further by considering hydrodynamic models, which follow the evolution of average quantities such as the local density or average velocity. Until recently, hydrodynamic models of interacting self-propelled particle systems were mostly derived on phenomenological considerations \cite{Baskaran_Marchetti_PRE08, Bertin_etal_JPA09, Ratushnaya_etal_PhysicaA07, Toner_Tu_PRL95, Toner_etal_AnnPhys05}. A series of works \cite{Degond_Motsch_M3AS08, Degond_etal_MAA13, Frouvelle_M3AS12} have firmly established the derivation of such hydrodynamic models from microscopic ones, and particularly of one of them, the 'Self-Organized Hydrodynamics' (SOH) (see the review \cite{Degond_etal_Schwartz13}). Within this framework, phase transitions have been analyzed \cite{Barbaro_Degond_DCDSB13, Degond_etal_JNonlinearSci13, Degond_etal_preprint13, Frouvelle_Liu_SIMA12} (see also the review \cite{Degond_etal_note_submitted13}). We wish to follow the same methodology here and derive hydrodynamic models of {\bf rotating} self-propelled particles interacting through local alignment. This work is focused on model derivation. So, we defer the analysis of phase transitions to future work. 

Situations where swarming agents are trapped in a rotation motion are not uncommon. A typical example is given by swimming agents such as bacteria or algae in a shear flow. In the case of elongated particles, the velocity shear induces a rotation of the particles in a motion named Jeffrey's orbits \cite{Jeffrey_ProcRSocLondonA22}. The combination of this effect with swimming leads bacteria to undergo a circular motion near boundaries \cite{Dunstan_etal_PhysFluids12, Lauga_etal_BiophysJ06}. This nurtures the so-called gyrotactic effect which is responsible for accumulation of phytoplankton in layers \cite{Durham_etal_Science09} and patches \cite{Durham_etal_PRL11}. Staying in the biological realm, we note that some strains of swarming bacteria exhibit circular motion and vortex formation \cite{Czirok_etal_PRE96}. In some circumstances, coordination of flagella beats leads sperm cells to self-organize in a collective formation of vortices \cite{Riedel_etal_Science05}. In a different context, roboticists are keen to find decentralized control algorithms of robot swarms inducing a collective circular motion of the swarm \cite{Paley_etal_IEEEControlSystemsMag07, Sepulchre_etal_IEEETransAutomaticControl08, Chen_Zhang_Automatica11}. Applications target the design of mobile sensor networks for mapping or monitoring. 

The goal of this paper is to provide a continuum description of these systems when the number of agents is large. We start by proposing an IBM which encompasses features (i) to (iv) above. This IBM combines the Kuramoto \cite{Kuramoto_Springer84} and Vicsek \cite{Vicsek_etal_PRL95} dynamics (see \cite{Acebron_etal_RevModPhys05} and \cite{Vicsek_Zafeiris_PhysRep12} for reviews on the Kuramoto and Vicsek models respectively). It borrows from the Kuramoto model the way the agents synchronize the phase of their rotation and from the Vicsek model the way this synchronization is coupled with the spatial localization of the agents. Indeed, agents look for neighbors, compute the average phase of their rotation motion and choose this phase as their target for their own phase. In the absence of proper rotation of the particles, one recovers exactly the Vicsek model in its time continuous form \cite{Czirok_etal_PRE96, Degond_Motsch_M3AS08}. By contrast, if the synchronization is global, i.e. the agents compute the average phase over the whole ensemble of particles, the original Kuramoto model is recovered. Previous works have acknowledged the proximity between the Kuramoto and Vicsek models, such as \cite{Chepizhko_Kulinskii_PhysicaA10, Ha_etal_QuarterlyApplMath11}. The present model is close to that proposed in \cite{Paley_etal_IEEEControlSystemsMag07, Sepulchre_etal_IEEETransAutomaticControl08}. A different, but related approach where the oscillators move diffusively in space, has been studied in \cite{Peruani_etal_NJP10}. But none of them have proposed a hydrodynamic description of a system of particles undergoing a combined Kuramoto-Vicsek dynamics. This is the goal pursued here. 

Similar to the present work, previous works have used circular motion as the free motion of the agents. In particular, the so-called `Persistent Turner' model has been proposed to describe the dynamics of fish \cite{Degond_Motsch_JSP08, Gautrais_etal_JMB09} and fish schools \cite{Degond_Motsch_JSP11, Gautrais_etal_PlosCB12}. However, there are significant differences. In the `Persistent Turner' model, the curvature of the motion undergoes stochastic changes. In the mean over time, the curvature is zero, and there is no preferred turning direction. By contrast, in the present work, the curvature is constant and so is its mean over time. Consequently, there is a definite preferred turning direction. These differences are significant and can be read on the structure of the resulting hydrodynamic models. 

After writing the combined Kuramoto-Vicsek IBM, we propose a mean-field kinetic description of this system by means of a Fokker-Planck type equation for the one-particle ensemble distribution function. After scaling the kinetic equation to non-dimensionless variables, we realize that two regimes are of interest. In the first one, the proper rotation of the particles is slow enough, so that the particles can reach an equilibrium under the combined influences of the alignment and noise without deviating from a straight line too much. In this regime, the hydrodynamic limit yields the SOH model \cite{Degond_Motsch_M3AS08, Degond_etal_MAA13, Frouvelle_M3AS12, Degond_etal_Schwartz13} with an additional source term in the velocity evolution equation stemming from the average proper rotation of the particle ensemble. This regime is called the slow angular velocity regime and the associated hydrodynamic models, the {\bf 'Self-Organized Hydrodynamics with proper Rotation (Small angular velocity case)' or SOHR-S}. 

Another regime is possible, where the proper rotation is of the same order as the alignment interaction and noise. This changes significantly the equilibrium velocity distribution of the particles. In order to maintain the propensity of the particles to align with the ensemble of neighboring particles, we are led to modify the definition of the direction to which elementary particles align. This modification is commented in great length in the corresponding section below. At this point, let us simply mention that this modification could account for the influence of volume exclusion interaction in the spirit of \cite{Baskaran_Marchetti_PRL10, Peruani_etal_PRE06}. In this regime, the obtained hydrodynamic model involves significant modifications compared with the previous SOH model and is called the {\bf 'Self-Organized Hydrodynamics with proper Rotation (Large angular velocity case)' or SOHR-L}.

The changes compared with the previous SOH model consist of two aspects. First, the velocity equation is coupled to the whole angular velocity distribution function (and not through simple moments such as the density or average angular momentum, by contrast with the SOHR-S model). Second, this equation involves additional terms which correspond to transport in the direction normal to the velocity, or off-diagonal terms in the pressure tensor. In spite of its complexity, the model is shown to be linearly well-posed when the angular velocity distribution function is an even function (i.e. there is no preferred turning direction when averaged over the particles). Also, the asymptotics for small angular velocities reduces the complexity of the system to that of three first order partial differential equations. More detailed analytical studies of this system are in progress.  

In both regimes, the derivation of hydrodynamic models is possible, in spite of the lack of momentum conservation. The lack of conservations is acknowledged (see e.g. the discussion in the introduction of \cite{Vicsek_Zafeiris_PhysRep12}) as one of the major differences and sources of analytical difficulties that complex systems in biology and social sciences present. The main contribution of previous works on the SOH model (see e.g. the review \cite{Degond_etal_Schwartz13}) has been to provide a way to bypass this lack of momentum conservation. The main tool for this is the concept of Generalized Collision Invariant (GCI). Again, this concept will be the key of the derivation of the SOHR models, in both the small and the large angular velocity cases. 

The paper is organized as follows. In section \ref{sec:IBM}, the IBM and its mean-field kinetic limit are introduced and scaling considerations are developed. Section \ref{sec:small} is devoted to the statement of the convergence of the mean-field kinetic model towards the hydrodynamic limit in the small angular velocity case. Some properties of the SOHR-S model are discussed. The case of the large angular velocity regime is then treated in section \ref{sec:large}. Section \ref{sec:large_properties} details some of the properties of the SOHR-L model, such as its linearized stability or its asymptotics in the small angular velocity limit. A conclusion is drawn in section \ref{sec:conclu}. Then, three appendices are devoted to the proofs of the formal convergence results towards the hydrodynamic limit in the small angular velocity case (Appendix \ref{sec:small_proof}) and in the large angular velocity case (Appendix \ref{sec:large_proof}) and to the formal asymptotics of the SOHR-L model when the angular velocities become small (Appendix \ref{sec:la_small_zeta_proof}). Finally Appendix \ref{sec:la_graphics} presents some graphical illustrations.

%%%%%%%%%%%%%%%%%%%%%%%%%%%%%%%%%%%%%%%%%%%%%%%%%%%%%%%%%%%%%%%%%%%%%%%%%%%%%%%%%%%%%%%%%%%%%%%%
%%%%%%%%%%%%%%%%%%%%%%%%%%%%%%%%%%%%%%%%%%%%%%%%%%%%%%%%%%%%%%%%%%%%%%%%%%%%%%%%%%%%%%%%%%%%%%%%
\setcounter{equation}{0}
\section{Individual-Based model, mean-field limit and scaling}
\label{sec:IBM}

We consider a system of $N$ particles or agents moving with constant speed $c$ in the two-dimensional plane ${\mathbb R}^2$. We denote by $(X_k(t),V_k(t))_{k=1, \ldots N}$ the positions and the normalized velocities of the particles, with $X_k(t) \in {\mathbb R}^2$ and $V_k(t) \in {\mathbb S}^1$, where ${\mathbb S}^1$ denotes the unit circle in ${\mathbb R}^2$. The actual velocities of the particles are $\tilde V_k = c V_k$. Each particle is subject to three different actions. The first one is a proper angular velocity $W_k$, which, in the absence of any other action, would result in a circular motion of radius $R_k = \frac{c}{|W_k|}$, rotating counter-clockwise if $W_k >0$ and clockwise if $W_k <0$. Then, each particle is subject to independent Brownian white noises $P_{V_k^\bot} \circ (\sqrt{2D} \,  d B^k_t)$ with uniform diffusivity $D$. The quantity $d B^k_t$ refers to the standard white noise in ${\mathbb R}^2$. It is projected onto a standard white noise on ${\mathbb S}^1$ thanks to the projection operator $P_{V_k^\bot}$. Denoting by $V_k^\bot$ the vector obtained from $V_k$ by a rotation of angle $\pi/2$, $P_{V_k^\bot}$ is the orthogonal projection onto the line generated by $V_k^\bot$, i.e. $P_{V_k^\bot} = V_k^\bot \otimes V_k^\bot = \mbox{Id} - V_k \otimes V_k$, where $\otimes$ denotes the tensor product of two vectors and Id is the identity matrix. The symbol '$\circ$' indicates that the corresponding stochastic differential equation is taken in the Stratonovich sense. The fact that the projection of a standard white noise in ${\mathbb R}^2$ onto the tangent line to the circle in the Stratonovich sense leads to a standard white noise on ${\mathbb S}^1$ can be found e.g. in \cite{Hsu_AMS02}. Finally, the particle velocities relax towards the neighbors' average velocity $\bar V_k$ with relaxation constant $\nu$. The quantity $\nu$ is also supposed uniform (i.e. all particles have identical $\nu$) and constant in time for simplicity. Following these rules, the particles evolve according to the following stochastic differential equations:
\begin{eqnarray}
&&\hspace{-1cm}
\frac{dX_k}{dt} = c V_k, \label{eq:IBM_x}\\
&&\hspace{-1cm}
dV_k =P_{V_k^\bot} \circ (\nu \, \bar V_k \, dt  + \sqrt{2D} \, dB_t) + W_k \, V_k^\bot \, dt, \label{eq:IBM_omega}
\end{eqnarray}

The vector $\bar V_k$ may be computed by different rules, leading to different types of models. For the time being, we assume that $\bar V_k$ is obtained by normalizing the average ${\mathcal J}_k$ of the velocities $V_j$ of the particles $j$ lying in a disk of given radius $R$ centered at $X_k$, i.e. 
\begin{eqnarray}
&&\hspace{-1cm} 
\bar V_k = \frac{{\mathcal J}_k}{|{\mathcal J}_k|}, \quad 
{\mathcal J}_k = \frac{1}{N} \sum_{j, \, |X_j - X_k| \leq R} V_j, \label{eq:IBM_bar_omega}
\end{eqnarray}

In the absence of self-rotation velocity $W_k = 0$, the system reduces to the time-continuous version of the Vicsek alignment model \cite{Vicsek_etal_PRL95} as proposed in \cite{Czirok_etal_PRE96, Degond_Motsch_M3AS08}. On the other hand, if the neighbor's average velocity is computed over all the particles, i.e. if (\ref{eq:IBM_bar_omega}) is replaced by 
\begin{eqnarray}
&&\hspace{-1cm} 
\bar V_k = \bar V  = \frac{{\mathcal J}}{|{\mathcal J}|}, \quad 
{\mathcal J} = \frac{1}{N} \sum_{j=1}^N V_j, \label{eq:IBM_bar_omega_kur}
\end{eqnarray}
then, the evolution of the velocities \, $(V_k)_{k=1,\ldots,N}$ \, does not depend on the positions $(X_k)_{k=1,\ldots,N}$ and the resulting system for $(V_k)_{k=1,\ldots,N}$ is nothing but the noisy Kuramoto model of oscillator synchronization \cite{Acebron_etal_RevModPhys05}. Indeed, considering the noiseless case $D=0$ for simplicity, we can write $V_k = (\cos \theta_k, \sin \theta_k)$, $\bar V = (\cos \bar \theta, \sin \bar \theta)$ and Eq. (\ref{eq:IBM_omega}) with (\ref{eq:IBM_bar_omega_kur}) can be written:
\begin{eqnarray*}
\frac{d\theta_k}{dt} &=& \nu \sin(\bar \theta - \theta_k)   + W_k \\
&=& \frac{\nu}{|{\mathcal J}|} \, \frac{1}{N} \sum_{j=1}^N \sin (\theta_j - \theta_k) + W_k. 
\end{eqnarray*}
This is the Kuramoto model with a coupling constant $K=\frac{\nu}{|{\mathcal J}|}$. In the standard Kuramoto model, the coupling constant $K$ is supposed independent of $|{\mathcal J}|$. The reason for taking $K=\frac{\nu}{|{\mathcal J}|}$ here is that the original time-continuous version of the Vicsek model as in \cite{Czirok_etal_PRE96, Degond_Motsch_M3AS08} corresponds to this choice. Additionally, with this choice, the macroscopic limit is simpler. 

In the context of the Vicsek model, the case where $\frac{\nu}{|{\mathcal J}|}$ is a constant (or more generally a smooth function of $|{\mathcal J}|$) has been studied in \cite{Degond_etal_JNonlinearSci13, Degond_etal_note_submitted13, Degond_etal_preprint13}. In this case, multiple equilibria and phase transitions may appear. Phase transitions are also seen in the Kuramoto model \cite{Acebron_etal_RevModPhys05, Bertini_etal_JSP10, Bertini_etal_arxiv12, Giacomin_etal_arxiv11, Giacomin_etal_SIMA12, Lucon_arxiv12}. This makes the physics more interesting but on the other hand, complicates the derivation of hydrodynamic models. Hence, in the present work, we keep the assumption of constant $\nu$ for the sake of simplicity and differ the study of the constant $\frac{\nu}{|{\mathcal J}|}$ case to future work.

In the limit of an infinite number of particles $N \to \infty$, the system can be described by the one-particle distribution function $f(x,v,W,t)$ where $(x,v,W)$ is the position in the phase space ${\mathbb R}^2 \times {\mathbb S}^1 \times {\mathbb R}$. The quantity $f(x,v,W,t) \, dx \, dv \, dW$ represents the probability of finding a particle in a neighborhood $dx \, d v \, dW$ of $(x,v,W)$. The evolution equation for $f$ deduced from system (\ref{eq:IBM_x}), (\ref{eq:IBM_omega}) (see e.g. \cite{Bolley_etal_AML11}) is given by the following Fokker-Planck equation: 
\begin{eqnarray}
&&\hspace{-1cm}
\partial_t f + c \nabla_x \cdot (v f) + \nabla_v \cdot (F_f \, f) - D \Delta_v f = 0, \label{eq:kinetic_f} \\
&&\hspace{-1cm}
F_f(x,v,W,t)= P_{v^\bot} (\nu \, \bar v_f(x,t)) + W v^\bot
, \label{eq:kinetic_F}
\end{eqnarray}
This equation expresses that the time derivative of $f$ is balanced by, on the one hand, first order fluxes in the $(x,v)$ space describing spatial transport by the velocity $c v$ (the second term) and velocity transport by the force $F_f$ (the third term) and by, on the other hand, velocity diffusion due to the Brownian noise (the fourth term). The operators $\nabla_v \cdot$ and $\Delta_v $ respectively stand for the divergence of tangent vector fields to ${\mathbb S}^1$ and the Laplace-Beltrami operator on ${\mathbb S}^1$. For later usage, we also introduce the symbol $\nabla_v$ which denotes the tangential gradient of scalar fields defined on ${\mathbb S}^1$. Let $\varphi (v)$ be a scalar function defined on ${\mathbb S}^1$ and let $\varphi (v) v^\bot$ a tangent vector field to ${\mathbb S}^1$. Denote by $\bar \varphi (\theta)$ the expression of $\varphi (v)$ in a polar coordinate system. Then, these operators are expressed as follows: 
$$\nabla_v \cdot (\varphi (v) v^\bot) = \partial_\theta \bar \varphi, \quad \nabla_v \varphi (v) = \partial_\theta \bar \varphi \, v^\bot, \quad \Delta_v \varphi (v) = \partial_\theta^2 \bar \varphi.$$ 
Eq. (\ref{eq:kinetic_F}) describes how the force term is computed. The first term describes the interaction force: it has has the form of a relaxation towards the neighbors' average velocity $\bar v_f(x,t)$ with a relaxation frequency $\nu$. The second term is the self-rotation force with angular velocity $W$. We note that there is no operator explicitly acting on the angular velocity $W$. Indeed, this quantity is supposed attached to each particle and invariant over time. System (\ref{eq:kinetic_f}), (\ref{eq:kinetic_F}) is supplemented with an initial condition $f_I(x,v,W) := f(x,v,W,t=0)$. 

We will present several ways of computing the neighbors' average velocity $\bar v_f(x,t)$. To make the model specific at this point, we simply consider the case where it is computed by the continuum counterpart of the discrete formula (\ref{eq:IBM_bar_omega}), namely:
\begin{eqnarray}
&&\hspace{-1cm}
\bar v_f(x,t) = \frac{{\mathcal J}_f(x,t)}{|{\mathcal J}_f(x,t)|},  \label{eq:kinetic_omega_bar} \\
&&\hspace{-1cm}
{\mathcal J}_f(x,t) = \int_{(y,v,W) \in {\mathbb R}^2 \times {\mathbb S}^1 \times {\mathbb R}} K \Big( \frac{|x-y|}{R} \Big) \, f(y,v,W,t) \, v \, dy \, dv \, dW. \label{eq:kinetic_Jf}
\end{eqnarray}
Here the summation of the neighbor's velocities over a disk centered at the location $x$ of the particle and of radius $R$ which was used in the discrete model (formula (\ref{eq:IBM_bar_omega})) is replaced by a more general formula involving a radially symmetric interaction kernel $K$. We recover an integration over such a disk if we choose $K(\xi) = \chi_{[0,1]} (\xi)$, with $\xi = \frac{|x-y|}{R}$ and $\chi_{[0,1]}$ is the indicator function of the interval $[0,1]$. For simplicity, we now normalize $K$ such that $\int_{{\mathbb R}^2} K(|x|) \, d x = 1$. The parameter $R$ will be referred to as the interaction range. 

In order to define the hydrodynamic scaling, we first non-dimensionalize the system. We introduce the time scale $t_0 = \nu^{-1}$ and the associated space scale $x_0 = c t_0 = c/\nu$. With these choices, the time unit is the time needed by a particle to adjust its velocity due to interactions with other particles (or mean interaction time) and the space unit is the mean distance traveled by the particles during the mean interaction time, i.e. the mean free path. We set $W_0$ the typical angular frequency. For instance, we can assign to $W_0$ the value $\bar W_1 + \bar W_2$ where $\bar W_1$ and $\bar W_2$ are respectively the mean and the standard deviation of $W$ over the initial probability distribution function $f_I \, dx \, dv \, dW$. Similarly, we introduce a distribution function scale $f_0 = \frac{1}{x_0^2 \, W_0}$ and a force scale $F_0 = \frac{1}{t_0}$.  

We introduce dimensionless variables $x = x_0 \, x'$, $t = t_0 \, t'$, $W = W_0 \, W'$, $f= f_0 f'$, $F_f = F_0 \, F'_{f'}$ as well as the following dimensionless parameters: 
\begin{eqnarray}
&&\hspace{-1cm}
d = \frac{D}{\nu}, \qquad \Upsilon = \frac{W_0}{\nu}, \qquad r = \frac{R \, \nu}{c}. \label{eq:ndparam}
\end{eqnarray}
These parameters are respectively the dimensionless diffusivity, the dimensionless intrinsic angular velocity and the dimensionless interaction range. The non-dimensionlized system solved by $f'(x',v,W',t')$ is written as follows (dropping the primes for simplicity): 
\begin{eqnarray}
&&\hspace{-1cm}
\partial_t f + \nabla_x \cdot (v f) + \nabla_v \cdot (F_f \, f) - d \Delta_v f = 0, \label{eq:ndkin_f} \\
&&\hspace{-1cm}
F_f(x,v,W,t)= P_{v^\bot} \bar v_f(x,t) + \Upsilon \, W v^\bot, \label{eq:ndkin_F}
\end{eqnarray}
where, in the simple example given above, the neighbors' average velocity is now given by 
\begin{eqnarray}
&&\hspace{-1cm}
\bar v_f(x,t) = \frac{{\mathcal J}_f(x,t)}{|{\mathcal J}_f(x,t)|},  \label{eq:ndkin_omega_bar} \\
&&\hspace{-1cm}
{\mathcal J}_f(x,t) = \int_{(y,v,W) \in {\mathbb R}^2 \times {\mathbb S}^1 \times {\mathbb R}} K \Big( \frac{|x-y|}{r} \Big) \, f(y,v,W,t) \, v \, dy \, dv \, dW. \label{eq:ndkin_Jf}
\end{eqnarray}

So far, the chosen time and space scales are microscopic ones: they are set up to describe the evolution of the system at the scale of the interactions between the agents. We are now interested by a description of the system at macroscopic scales, i.e. at scales which are described by units $\tilde x_0 = \frac{x_0}{\varepsilon}$ and $\tilde t_0 =  \frac{t_0}{\varepsilon}$ where $\varepsilon \ll 1$ is a small parameter. By changing these units, we correspondingly change the variables $x$ and $t$ and the unknown $f$ to new variables and unknowns $\tilde x = \varepsilon \, x$, $\tilde t = \varepsilon t$, $\tilde f = \frac{f}{\varepsilon^2}$. In performing this change of variables, we must state how the dimensionless parameters (\ref{eq:ndparam}) behave as $\varepsilon \to 0$. We assume that $d = {\mathcal O}(1)$ and $r = {\mathcal O}(1)$ as $\varepsilon \to 0$, and for simplicity, we assume that $d$ and $r$ remain constant. By contrast, we will investigate two different scaling assumptions for $\Upsilon$ and we define a new parameter $\eta = \frac{\varepsilon}{\Upsilon}$. After changing to the macroscopic variables $\tilde x$, $\tilde t$ the system reads (dropping the tildes for simplicity): 
\begin{eqnarray}
&&\hspace{-1cm}
\partial_t f^\varepsilon + \nabla_x \cdot (v f^\varepsilon) = \frac{1}{\varepsilon} \big( - \nabla_v \cdot (P_{v^\bot} \bar v^\varepsilon_{f^\varepsilon} \, f^\varepsilon) + d \Delta_v f^\varepsilon \big) - \frac{1}{\eta} W \nabla_v \cdot (v^\bot f^\varepsilon), \label{eq:skin_f}
\end{eqnarray}
where again in the simplest case, the neighbors' average velocity is given by 
\begin{eqnarray}
&&\hspace{-1cm}
\bar v^\varepsilon_f(x,t) = \frac{{\mathcal J}^\varepsilon_f(x,t)}{|{\mathcal J}^\varepsilon_f(x,t)|},  \label{eq:skin_omega_bar} \\
&&\hspace{-1cm}
{\mathcal J}^\varepsilon_f(x,t) = \int_{(y,v,W) \in {\mathbb R}^2 \times {\mathbb S}^1 \times {\mathbb R}} K \Big( \frac{|x-y|}{\varepsilon r} \Big)  \, f(y,v,W,t) \, v\, dy \, dv \, dW. \label{eq:skin_Jf}
\end{eqnarray}
Next, by Taylor expansion and owing to the rotational symmetry of the function $x \in {\mathbb R}^2 \mapsto K(|x|)$, we have \cite{Degond_Motsch_M3AS08}: 
\begin{eqnarray}
&&\hspace{-1cm}
\bar v^\varepsilon_f(x,t) = \Omega_f(x,t) + {\mathcal O}(\varepsilon^2), \qquad \Omega_f(x,t) = \frac{J_f(x,t)}{|J_f(x,t)|} ,  \label{eq:omega_bar_expan} \\
&&\hspace{-1cm}
J_f(x,t) = \int_{(v,W) \in {\mathbb S}^1 \times {\mathbb R}}  f(x,v,W,t) \, v \, dv \, dW. \label{eq:Jf_expan}
\end{eqnarray}
In other words, up to ${\mathcal O}(\varepsilon^2)$ terms, the interaction force is given by a local expression, involving only the distribution function $f$ at position $x$. The quantity $J_f(x,t)$ is the local particle flux at point $x$ and time~$t$. By contrast, the expression (\ref{eq:skin_omega_bar}), (\ref{eq:skin_Jf}) of $\bar v^\varepsilon_f$ is spatially non-local: it involves a convolution of $f$ with respect to the non-local kernel $K$. We now omit the ${\mathcal O}(\varepsilon^2)$ terms as they have no contribution to the hydrodynamic limit at leading order (which is what we are interested in). 

The remainder of this work is concerned with the formal limit $\varepsilon \to 0$ of the following perturbation problem: 
\begin{eqnarray}
&&\hspace{-1cm}
\partial_t f^\varepsilon + \nabla_x \cdot (v f^\varepsilon) = \frac{1}{\varepsilon} \big( - \nabla_v \cdot (P_{v^\bot} \Omega_{f^\varepsilon} \, f^\varepsilon) + d \Delta_v f^\varepsilon \big) - \frac{1}{\eta} W \nabla_v \cdot (v^\bot f^\varepsilon), \label{eq:finkin_f} \\
&&\hspace{-1cm}
\Omega_f(x,t) = \frac{J_f(x,t)}{|J_f(x,t)|} ,  \quad J_f(x,t) = \int_{(v,W) \in {\mathbb S}^1 \times {\mathbb R}}  f(x,v,W,t) \, v \, dv \, dW. \label{eq:finkin_Omega}
\end{eqnarray}
We will be interested in the following two scaling assumptions for $\eta$
\begin{itemize}
\item[(i)] Small angular velocities: $\eta = {\mathcal O}(1)$. In this regime, the characteristic angular velocity satisfies $\Upsilon =  {\mathcal O}(\varepsilon)$. It takes the particles a macroscopic time interval to perform a finite angle rotation. 
\item[(ii)] Large angular velocities: $\eta = {\mathcal O}(\varepsilon)$. In this case, the characteristic angular velocity satisfies $\Upsilon =  {\mathcal O}(1)$. It takes the particles a microscopic time interval to perform finite angle rotations. Over a macroscopic time interval, the number of rotations is ${\mathcal O}\big( \frac{1}{\varepsilon}\big)$. 
\end{itemize}
We expect that case (i) is just a perturbation of the case where there is no proper rotation, and which has previously been investigated in \cite{Degond_Motsch_M3AS08}. On the other hand, case (ii) involves a larger modification and we expect that significant new behaviors are captured. However, we will see that case (ii) requires a modification of the way the agents' turning velocity is computed. Indeed, the agents need to take their proper angular velocity into account in the evaluation of the turning velocity that produces alignment with their neighbors. Therefore, according to whether that proper velocity goes along or against their will, the agents need to achieve smaller or larger turning. Precisely, the changes to Eq. (\ref{eq:finkin_Omega}) that are needed will be described in greater detail below. The next section is devoted to the investigation of case (i).

%%%%%%%%%%%%%%%%%%%%%%%%%%%%%%%%%%%%%%%%%%%%%%%%%%%%%%%%%%%%%%%%%%%%%%%%%%%%%%%%%%%%%%%%%%%%%%%%
%%%%%%%%%%%%%%%%%%%%%%%%%%%%%%%%%%%%%%%%%%%%%%%%%%%%%%%%%%%%%%%%%%%%%%%%%%%%%%%%%%%%%%%%%%%%%%%%
\setcounter{equation}{0}
\section{Small angular velocities}
\label{sec:small}

In the case of small angular velocities, we have $\eta = {\mathcal O}(1)$. We make $\eta = 1$ for simplicity. The problem is now written: 
\begin{eqnarray}
&&\hspace{-1cm}
\partial_t f^\varepsilon + \nabla_x \cdot (v f^\varepsilon) + W \nabla_v \cdot (v^\bot f^\varepsilon)= \frac{1}{\varepsilon} Q(f^\varepsilon),  
\label{eq:sm_smallkin_f} 
\end{eqnarray}
where the 'collision operator' $Q(f)$ is given by: 
\begin{eqnarray}
&&\hspace{-1cm}
Q(f) =  - \nabla_v \cdot (P_{v^\bot} \Omega_{f} \, f) + d \Delta_v f  , 
\label{eq:sm_smallkin_coll_op} \\
&&\hspace{-1cm}
\Omega_f(x,t) = \frac{J_f(x,t)}{|J_f(x,t)|} ,  \quad J_f(x,t) = \int_{(v,W) \in {\mathbb S}^1 \times {\mathbb R}}  f(x,v,W,t) \, v \, dv \, dW. 
\label{eq:sm_smallkin_Omega}
\end{eqnarray}

The formal limit $\varepsilon \to 0$ has been established in \cite{Degond_Motsch_M3AS08, Degond_etal_MAA13} when there is no self-rotation term $W \nabla_v \cdot (v^\bot f^\varepsilon)$ and no dependence of $f$ upon $W$. The present analysis is a somewhat straightforward extension of this earlier work. Before stating the theorem, we need to recall the definition of the von Mises-Fisher (VMF) distribution $M_\Omega(v)$. Its expression is given by: 
\begin{eqnarray}
&&\hspace{-1cm}
M_\Omega(v) = Z_d^{-1} \, \exp \big( \frac{v \cdot \Omega}{d} \big) , \quad \quad Z_d = \int_{v \in {\mathbb S}^1} \exp \big( \frac{v \cdot \Omega}{d} \big)  \, dv. 
\label{eq:sm_smallkin_VMF}
\end{eqnarray}
By construction, $M_\Omega(v)$ is a probability density and due to rotational symmetry, the constant $Z_d$ does not depend on $\Omega$. The flux of the VMF distribution is given by: 
\begin{equation}
\int_{v \in {\mathbb S}^1} M_\Omega(v) \, v  \, dv = c_1 \, \Omega, \quad \quad c_1 = c_1(d) = \frac{\int_{v \in {\mathbb S}^1} \exp \big( \frac{v \cdot \Omega}{d} \big)  (v \cdot \Omega) \, dv}{\int_{v \in {\mathbb S}^1} \exp \big( \frac{v \cdot \Omega}{d} \big)  \, dv}. 
\label{eq:sm_smallkin_cur_equi}
\end{equation}
The parameter $c_1(d)$ does not depend on $\Omega$. It is given by 
\begin{eqnarray*}
c_1(d) = \frac{\int_0^\pi e^{\frac{\cos \theta}{d}} \, \cos \theta \, d \theta}{\int_0^\pi e^{\frac{\cos \theta}{d}} \, d \theta} = \frac{I_1\big(\frac{1}{d} \big)}{I_0\big(\frac{1}{d} \big)}, 
\end{eqnarray*}
where $\beta \in {\mathbb R} \to I_k(\beta) \in {\mathbb R}$ is the modified Bessel function:
$$ I_k(\beta) = \frac{1}{\pi} \int_0^{\pi} \exp\{ \beta \, \cos \theta \} \, \cos (k \, \theta) \, d \theta, \quad \forall \beta \in {\mathbb R}, \quad  \forall k \in {\mathbb N}.  $$
It verifies $0 \leq c_1(d) \leq 1$ and is a strictly decreasing function of $d \in [0,\infty)$. When $c_1$ is small, the VMF distribution is close to the uniform distribution. By contrast, when $c_1$ is close to $1$, the VMF distribution is close to the Dirac delta at $v = \Omega$. The parameter $c_1$ measures the degree of alignment of the VMF distribution about the direction of $\Omega$, hence its name of 'order parameter'.  

Now, we can state the theorem which establishes the limit $\varepsilon \to 0$ of (\ref{eq:sm_smallkin_f}). 

\begin{theorem}
We assume that the limit $f^0 = \lim_{\varepsilon \to 0} f^\varepsilon$ exists and that the convergence is as regular as needed  (i.e. occurs in functional spaces that allow the rigorous justification of all the computations below). Then, we have 
\begin{equation} 
f^0 (x,v, W,t) = \rho_W(x,t) \, M_{\Omega(x,t)}(v). 
\label{eq:sm_smallkin_f0}
\end{equation}
where, for any $(x,t)$, the function $W \in {\mathbb R}  \to \rho_W(x,t) \in {\mathbb R}$ belongs to $L^1({\mathbb R})$ and has first moment finite, and the vector $\Omega(x,t)$ belongs to ${\mathbb S}^1$. The functions $\rho_W(x,t)$ and $\Omega(x,t)$ satisfy the following system of hydrodynamic equations: 
\begin{eqnarray}
&&\hspace{-1cm}
\partial_t \rho_W + \nabla_x \cdot (c_1 \rho_W \Omega) = 0,  \quad \forall W \in {\mathbb R}, 
\label{eq:sm_smallhydro_rhoW} \\
&&\hspace{-1cm}
\rho \, \big( \partial_t \Omega + c_2 \, (\Omega \cdot \nabla_x)  \Omega - Y \Omega^\bot \big) + d \, P_{\Omega^\bot} \nabla_x \rho  = 0, 
\label{eq:sm_smallhydro_Omega} \\
&&\hspace{-1cm}
\rho(x,t) = \int_{W \in {\mathbb R}} \rho_W(x,t) \, dW, \qquad (\rho Y)(x,t) = \int_{W \in {\mathbb R}} \rho_W(x,t) \, W \, dW. 
\label{eq:sm_smallhydro_moments} 
\end{eqnarray}
The constants $c_1$, $c_2$ are respectively given by formulas (\ref{eq:sm_smallkin_cur_equi}) and (\ref{eq:sm_express_c2}) in Appendix \ref{sec:small_proof} below. 
\label{thm:small_hydro}
\end{theorem}

The proof of Theorem \ref{thm:small_hydro} is developed in Appendix \ref{sec:small_proof}. We now discuss the significance of the results. Eq. (\ref{eq:sm_smallhydro_rhoW}) is a continuity equation for the density of particles of given proper angular velocity $W$. Indeed, since the interactions do not modify the proper angular velocities of the particles, we must have an equation expressing the conservation of particles for each of these velocities $W$. However, the self alignment force modifies the actual direction of motion $v$ of the particles. This interaction couples particles with different proper angular velocities. Therefore, the mean direction of motion $\Omega$ is common to all particles (and consequently, does not depend on $W$) and obeys a balance equation which bears similarities with the gas dynamics momentum conservation equations. 

Since $c_1$ and $\Omega$ do not depend on $W$, the dependence on $W$ in eq. (\ref{eq:sm_smallhydro_rhoW}) can be integrated out, which leads to the following system of equations:  
\begin{eqnarray}
&&\hspace{-1cm}
\partial_t \rho + \nabla_x \cdot (c_1 \rho \Omega) = 0.  
\label{eq:sm_smallhydro_rho} \\
&&\hspace{-1cm}
\partial_t (\rho Y) + \nabla_x \cdot (c_1 \rho Y \Omega) = 0.  
\label{eq:sm_smallhydro_Y} \\
&&\hspace{-1cm}
\rho \, \big( \partial_t \Omega + c_2 \, (\Omega \cdot \nabla_x)  \Omega - Y \Omega^\bot ) + d \, P_{\Omega^\bot} \nabla_x \rho  = 0, 
\label{eq:sm_smallhydro_Omega_bis}
\end{eqnarray}
Therefore, $\rho$, $Y$ and $\Omega$ can first be computed by solving the system (\ref{eq:sm_smallhydro_rho}), (\ref{eq:sm_smallhydro_Y}), (\ref{eq:sm_smallhydro_Omega_bis}). Once $\Omega$ is known, eq. (\ref{eq:sm_smallhydro_rhoW}) is just a transport equation with given coefficients, which can be easily integrated (provided that the vector field $\Omega$ is smooth). Eq. (\ref{eq:sm_smallhydro_rho}) expresses the conservation of the total density of particles (i.e. integrated with respect to $W \in {\mathbb R}$), while (\ref{eq:sm_smallhydro_Y}) expresses the conservation of the 'angular momentum density' $\rho Y$. Using the mass conservation eq. (\ref{eq:sm_smallhydro_rho}), Eq. (\ref{eq:sm_smallhydro_Y}) can be rewritten (for smooth solutions) as a transport equation for the 'average rotation velocity' $Y$: 
\begin{eqnarray}
&&\hspace{-1cm}
\partial_t Y + c_1 \, \rho \, \Omega \cdot \nabla_x \, Y = 0,  
\label{eq:sm_smallhydro_Y_2} 
\end{eqnarray}
which simply expresses that the average rotation velocity $Y$ is convected at the flow speed~$c_1 \Omega$. 

Suppose that $Y_{t=0}=0$. Then, by (\ref{eq:sm_smallhydro_Y_2}), we have $Y(x,t) \equiv 0$ for all $(x,t) \in {\mathbb R}^2 \times [0,\infty)$. In this case, the system reduces to the following one: 
\begin{eqnarray}
&&\hspace{-1cm}
\partial_t \rho + \nabla_x \cdot (c_1 \rho \Omega) = 0.  
\label{eq:sm_smallhydro_rho_2} \\
&&\hspace{-1cm}
\rho \, \big( \partial_t \Omega + c_2 \, (\Omega \cdot \nabla_x)  \Omega) + d \, P_{\Omega^\bot} \nabla_x \rho  = 0, 
\label{eq:sm_smallhydro_Omega_bis_2}
\end{eqnarray}
which has been studied in earlier work \cite{Degond_etal_Schwartz13, Degond_Motsch_M3AS08, Degond_etal_MAA13, Motsch_Navoret_MMS11}. This system is referred to as the {\bf 'Self-Organized Hydrodynamics' (SOH)}. As mentioned above, it bears similarities with the isothermal compressible gas dynamics equations, but differs from it by several aspects, which have been developed in earlier work (see e.g. the review \cite{Degond_etal_Schwartz13}). These are: 
\begin{itemize}
\item[(i)] The mean velocity $\Omega$ is a vector of unit norm (specifically, it is the direction of the mean velocity rather than the mean velocity itself). 
\item[(ii)] The projection operator $P_{\Omega^\bot}$ multiplies the pressure gradient term $d \nabla_x \rho$. It is required to maintain the constraint that $|\Omega|=1$. Indeed, multiplying scalarly (\ref{eq:sm_smallhydro_Omega}) by $\Omega$, we realize that
$(\partial_t + c_2 \Omega \cdot \nabla_x ) |\Omega|^2 = 0$. Therefore, if $|\Omega| = 1$ uniformly at $t=0$, it stays of unit norm at all times. The projection operator $P_{\Omega^\bot}$ brings a non-conservative term in this equation. Hence, (\ref{eq:sm_smallhydro_Omega}) is not a conservation equation: it does not express any momentum balance. 
\item[(iii)] The convection velocity of $\Omega$ is $c_2$ and is different from the convection velocity $c_1$ of $\rho$. In classical fluids, these two velocities are equal. This results from the Galilean invariance of the gas dynamics system. Here, the system is not Galilean invariant (the velocities are normalized to $1$: this property is not invariant under Galilean transforms) and consequently, these two convection velocities may differ. The loss of Galilean invariance by fluid models of self-propelled particles has been noted earlier in \cite{Toner_etal_AnnPhys05, Tu_etal_PRL98}. As a consequence, in such fluids, the propagation of sound is anisotropic \cite{Tu_etal_PRL98}.  
\end{itemize}

The model with non-vanishing average rotation velocity (\ref{eq:sm_smallhydro_rho})-(\ref{eq:sm_smallhydro_Omega_bis}) appears as an enrichment of the standard SOH model by the following two aspects: 
\begin{itemize}
\item[(i)] An additional term, namely $-Y \Omega^\bot $, is present in the velocity evolution eq. (\ref{eq:sm_smallhydro_Omega_bis}). This term expresses how the self-rotation of the particles influences the evolution of the mean velocity direction $\Omega$. Quite naturally, it depends on the angular momentum density $\rho Y$ which provides the contribution of the proper angular rotation of the particles to the evolution of the mean velocity.
\item[(ii)] An additional equation, namely (\ref{eq:sm_smallhydro_Y}) (or (\ref{eq:sm_smallhydro_Y_2}) in non-conservative form) is added to the system. It shows that the average angular velocity $Y$ is passively transported by the flow velocity $c_1 \Omega$. 
\end{itemize}

This model will be referred to as the {\bf 'Self-Organized Hydrodynamics with proper Rotation (small angular velocity case)' or SOHR-S}. 

In \cite{Degond_Motsch_M3AS08, Degond_etal_MAA13}, it is shown that the SOH model (\ref{eq:sm_smallhydro_rho_2}), (\ref{eq:sm_smallhydro_Omega_bis_2}) is hyperbolic. Its two eigenvalues evaluated at a state $(\rho, \Omega)$ are given by 
\begin{eqnarray} 
\gamma_\pm = \frac{1}{2} \big[ (c_1+c_2) \cos \theta \pm \big( (c_2-c_1)^2 \cos^2 \theta + 4 d \sin^2 \theta
\big)^{1/2} 
\big], 
\label{eq:SOH_eigenvalues} 
\end{eqnarray}
where $\Omega = (\cos \theta, \sin \theta)^T$ and the exponent 'T' denotes the transpose of a vector. Apart from additional zero-th order terms, the SOHR-S model is derived from the SOH model by the addition of the convection equation (\ref{eq:sm_smallhydro_Y_2}) with convection velocity $c_1 \Omega$. It is a hyperbolic problem, whose eigenvalues consist of the two eigenvalues  (\ref{eq:SOH_eigenvalues}) of the SOH model on the one hand, and of the convection speed $c_1 \cos \theta$ of the additional equation (\ref{eq:sm_smallhydro_Y_2}) on the other hand. These three eigenvalues are real and distinct, except in the case $\theta = 0$. Therefore, the problem is strictly hyperbolic in most of the domain where the state variables $(\rho, \Omega)$ are defined. This gives a good indication that at least local well-posedness of the SOHR-S model can be achieved.

%%%%%%%%%%%%%%%%%%%%%%%%%%%%%%%%%%%%%%%%%%%%%%%%%%%%%%%%%%%%%%%%%%%%%%%%%%%%%%%%%%%%%%%%%%%%%%%%
%%%%%%%%%%%%%%%%%%%%%%%%%%%%%%%%%%%%%%%%%%%%%%%%%%%%%%%%%%%%%%%%%%%%%%%%%%%%%%%%%%%%%%%%%%%%%%%%
\setcounter{equation}{0}
\section{Large angular velocities}
\label{sec:large}

Now, we investigate the case of large proper angular velocities, i.e. $\eta = {\mathcal O}(\varepsilon)$. We make $\eta = \varepsilon$ for simplicity. The problem is now written: 
\begin{eqnarray}
&&\hspace{-1cm}
\partial_t f^\varepsilon + \nabla_x \cdot (v f^\varepsilon) = \frac{1}{\varepsilon} \big( - \nabla_v \cdot (P_{v^\bot} \omega_{f^\varepsilon} (W) \, f^\varepsilon) - W \nabla_v \cdot (v^\bot f^\varepsilon) + d \Delta_v f^\varepsilon  \big), 
\label{eq:la_kin_f_0}
\end{eqnarray}
Now, by contrast to the small angular velocity case (section \ref{sec:small}), we abandon the hypothesis that $\omega_f = \Omega_f$, where we recall that (see \ref{eq:sm_smallkin_Omega}):
\begin{eqnarray}
&&\hspace{-1cm}
\Omega_f(x,t) = \frac{J_f(x,t)}{|J_f(x,t)|} ,  \quad J_f(x,t) = \int_{(v,W) \in {\mathbb S}^1 \times {\mathbb R}}  f(x,v,W,t) \, v \, dv \, dW. 
\label{eq:la_kin_Omega}
\end{eqnarray}
Indeed, the agents' proper angular velocity being large, it influences their evaluation of the turning velocity that produces alignment with their neighbors. According to the situation, the proper angular velocity goes along or against the turning direction they want to achieve. Therefore, the agents need to compensate for it by realizing smaller or larger turning speeds. This results in a prescription for $\omega_f$ which is different from $\Omega_f$ and which requires $\omega_f$ to be dependent of $W$, as indicated in (\ref{eq:la_kin_f_0}).

The precise determination of $\omega_f$ requires several steps. Before going into this determination, we write (\ref{eq:la_kin_f_0}) as follows: 
\begin{eqnarray}
&&\hspace{-1cm}
\partial_t f^\varepsilon + \nabla_x \cdot (v f^\varepsilon) = \frac{1}{\varepsilon} \tilde Q(f^\varepsilon),  
\label{eq:la_kin_f} 
\end{eqnarray}
where $\tilde Q(f)$ is a new collision operator given by: 
\begin{eqnarray}
&&\hspace{-1cm}
\tilde Q(f) =  \tilde {\mathcal Q}_{\omega_f} (f),  
\label{eq:la_kin_coll_op} 
\end{eqnarray}
where $\omega_f$: $W \in {\mathbb R} \to \omega_f(W) \in {\mathbb S}^1$ is the function to be determined below and where, for any given function $\omega$: $W \in {\mathbb R} \to \omega(W) \in {\mathbb S}^1$, we define:
\begin{eqnarray}
&&\hspace{-1cm}
\tilde {\mathcal Q}_{\omega} (f) := - \nabla_v \cdot (F_\omega \, f) + d \Delta_v f  , 
\label{eq:la_kin_coll_op_omega} \\
&&\hspace{-1cm}
F_\omega (v,W) : = P_{v^\bot} \omega(W) + W v^\bot.  
\label{eq:la_F_omega} 
\end{eqnarray}

\bigskip
We define $\tilde {\mathcal E}_\omega$, the  set of equilibria of $\tilde {\mathcal Q}_\omega$, as follows: 

\begin{definition} 
Let $\omega$: $W \in {\mathbb R} \to \omega(W) \in {\mathbb S}^1$ be given. The set $\tilde {\mathcal E}_\omega$ of equilibria of $\tilde {\mathcal Q}_\omega$ is defined by 
$$\tilde {\mathcal E}_\omega = \big\{ f \in L^1({\mathbb R}, C^2({\mathbb S}^1)) \, \, | \, \, f \geq 0 \, \,  \mbox{ and } \, \, \tilde {\mathcal Q}_{\omega} (f) = 0 \big\} .$$ 
\label{def:la_equi}
\end{definition}

\noindent
To determine $\tilde {\mathcal E}_\omega$, we first define what are the analogs of the von Mises-Fisher distributions in the present case. The existence of these objects requires the following preliminary lemma.

\begin{lemma}
Let $W \in {\mathbb R}$ be given. There exists a unique $2\pi$-periodic solution of the following problem: 
\begin{eqnarray}
&&\hspace{-1cm}
\Phi_W''(\theta) - \frac{1}{d} \big( (W - \sin \theta) \Phi_W \big)'(\theta) = 0, \qquad \int_0^{2 \pi} \Phi_W(\theta) \, d \theta = 1,
\label{eq:la_def_Phi}
\end{eqnarray}
where the primes denote derivatives with respect to $\theta$. We denote by $\Phi_W$ this unique solution. It is positive and it belongs to $C^\infty({\mathbb S}^1)$. 
\label{lem:la_exist_Phi}
\end{lemma} 

\noindent
We can now define the analogs of the von Mises-Fisher distributions:

\begin{definition}
Let $\Phi_W$ be the function defined in the previous lemma. Let $\omega$: $W \in {\mathbb R} \to \omega(W) \in {\mathbb S}^1$ be fixed. We define $\tilde M_\omega$ such that:
\begin{eqnarray}
&&\hspace{-1cm}
\tilde M_\omega (v,W) = \Phi_W(\theta), \quad \mbox{ with } \quad \theta = \widehat{(\omega(W),v)}. 
\label{eq:la_def_Mom}
\end{eqnarray}
For any given $W \in {\mathbb R}$, the distribution $\tilde M_\omega (v,W) \, dv$ is a probability measure on ${\mathbb S}^1$. We refer to it as the 'Generalized von Mises-Fisher' (GVM) distribution. 
\label{def:la_VMF_generalise}
\end{definition}

\noindent
Thanks to the definition of $\tilde M_\omega$, we can describe the set~$\tilde {\mathcal E}_\omega$, as done in the lemma just below:

\begin{lemma}
The set $\tilde {\mathcal E}_\omega$ is the set of all functions of the form 
\begin{eqnarray}
&&\hspace{-1cm}
(v,W) \mapsto \rho_W \tilde M_\omega(v,W), 
\label{eq:la_def_equ_tildeQom}
\end{eqnarray}
where the function $W \mapsto \rho_W \in {\mathbb R}_+$ is arbitrary in $L^1({\mathbb R})$. 
\label{lem:la_equilibria_Qom}
\end{lemma}

\medskip
We now define the direction of the flux associated to a GVM equilibrium $\tilde M_\omega$: 

\begin{definition}
Given $\omega$: $W \in {\mathbb R} \to \omega(W) \in {\mathbb S}^1$ and $W \in {\mathbb R}$, we define:
\begin{eqnarray}
&&\hspace{-1cm}
u_\omega(W) = \int_{v \in {\mathbb S}^1} \tilde M_\omega(v,W) \, v \, dv, 
\label{eq:la_def_u_tildeom_1} \\
&&\hspace{-1cm}
\Psi_\omega (W) = \frac{u_\omega(W)}{|u_\omega(W)|}, \qquad \tilde c_{1 \, \omega}(W) = |u_\omega(W)|. 
\label{eq:la_def_u_tildeom_2}
\end{eqnarray}
We have $u_\omega(W) \in {\mathbb R}^2$, $\Psi_\omega(W) \in {\mathbb S}^1$. The vector $\Psi_\omega (W)$ is the direction of the GVM $\tilde M_\omega$ for a given angular rotation $W$ and the real number $\tilde c_{1 \, \omega}(W)$ is its order parameter for this angular rotation (again, we have $0 \leq \tilde c_{1 \, \omega}(W) \leq 1$).
\label{def:la_fluxes}
\end{definition} 

\noindent
We stress the fact that $\Psi_\omega (W)  \not = \omega$ unless $W=0$. This is in marked contrast with the small angular velocity case, where the direction of the VMF distribution $M_\Omega$ is precisely equal to $\Omega$. This is the reason why, in the present case, we cannot set $\omega_f = \Omega_f$ (we recall that, for a given distribution $f$, the direction of the local flux $\Omega_f$ is given by (\ref{eq:la_kin_Omega})). Indeed, the 'consistency relation' that the direction of the equilibrium $\tilde M_{\Omega_f}$ should be $\Omega_f$ would not be realized. So, for a given local velocity direction $\Omega_f$, we will have to look for~$\omega_f(W)$ which realizes that, for any value of the angular velocity $W$, the direction of the associated GVM $\tilde M_{\omega_f}$ is equal to $\Omega_f$, i.e. $\Psi_{\omega_f} (W) = \Omega_f$. From the present considerations, we will have $\omega_f(W) \not = \Omega_f$, unless $W=0$. To do this, we have a first steps to go. For later usage, we first state the following auxiliary lemma:

\begin{lemma}
The real number $\tilde c_{1 \, \omega}(W)$ does not depend on $\omega$ and is denoted below $\tilde c_1(W)$. 
\label{lem:la_c1_indep_om}
\end{lemma}

\noindent
\noindent
Now, as developed above, for a fixed direction $\Omega$, we are interested in finding a function $\omega$ such that the direction $\Psi_\omega (W)$ of $\tilde M_\omega$ coincides with $\Omega$, for all angular velocities $W$. Such an $\omega$ can be uniquely determined, as the lemma below shows. 

\begin{lemma}
Let $\Omega \in {\mathbb S}^1$. Then, the equation $\Psi_\omega (W) = \Omega$, $\forall W \in {\mathbb R}$, determines a unique function $\omega$: $W \in {\mathbb R} \mapsto \omega(W) \in {\mathbb S}^1$. We denote this unique solution by $\omega_\Omega$. By definition, we have 
\begin{eqnarray}
&&\hspace{-1cm}
\Psi_{\omega_\Omega}(W) = \Omega, \quad \forall W \in {\mathbb R}.  
\label{eq:la_tildeom=Om}
\end{eqnarray}
\label{lem:la_exist_omOM}
\end{lemma}

\noindent
Now, as explained above, we define $\omega_f$  such that the direction $\Psi_{\omega_f}(W)$ of the associated GVM $\tilde M_{\omega_f}$ coincides with the local flux $\Omega_f$ for all values of the angular velocity $W \in {\mathbb R}$. This leads to the following definition:

\begin{definition}
Given a distribution function $f(\omega,W)$, we define $\omega_f$ by:
\begin{eqnarray}
&&\hspace{-1cm}
\omega_f = \omega_{\Omega_f},  
\label{eq:la_def_omf}
\end{eqnarray}
i.e. we have, 
\begin{eqnarray}
&&\hspace{-1cm}
\Psi_{\omega_f}(W) = \Omega_f, \quad \forall W \in {\mathbb R}.  
\label{eq:la_tildeom=Omf}
\end{eqnarray}
\label{def:la_def_omf}
\end{definition}

\noindent
The proofs of Lemmas \ref{lem:la_exist_Phi} to \ref{lem:la_exist_omOM} are given in appendix \ref{sec:large_proof}.  

\medskip
We now comment on the rationale for the definition of $\omega_f$. The Individual-Based model whose mean-field limit gives rise to the kinetic equation (\ref{eq:la_kin_f_0}) is obviously as follows (with the notations of section \ref{sec:IBM}): 
\begin{eqnarray}
&&\hspace{-1cm}
\frac{dX_k}{dt} = c V_k, 
\label{eq:la_IBM_x}\\
&&\hspace{-1cm}
dV_k =P_{V_k^\bot} \circ (\omega_{\bar V_k}(W_k) \, dt  + \sqrt{2D} \, dB_t) + W_k \, V_k^\bot \, dt. 
\label{eq:la_IBM_omega}
\end{eqnarray}
Here, $\omega_{\bar V_k}$ is the function defined by (\ref{eq:la_tildeom=Om}) where $\Omega$ is substituted by $\bar V_k$. The goal is to model a relaxation dynamic towards the local mean alignment direction, i.e. the direction $\bar V_k$. For this to happen, the particles have to choose the alignment force $P_{V_k^\bot} \omega_{\bar V_k}$ in a proper way. Because of the self-rotation velocity $W_k$, this force cannot be equal to $P_{V_k^\bot} \bar V_k$. Indeed, if this were the case, the relaxation force would vanish when $V_k = \bar V_k$ and could not compensate for the self-rotation force $W_k \, V_k^\bot$. In the absence of noise, the alignment force $P_{V_k^\bot} \omega_{\bar V_k}$ which compensates for self rotation is given by 
$$P_{V_k^\bot} \omega_{\bar V_k} + W_k \, V_k^\bot = P_{V_k^\bot} \bar V_k, $$ 
(which has a solution $\omega_{\bar V_k}$ only in a finite range of values of $W_k$). In the presence of noise, the alignment force which compensates for self-rotation cannot be computed a priori. To propose an explicit value of $\omega_{\bar V_k}$, we assume that the distribution of the particles in $(v,W)$-space is locally at equilibrium, i.e. is a GVM distribution $\tilde M_{\omega_{\bar V_k}}$. Then, the alignment force $P_{V_k^\bot} \omega_{\bar V_k}$ vanishes when $V_k$ is equal to $\omega_{\bar V_k}$, i.e.  when $V_k$ coincides with the direction $\omega_{\bar V_k}$ such that there is no action on the particles when they are distributed according to a GVM. Indeed, when $V_k = \omega_{\bar V_k}$, the right-hand side of (\ref{eq:la_IBM_omega}) is zero on the average in the sense that the associated Fokker-Planck operator resulting from applying the Ito formula to (\ref{eq:la_IBM_omega}) vanishes (which is what saying that the particle distribution is a GVM means). This means that the relaxation has been achieved 'statistically'. Once translated in the mean-field framework of (\ref{eq:la_kin_f_0}), this leads to our definition (\ref{eq:la_tildeom=Omf}). 

Obviously, the use of the equilibrium to compute $\omega_f$ restricts the applicability of this model to a situation close to such an equilibrium. Since the goal is precisely to explore the hydrodynamic regime which prevails in such situations of closeness to equilibrium, this approach is still consistent. Another question is about the likeliness that agents are able to perform such a complicated computation. However, we can think that this dynamic is a simple outcome of collisions between the particles. Imagine a set of self-rotating robots with elongated shapes. The volume-exclusion interaction between elongated self-propelled objects through hard-core collisions result in an alignment dynamic, as already shown in e.g.  \cite{Baskaran_Marchetti_PRE08, Czirok_Vicsek_PhysicaA00, Henkes_etal_PRE11, Peruani_etal_PRE06}. Therefore, the 'computation' of the magnitude of the self-alignment force may be just an outcome of an equilibration between the self rotation force and the pressure exerted by the neighboring agents through the collisions.

\bigskip
The goal is now to investigate the limit $\varepsilon \to 0$ of the solution of (\ref{eq:la_kin_f}). More precisely, we show the: 

\begin{theorem}
Let $f^\varepsilon$ be a solution of (\ref{eq:la_kin_f_0}) with $\omega_f$ given by (\ref{eq:la_def_omf}). We assume that the limit $f^0 = \lim_{\varepsilon \to 0} f^\varepsilon$ exists and that the convergence is as regular as needed. Then, we have 
\begin{equation} 
f^0 (x,v, W,t) = \rho_W(x,t) \, \tilde M_{\omega_{\Omega(x,t)}}(v,w). 
\label{eq:la_smallkin_f0}
\end{equation}
where, for any $(x,t)$, the function $W \in {\mathbb R}  \to \rho_W(x,t) \in {\mathbb R}$ belongs to $L^1({\mathbb R})$ and the vector $\Omega(x,t)$ belongs to ${\mathbb S}^1$. The functions $\rho_W(x,t)$ and $\Omega(x,t)$ satisfy the following system of hydrodynamic equations: 
\begin{eqnarray}
&&\hspace{-1cm}
\partial_t \rho_W + \nabla_x \cdot (\tilde c_1 \rho_W \Omega) = 0,  \quad \forall W \in {\mathbb R}, 
\label{eq:la_smallhydro_rhoW} \\
&&\hspace{-1cm}
m_1[\rho_W] \, \Omega_t + m_2[\rho_W] \, (\Omega \cdot \nabla_x) \Omega + m_3[\rho_W] \,  (\Omega^\bot \cdot \nabla_x) \Omega  \nonumber \\
&&\hspace{0cm}
+ \Omega^\bot \, \big(\, m_4[\rho_W] \, (\nabla_x \cdot \Omega) + (\Omega^\bot \cdot \nabla_x)  \, m_5[\rho_W]  + (\Omega \cdot \nabla_x) \, m_6[\rho_W] \, 
\big) = 0, 
\label{eq:la_smallhydro_Omega} 
\end{eqnarray}
where $m_1[\rho_W]$, \ldots, $m_6[\rho_W]$ are moments of $\rho_W$ given by formulas (\ref{eq:la_mom_rhoW}) in Appendix \ref{sec:large_proof} below. 
\label{thm:la_hydro}
\end{theorem}

\medskip
Eq. (\ref{eq:la_smallhydro_rhoW}) expresses the conservation of particles of given angular momentum $W$, exactly in the same way as in the small angular velocity case (see Eq. (\ref{eq:sm_smallhydro_rhoW})). The velocity evolution eq. (\ref{eq:la_smallhydro_Omega}) has also a similar structure (see Eq. (\ref{eq:sm_smallhydro_Omega})) but contains more terms. The analog terms to those of (\ref{eq:sm_smallhydro_Omega}) are the first term (corresponding to the first term of (\ref{eq:sm_smallhydro_Omega})), the second one (corresponding to the second term of (\ref{eq:sm_smallhydro_Omega})) and the fifth one (corresponding to the fourth term of (\ref{eq:sm_smallhydro_Omega})). The difference is the replacement of $\rho$, which appears in the three terms of  (\ref{eq:sm_smallhydro_Omega}) by three different moments of $\rho_W$. This is a consequence of the dependence of the GVM $\tilde M_{\omega_\Omega}$ and the GCI $\tilde \chi_\Omega$ (which will be found in section \ref{subsec:la_GCI}) on the angular velocity $W$. There was no such dependence of the VMF $M_\Omega$ and of the GCI $\chi_\Omega$ in the small angular velocity case. 

The third term of (\ref{eq:sm_smallhydro_Omega}) which originated from the particle self-rotation disappears in the large angular velocity case investigated here, but three new terms appear. The third term of (\ref{eq:la_smallhydro_Omega}) describes transport in the direction perpendicular to the mean velocity $\Omega$. The direction of transport is determined by the sign of $m_3$. The fourth term is a contribution of the compressibility of the velocity field to its transport: regions of compression or rarefaction induce rotation of the velocity field in one direction or the other one according to the sign of $m_4$. Finally, the sixth term is an off-diagonal term in the pressure tensor, where gradients of the moment $m_6$ of the density distribution $\rho_W$ induce rotation of the velocity field. All these three terms obviously translate the average influence of the individual particle self-rotation.

By analogy with the previous model, this model will be referred to as the {\bf 'Self-Organized Hydrodynamics with proper Rotation (Large angular velocity case)' or \mbox{SOHR-L}}. 

The proof of Theorem \ref{thm:la_hydro} follows the same structure as the small angular velocity case. We start with the definition of the equilibria, followed by the determination of the GCI. We end up with the convergence $\varepsilon \to 0$.

%%%%%%%%%%%%%%%%%%%%%%%%%%%%%%%%%%%%%%%%%%%%%%%%%%%%%%%%%%%%%%%%%%%%%%%%%%%%%%%%%%%%%%%%%%%%%%%%
%%%%%%%%%%%%%%%%%%%%%%%%%%%%%%%%%%%%%%%%%%%%%%%%%%%%%%%%%%%%%%%%%%%%%%%%%%%%%%%%%%%%%%%%%%%%%%%%
%%%%%%%%%%%%%%%%%%%%%%%%%%%%%%%%%%%%%%%%%%%%%%%%%%%%%%%%%%%%%%%%%%%%%%%%%%%%%%%%%%%%%%%%%%%%%%%%
%%%%%%%%%%%%%%%%%%%%%%%%%%%%%%%%%%%%%%%%%%%%%%%%%%%%%%%%%%%%%%%%%%%%%%%%%%%%%%%%%%%%%%%%%%%%%%%%
\setcounter{equation}{0}
\section{Properties of the SOHR-L hydrodynamic model}
\label{sec:large_properties}

We investigate some properties of the SOHR-L hydrodynamic model (\ref{eq:la_smallhydro_rhoW}), (\ref{eq:la_smallhydro_Omega}). In a first section, we study its linearized stability about a uniform steady-state. For the sake of simplicity, we restrict ourselves to the case where the unperturbed density distribution $\rho_W$ is even in $W$ (which means that there are as many particles rotating in the clockwise direction with angular speed $|W|$ as particles rotating counter-clockwise with the same angular speed). In this case, we prove the linearized stability of the model. This is a good indication of the well-posedness of the SOHR-L model in this case, although a rigorous proof of this fact is still lacking. The investigation of the linearized stability of the SOHR-L model in the general case is deferred to future work. 

In a second section, we investigate the asymptotics of the SOHR-L model  (as well as that of the SOHR-S model) when $W$ is small. We show that both models reduce to the SOH model (\ref{eq:sm_smallhydro_rho_2}), (\ref{eq:sm_smallhydro_Omega_bis_2}) in this limit, but with different coefficients. We also establish the asymptotics of the SOHR-L model to second order for small $W$ and compare the resulting model to the SOHR-S model.

%%%%%%%%%%%%%%%%%%%%%%%%%%%%%%%%%%%%%%%%%%%%%%%%%%%%%%%%%%%%%%%%%%%%%%%%%%%%%%%%%%%%%%%%%%%%%%%%
%%%%%%%%%%%%%%%%%%%%%%%%%%%%%%%%%%%%%%%%%%%%%%%%%%%%%%%%%%%%%%%%%%%%%%%%%%%%%%%%%%%%%%%%%%%%%%%%
\subsection{Linearized stability of the SOHR-L system}
\label{subsec:la_linearized}

We first consider a pair $(\rho_{0W}, \Omega_0)$ such that (i) $\rho_{0W}$ and $\Omega_0$ are independent of $x$, (ii) the function $W \in {\mathbb R} \mapsto \rho_{0W}$ belongs to $L^1({\mathbb R})$, (iii) $\rho_{0W} \geq 0$, (iv) all the moments $m_K[\rho_{0W}]$, $k=1, \ldots, 6$ exist,  (v) $|\Omega_0| = 1$. Such a pair $(\rho_{0W}, \Omega_0)$ is a steady-state of the SOHR-L system (\ref{eq:la_smallhydro_rhoW}), (\ref{eq:la_smallhydro_Omega}). The goal of this section is to study the linearized stability of the SOHR-L system about such a uniform steady-state. 

We linearize the system. We introduce a small parameter $\delta \ll 1$ and look for solutions such that 
\begin{eqnarray}
\rho_W (x,t)= \rho_{0W} + \delta \, \rho_{1W}(x,t) + {\mathcal O}(\delta^2), \qquad \Omega(x,t) = \Omega_0 + \delta \, \Omega_1 + {\mathcal O}(\delta^2).
\label{eq:la_lin_exp}
\end{eqnarray}
The constraint $|\Omega(x,t)| = 1$ translates into the constraint
\begin{eqnarray}
\Omega_0 \cdot \Omega_1 = 0. 
\label{eq:la_lin_constraint}
\end{eqnarray}
The linearized system obtained by introducing (\ref{eq:la_lin_exp}) into (\ref{eq:la_smallhydro_rhoW}), (\ref{eq:la_smallhydro_Omega}) and neglecting terms of order ${\mathcal O}(\delta^2)$ reads as follows: 
\begin{eqnarray}
&&\hspace{-1cm}
\partial_t \rho_{W} + \tilde c_1 \, \Omega_0 \cdot \nabla_x \rho_{W} + \tilde c_1 \, \rho_{0W} \, \nabla_x \cdot \Omega= 0,  \quad \forall W \in {\mathbb R}, 
\label{eq:la_lin_hydro_rhoW} \\
&&\hspace{-1cm}
m_1 \, \Omega_t + m_2 \, (\Omega_0 \cdot \nabla_x) \Omega + m_3 \,  (\Omega_0^\bot \cdot \nabla_x) \Omega  \nonumber \\
&&\hspace{0cm}
+ \Omega_0^\bot \, \big(\, m_4 \, (\nabla_x \cdot \Omega) +  (\Omega_0^\bot \cdot \nabla_x) \, m_5[\rho_{W}]  + (\Omega_0 \cdot \nabla_x) \, m_6[\rho_{W}] \, 
\big) = 0, 
\label{eq:la_lin_hydro_Omega} 
\end{eqnarray}
where $m_1$, \ldots $m_4$ are evaluated on $\rho_{0W}$ except otherwise stated and where the index '$1$' on the perturbation is omitted for the sake of clarity. Next, we consider plane-wave solutions:
\begin{eqnarray}
\rho_W (x,t) = \tilde \rho_W \, e^{i(x \cdot \xi - \mu t)}, \qquad \Omega = \tilde \Omega \, e^{i(x \cdot \xi - \mu t)},
\label{eq:la_lin_plane_wave} 
\end{eqnarray}
where $\tilde \rho_W$, $\tilde \Omega$ are the wave amplitudes, $\xi  \in {\mathbb R}$ is the wave-number and $\mu \in {\mathbb C}$ is the frequency. Here, $x \in {\mathbb R}$ is a one-dimensional spatial variable, corresponding to the direction of propagation of the plane wave. Indeed, the SOHR-L being invariant under rotations, the plane-wave analysis is independent of the choice of the direction of propagation. 
We let $\Omega_0 = (\cos \theta, \sin \theta)$. 
The constraint (\ref{eq:la_lin_constraint}) translates into $\Omega_0 \cdot \tilde \Omega = 0$, i.e. $\tilde \Omega = \tilde \sigma ( - \sin \theta, \cos \theta)$ with $\tilde \sigma \in {\mathbb R}$. Inserting (\ref{eq:la_lin_plane_wave}) into (\ref{eq:la_lin_hydro_rhoW}), (\ref{eq:la_lin_hydro_Omega}), we get (again, omitting the tildes on $\rho_W$ and $\Omega$ for the sake of clarity):
\begin{eqnarray}
&&\hspace{-1cm}
\big(  - \mu + \tilde c_1 \, \xi \, \cos \theta \, \big) \, \rho_{W} - \tilde c_1 \, \rho_{0W} \, \xi \, \sin \theta \, \sigma = 0,  \quad \forall W \in {\mathbb R}, 
\label{eq:la_lin_plane_rhoW} \\
&&\hspace{-1cm}
\big(  - \mu \, m_1 + m_2 \, \xi \, \cos \theta - (m_3 + m_4) \,  \xi \, \sin \theta \, \big) \, \sigma \nonumber \\
&&\hspace{2cm}
- \xi \, \sin \theta \, m_5[\rho_{W}]  + \xi \, \cos \theta \,  m_6[\rho_{W}] = 0. 
\label{eq:la_lin_plane_Omega} 
\end{eqnarray}
From (\ref{eq:la_lin_plane_rhoW}), we get:
\begin{eqnarray}
&&\hspace{-1cm}
\rho_{W} = \frac{\tilde c_1 \, \rho_{0W}}{- \mu + \tilde c_1 \, \xi \, \cos \theta} \, \xi \, \sin \theta \, \, \sigma ,  \quad \forall W \in {\mathbb R}. 
\label{eq:la_lin_plane_rhoW_2} 
\end{eqnarray} 
Therefore, 
\begin{eqnarray}
&&\hspace{-1cm}
m_k[\rho_W] = m_k \Big[  \frac{\tilde c_1 \, \rho_{0W}}{- \mu + \tilde c_1 \, \xi \, \cos \theta} \Big] \, \xi \, \sin \theta \,\, \sigma, \quad k=5, \, 6. 
\label{eq:la_lin_plane_rhoW_3} 
\end{eqnarray} 
Inserting (\ref{eq:la_lin_plane_rhoW_3} ) into (\ref{eq:la_lin_plane_Omega}), we get a non-trivial solution $\sigma$ if and only if the following dispersion relation is satisfied: 
\begin{eqnarray}
&&\hspace{-1cm}
- \mu \, m_1 + m_2 \, \xi \, \cos \theta - (m_3 + m_4) \,  \xi \, \sin \theta  \nonumber \\
&&\hspace{0cm}
- m_5 \Big[  \frac{\tilde c_1 \, \rho_{0W}}{- \mu + \tilde c_1 \, \xi \, \cos \theta} \Big] \xi^2 \, \sin^2 \theta  + m_6 \Big[  \frac{\tilde c_1 \, \rho_{0W}}{- \mu + \tilde c_1 \, \xi \, \cos \theta} \Big] \xi^2 \, \cos \theta \, \sin \theta  = 0. 
\label{eq:la_lin_plane_dispersion} 
\end{eqnarray}

Next, we seek some simplifications in the case where the function $W \in {\mathbb R} \mapsto \rho_{0W}$ is even. For this purpose, we will need the following lemma about the evenness/oddness of the coefficients $a_k$, $k=1,\ldots,6$ of the corresponding moments $m_k$. 

\begin{lemma}
(i) We have: 
\begin{eqnarray}
&&\hspace{-1cm}
\Phi_{-W}(\theta) = \Phi_W(-\theta), \qquad X_{-W}(\theta) = - X_W(- \theta),  
\label{eq:la_parity_phi_W}
\end{eqnarray}
where $\Phi_W$ is defined by (\ref{eq:la_def_Phi}) and $X_W$ by (\ref{eq:la_def_X}).

\medskip
\noindent
(ii) The following functions of $W$ are even: $\tilde c_1$, $\lambda$, $a_1$, $a_2$, $a_5$ (see (\ref{eq:la_def_u_tildeom_2}), (\ref{eq:la_def_lambda}), (\ref{eq:def_a1}), (\ref{eq:def_a2}), (\ref{eq:def_a5}) for the definitions of these functions).    

\medskip
\noindent
(iii) The following functions of $W$ are odd: $C$, $\psi$, $a_3$, $a_4$, $a_6$ (see (\ref{eq:la_Phi_deriv}), (\ref{eq:la_psi(W)_def}), (\ref{eq:def_a3}), (\ref{eq:def_a4}), (\ref{eq:def_a6}) for the definitions of these functions).   

\label{lem:la_even_a}
\end{lemma}

\medskip
\noindent
{\bf Proof.} (i) We form Eqs. (\ref{eq:la_def_Phi}) and (\ref{eq:la_def_X}) for $\Phi_W(-\theta)$ and $X_W(-\theta)$. By changing $W$ into $-W$, we recover the same equations for  $\Phi_{-W}(\theta)$ and $-X_{-W}(\theta)$  respectively, which shows (\ref{eq:la_parity_phi_W}). 

\noindent
(ii) and (iii) By (\ref{eq:la_def_u_tildeom_2}) $\tilde c_1$ is clearly even. By writing (\ref{eq:la_Phi_deriv}) at $- \theta$ and for $-W$ and using the first equation of (\ref{eq:la_parity_phi_W}), we get that $C$ is odd. Now, using the first equation of (\ref{eq:la_parity_phi_W}) into (\ref{eq:la_Psi(W)_deriv}) and changing $\theta$ into $-\theta$, we get that $\Psi_\omega(-W)$ is the symmetric of $\Psi_\omega(W)$ about the line spanned by $\omega$. As an immediate consequence, $\psi$ is even. Changing $\theta$ into $-\theta$ in (\ref{eq:la_def_lambda}) and using the first equation of (\ref{eq:la_parity_phi_W}), the evenness of $\tilde c_1$ and the oddness of $\psi$, we get that $\lambda$ is even. By similar considerations, we get that $a_1$, $a_2$, $a_5$
are even and $a_3$, $a_4$ and $a_6$ are odd. \endproof

\medskip
Now, we assume that $\rho_{0W}$ is even with respect to $W$. Then, $\frac{\tilde c_1 \, \rho_{0W}}{- \mu + \tilde c_1 \, \xi \, \cos \theta}$ is also even with respect to $W$. Therefore, the coefficients $m_3$, $m_4$ and $m_6 [  \frac{\tilde c_1 \, \rho_{0W}}{- \mu + \tilde c_1 \, \xi \, \cos \theta} ]$ vanish in (\ref{eq:la_lin_plane_dispersion}), as the result of the integration of an odd coefficient of $W$ against the even function $\rho_{0W}$. The resulting dispersion relation is written: 
\begin{eqnarray}
&&\hspace{-1cm}
- \mu \, m_1 + m_2 \, \xi \, \cos \theta  - m_5 \Big[  \frac{\tilde c_1 \, \rho_{0W}}{- \mu + \tilde c_1 \, \xi \, \cos \theta} \Big] \xi^2 \, \sin^2 \theta  = 0. 
\label{eq:la_lin_plane_dispersion_even} 
\end{eqnarray}
We now show that for all $\xi \in {\mathbb R}$ and $\theta \in [0,2\pi]$, the roots $\mu$ of (\ref{eq:la_lin_plane_dispersion_even}) can only be real, which proves the linearized stability of the system. Indeed, suppose that $\mu = \alpha + i \beta$ with $\alpha, \, \beta \in {\mathbb R}$, taking the imaginary part of (\ref{eq:la_lin_plane_dispersion_even}), we get 
\begin{eqnarray}
&&\hspace{-1cm}
- \beta \, m_1   - \beta m_5 \Big[  \frac{\tilde c_1 \, \rho_{0W}}{(- \alpha + \tilde c_1 \, \xi \, \cos \theta)^2 + \beta^2} \Big] \xi^2 \, \sin^2 \theta  = 0. 
\label{eq:la_lin_plane_dispersion_imag} 
\end{eqnarray}
If $\beta \not = 0$, we deduce from (\ref{eq:la_lin_plane_dispersion_imag}) that:
\begin{eqnarray}
&&\hspace{-1cm}
m_1   = - m_5 \Big[  \frac{\tilde c_1 \, \rho_{0W}}{(- \alpha + \tilde c_1 \, \xi \, \cos \theta)^2 + \beta^2} \Big] \xi^2 \, \sin^2 \theta . 
\label{eq:la_lin_plane_dispersion_imag_2} 
\end{eqnarray}
Numerically, we realize below that the coefficients $a_1$ and $a_5$ are non negative (see Appendix \ref{sec:la_graphics}). Since we know that $\tilde c_1$ is also non negative, (\ref{eq:la_lin_plane_dispersion_imag_2}) cannot have any root. Thus, $\beta = 0$. We summarize this in the following result: 

\begin{proposition}
Consider a uniform steady-state $(\rho_{0W},\Omega_0)$ where $\rho_{0W} \geq 0$ is such that $(1+|W|)^k \rho_{0W}$ is integrable for $k$ large enough, and where $|\Omega_0|=1$. We assume that the coefficient $a_1$ and $a_5$ given by (\ref{eq:def_a1}) and (\ref{eq:def_a1}) are positive (and this is verified numerically). If $\rho_{0W}$ is even with respect to $W$, the SOHR-L model (\ref{eq:la_smallhydro_rhoW}), (\ref{eq:la_smallhydro_Omega}) is linearly stable about this steady-state. 
\label{prop:la_lin_stab}
\end{proposition}

This linear stability result is a first step towards a local-in-time existence result for the full SOHR-L system. Proving such an existence result is outside the scope of the present paper. 

\begin{remark}
in the special cases $\theta = 0$ (the plane-wave perturbation propagates in the same direction as the unperturbed velocity field $\Omega_0$) or $\theta = \frac{\pi}{2}$ (the propagation direction is perpendicular to it), the dispersion relation (\ref{eq:la_lin_plane_dispersion_even}) can be solved explicitly: 
\begin{itemize}
\item[(i)] Case $\theta = 0$. Then the dispersion relation reduces to 
$$ \mu = \frac{m_2}{m_1} \, \xi. $$
This corresponds to a pure convection wave of $\Omega$ in the $x$-direction. It comes from the convection operator:  
\begin{eqnarray*}
&&\hspace{-1cm}
m_1[\rho_W] \, \Omega_t + m_2[\rho_W] \, (\Omega \cdot \nabla_x) \Omega . 
\end{eqnarray*}
\item[(ii)] Case $\theta = \frac{\pi}{2}$. Then, the dispersion relation reads:
$$ \mu = \Big( \frac{m_5[\tilde c_1 \rho_{0W}]}{m_1} \Big)^{1/2} \, |\xi|. $$
This corresponds to acoustic waves propagating symmetrically in both the positive and negative directions. They come from the acoustic operator: 
\begin{eqnarray*}
&&\hspace{-1cm}
m_1[\rho_W] \, \Omega_t + \Omega^\bot \,  (\Omega^\bot \cdot \nabla_x)  \, m_5[\rho_W] . 
\end{eqnarray*}
\end{itemize}
\label{rem:la_special_cases}
\end{remark}

%%%%%%%%%%%%%%%%%%%%%%%%%%%%%%%%%%%%%%%%%%%%%%%%%%%%%%%%%%%%%%%%%%%%%%%%%%%%%%%%%%%%%%%%%%%%%%%%
%%%%%%%%%%%%%%%%%%%%%%%%%%%%%%%%%%%%%%%%%%%%%%%%%%%%%%%%%%%%%%%%%%%%%%%%%%%%%%%%%%%%%%%%%%%%%%%%
\subsection{Small angular velocity limit of the SOHR-L model}
\label{subsec:la_asymptot_small}

In this section, we study the asymptotics of the SOHR-L model (\ref{eq:la_smallhydro_rhoW}), (\ref{eq:la_smallhydro_Omega}) when the angular velocity is small. For this purpose, we change the scaling $\eta = \varepsilon$ which was made at the beginning of section \ref{sec:large} into $\eta = \varepsilon/\zeta $. We first keep $\zeta = {\mathcal O}(1)$ when performing the limit $\varepsilon \to 0$. The resulting model is the SOHR-L model (\ref{eq:la_smallhydro_rhoW}), (\ref{eq:la_smallhydro_Omega}), where now, the moments $m_k[\rho_W]$ (see (\ref{eq:la_mom_rhoW})) and the associated coefficients $a_k$ (see (\ref{eq:def_a1}) to (\ref{eq:def_a6})) depend on the parameter $\zeta$. In a second step, we investigate the limit $\zeta \to 0$ in this SOHR-L model with $\zeta$-dependent coefficients. 

\bigskip
\noindent
{\bf First step: limit $\varepsilon \to 0$. Derivation of the SOHR-L model with $\zeta$-dependent coefficients.} Introducing the parameter $\zeta$ transforms (\ref{eq:la_kin_f_0}) into
\begin{eqnarray}
&&\hspace{-1.5cm}
\partial_t f^\varepsilon + \nabla_x \cdot (v f^\varepsilon) = \frac{1}{\varepsilon} \big( - \nabla_v \cdot (P_{v^\bot} \omega_{f^\varepsilon} (W) \, f^\varepsilon) - \zeta \, W \nabla_v \cdot (v^\bot f^\varepsilon) + d \Delta_v f^\varepsilon  \big). 
\label{eq:la_kin_f_zeta}
\end{eqnarray}
It is an easy matter to show that the associated equilibria are of the form $\rho_W \, \tilde M_{\omega_\Omega} (v, \zeta W)$ where $\rho_W$ and $\Omega$ are arbitrary and $\tilde M_{\omega_\Omega} (v, W)$ is the GVM defined at Definition \ref{def:la_VMF_generalise}. In particular, we can write 
\begin{eqnarray}
&&\hspace{-1cm}
\tilde M_{\omega_\Omega} (v,\zeta W) = \Phi_{\zeta W}(\theta), \quad \mbox{ with } \quad \theta = \widehat{(\omega_\Omega(\zeta W),v)}. 
\label{eq:la_def_Mom_zeta}
\end{eqnarray}
Similarly, the GCI are of the form $\beta \tilde \chi_\Omega(v,\zeta W) + \phi(W)$, where $\beta$ and $\phi$ are arbitrary and $\tilde \chi_\Omega(v,W)$ is the GCI defined in Prop. \ref{prop:la_exist_GCI}. Thus, 
\begin{eqnarray}
&&\hspace{-1cm}
\tilde \chi_\Omega(v,\zeta W) = X_{\zeta W}(\theta), 
\label{eq:la_def_chi_zeta}
\end{eqnarray}
with the same definition of $\theta$. It follows that $f^0 = \lim_{\varepsilon \to 0} f^\varepsilon$ where $f^\varepsilon$ is the solution of (\ref{eq:la_kin_f_zeta}) is given by 
$$ f^0(x,v,W,t) = \rho_W(x,t) \, \tilde M_{\omega^\zeta_{\Omega(x,t)}} (v, \zeta \, W) \, $$
where 
\begin{eqnarray}
&&\hspace{-1cm}
\omega^\zeta_{\Omega} (W) = \omega_\Omega(\zeta \, W).  
\label{eq:la_om_zeta}
\end{eqnarray}
The functions $\rho_W(x,t)$ and $\Omega(x,t)$ satisfy the system (\ref{eq:la_smallhydro_rhoW}), (\ref{eq:la_smallhydro_Omega}), with coefficients $\tilde c_1^\zeta$, $m_k^\zeta[\rho_W]$ such that
\begin{eqnarray}
&&\hspace{-1cm}
\tilde c_1^\zeta (W) = \tilde c_1 (\zeta W), \quad  m_k^\zeta[\rho_W] = \int_{w \in {\mathbb R}}  a_k(\zeta W) \, \rho_W \, dW, \quad k=1, \ldots , 6 . 
\label{eq:la_mom_rhoW_zeta}
\end{eqnarray}

\bigskip
\noindent
{\bf Second step: limit $\zeta \to 0$ in the SOHR-L model with $\zeta$-dependent coefficients.} We can now state the following proposition, whose proof can be found in Appendix \ref{sec:la_small_zeta_proof}:

\begin{proposition}
The formal small angular velocity limit $\zeta \to 0$ of the SOHR-L model (\ref{eq:la_smallhydro_rhoW}), (\ref{eq:la_smallhydro_Omega}) with $\zeta$-dependent coefficients is the model
\begin{eqnarray}
&&\hspace{-1cm}
\partial_t \rho_W + \nabla_x \cdot (c_1 \rho_W \Omega) = 0,  \quad \forall W \in {\mathbb R}, 
\label{eq:la_limSOHRL_rhoW} \\
&&\hspace{-1cm}
\rho \, \big( \partial_t \Omega + c_2 \, (\Omega \cdot \nabla_x)  \Omega \big) + c_5 \, P_{\Omega^\bot} \nabla_x \rho  = 0, 
\label{eq:la_limSOHRL_Omega}, 
\end{eqnarray}
with $\rho$ given by (\ref{eq:sm_smallhydro_moments}),  $c_2$ by (\ref{eq:sm_express_c2}) and $c_5$ by
\begin{eqnarray}
&&\hspace{-1cm}
c_5 = \frac{\int_0^{2 \pi} e^{\frac{\cos \theta}{d}} \, \sin^2 \theta \, d \theta}{\int_0^{2 \pi} e^{\frac{\cos \theta}{d}} \, \cos \theta \, d \theta} = \frac{1}{2} \, \frac{ I_0 \big( \frac{1}{d} \big) - I_2 \big( \frac{1}{d}\big) }{ I_1 \big( \frac{1}{d} \big) }
\label{eq:la_limSOHRL_Theta} 
\end{eqnarray}
\label{prop:la_limit_SOHRL_small_vel}
\end{proposition}

\smallskip
The same study can be performed in the small angular velocity case. Replacing $W$ by $\zeta W$ in the kinetic equation (\ref{eq:sm_smallkin_f}) and performing the limit $\varepsilon \to 0$ keeping $\zeta$ fixed leads to the SOHR-S system (\ref{eq:sm_smallhydro_rhoW}), (\ref{eq:sm_smallhydro_Omega}) with a factor $\zeta$ multiplying the term $Y \Omega^\bot$ in (\ref{eq:sm_smallhydro_Omega}). Therefore, the limit $\zeta \to 0$ in the SOHR-S system with $\zeta$-dependent parameters is immediate and leads to the system: 
\begin{eqnarray}
&&\hspace{-1cm}
\partial_t \rho_W + \nabla_x \cdot (c_1 \rho_W \Omega) = 0,  \quad \forall W \in {\mathbb R}, 
\label{eq:la_limSOHRS_rhoW} \\
&&\hspace{-1cm}
\rho \, \big( \partial_t \Omega + c_2 \, (\Omega \cdot \nabla_x)  \Omega \big) + d \, P_{\Omega^\bot} \nabla_x \rho  = 0, 
\label{eq:la_limSOHRS_Omega}, 
\end{eqnarray}
we see that the structure of this system is the same as that of (\ref{eq:la_limSOHRL_rhoW}), (\ref{eq:la_limSOHRL_Omega}). However, the coefficients of the pressure term $P_{\Omega^\bot} \nabla_x \rho$ of the two systems are different. While it is simply the noise coefficient $d$ in the SOHR-S case, it is equal to a new coefficient $c_5$ in the SOHR-L case. Therefore, even for very small angular velocities, the two systems do not coincide. This is due to the different ways of computing the interaction force. 

Like in the case of the SOHR-S model, the density equations (\ref{eq:la_limSOHRL_rhoW}) or (\ref{eq:la_limSOHRS_rhoW}) can be integrated with respect to $W$, since $c_1$ does not depend on $W$. In
both cases, the resulting system is nothing but the standard SOH model (\ref{eq:sm_smallhydro_rho_2}), (\ref{eq:sm_smallhydro_Omega_bis_2}) (see section \ref{sec:small}). However, again, the coefficients of the pressure term $P_{\Omega^\bot} \nabla_x \rho$ in the velocity eq. (\ref{eq:sm_smallhydro_Omega_bis_2}) differ. It is indeed equal to $d$ in the case of the SOHR-S model (\ref{eq:la_limSOHRS_rhoW}), (\ref{eq:la_limSOHRS_Omega}), while it is equal to $c_5$ in the case of the SOHR-L model (\ref{eq:la_limSOHRL_rhoW}), (\ref{eq:la_limSOHRL_Omega}). 

\medskip
\noindent
{\bf Approximation up to ${\mathcal O}(\zeta^2)$ of the SOHR-L model in the limit $\zeta \to 0$.} Proposition \ref{prop:la_limit_SOHRL_small_vel} shows that the small angular velocity limit of the SOHR-L model leads to the standard SOH Model (with slightly modified coefficients) for the total density $\rho$ and velocity direction $\Omega$. Therefore, information about the self-rotation of the particles is lost. Indeed, since the SOH model also describes particles with no self-rotation \cite{Degond_Motsch_M3AS08}, one cannot distinguish any influence of the particle self-rotation by looking at it. In order to retain some of the influence of the self-rotation of the particles in this limit, it is interesting to compute the first-order correction terms in ${\mathcal O}(\zeta)$. In this way, we will get the corrections to the SOH model induced by the self-rotation. The resulting model is stated in the following proposition, whose proof is sketched in Appendix \ref{sec:la_small_zeta_proof}: 

\begin{proposition}
The ${\mathcal O}(\zeta^2)$ approximation of the SOHR-L model (\ref{eq:la_smallhydro_rhoW}), (\ref{eq:la_smallhydro_Omega}) with $\zeta$-dependent coefficients, in the limit $\zeta \to 0$, is the model
\begin{eqnarray}
&&\hspace{-1cm}
\partial_t \rho_W + \nabla_x \cdot (c_1 \rho_W \Omega) = 0,  \quad \forall W \in {\mathbb R}, 
\label{eq:la_limSOHRL_forder_rhoW} \\
&&\hspace{-1cm}
\rho \, \big( \partial_t \Omega + c_2 \, (\Omega \cdot \nabla_x)  \Omega \big) + c_5 \, P_{\Omega^\bot} \nabla_x \rho  \nonumber \\
&&\hspace{0cm}
+ \zeta \, \rho \, Y \, \big( c_3 \, (\Omega^\bot \cdot \nabla_x) \Omega + c_4 \, (\nabla_x \cdot \Omega) \, \Omega^\bot \big) + \zeta \, c_6 (\Omega \cdot \nabla_x) (\rho Y) \, \,  \Omega^\bot
= 0, 
\label{eq:la_limSOHRL_forder_Omega} 
\end{eqnarray}
with $\rho$ and $\rho Y$ given by (\ref{eq:sm_smallhydro_moments}),  $c_2$ by (\ref{eq:sm_express_c2}),  $c_5$ by (\ref{eq:la_limSOHRL_Theta}) and $c_k = \frac{a_k^1}{a_1(0)}$, $k = 3, \, 4, \, 6$, $a_1(0)$ being given by (\ref{eq:la_a1_zeta=0}).   
\label{prop:la_expan_forder}
\end{proposition}

Here, compared to the SOHR-S system (\ref{eq:sm_smallhydro_rhoW}), (\ref{eq:sm_smallhydro_Omega}), the particle self-rotation introduces structurally different terms. In the SOHR-S system, self-rotation is taken into account through the source term $- Y \Omega^\bot$ in the velocity direction equation (\ref{eq:sm_smallhydro_Omega}). This term corresponds to an acceleration in the direction of the average self-rotation and proportional to it. In the system issued from the SOHR-L model (\ref{eq:la_limSOHRL_forder_rhoW}), (\ref{eq:la_limSOHRL_forder_Omega}), self-rotation introduces differential terms. The first two ones (those multiplied by $c_3$ and $c_4$) are proportional to both, the average self-rotation $Y$ and differential terms acting on the velocity direction $\Omega$ (namely $(\Omega^\bot \cdot \nabla_x) \Omega$ and $(\nabla_x \cdot \Omega)$). So, in the case of a uniform vector field $\Omega$, these two terms would not induce any acceleration, by contrast to what happens in the SOHR-S system. The operator $(\Omega^\bot \cdot \nabla_x) \Omega$ produces an acceleration if the vector fields varies in the direction normal to itself. Regions of compression or rarefaction also give rise to an acceleration due to the term $(\nabla_x \cdot \Omega)$. The last term (multiplied by $c_6$) is proportional to the gradient of the average angular momentum $\rho Y$ in the direction of $\Omega$. Therefore, variations of the average angular momentum in the direction of the flow produce an acceleration term as well. Again, in the case where $\rho Y$ is  uniform, this acceleration term vanishes, by contrast to what happens in the case of the SOHR-S system. 

One can interpret this difference as follows. In the kinetic equation leading to the SOHR-L system (\ref{eq:la_kin_f_0}), the particle acceleration $P_{v^\bot} \omega_{\Omega_f}$ is modified compared to that used in the kinetic equation leading to the SOHR-S system (\ref{eq:sm_smallkin_f}), namely $P_{v^\bot} \Omega_f$. The use of $\omega_{\Omega_f}$ instead of $\Omega_f$ introduces some kind of compensation for the self rotation $W v^\bot$ and reduces its influence. This is why, in the hydrodynamic model (\ref{eq:la_limSOHRL_forder_rhoW}), (\ref{eq:la_limSOHRL_forder_Omega}), self-rotation appears through differential terms instead of source terms like in the SOHR-S model. In a spatially homogeneous situation, where $\rho$ and $\Omega$ are uniform, the compensation of self-rotation by the use of $\omega_{\Omega_f}$ in the acceleration is total, and there is no influence of self-rotation in the hydrodynamic model. By contrast, in the SOHR-S case, even in the spatially homogeneous situation, there cannot be any compensation, and the influence of self-rotation in the hydrodynamic model persists.

%%%%%%%%%%%%%%%%%%%%%%%%%%%%%%%%%%%%%%%%%%%%%%%%%%%%%%%%%%%%%%%%%%%%%%%%%%%%%%%%%%%%%%%%%%%%%%%%
%%%%%%%%%%%%%%%%%%%%%%%%%%%%%%%%%%%%%%%%%%%%%%%%%%%%%%%%%%%%%%%%%%%%%%%%%%%%%%%%%%%%%%%%%%%%%%%%
%%%%%%%%%%%%%%%%%%%%%%%%%%%%%%%%%%%%%%%%%%%%%%%%%%%%%%%%%%%%%%%%%%%%%%%%%%%%%%%%%%%%%%%%%%%%%%%%
%%%%%%%%%%%%%%%%%%%%%%%%%%%%%%%%%%%%%%%%%%%%%%%%%%%%%%%%%%%%%%%%%%%%%%%%%%%%%%%%%%%%%%%%%%%%%%%%
\setcounter{equation}{0}
\section{Conclusion and perspectives}
\label{sec:conclu}

In this paper, we have derived hydrodynamic models for a system of noisy self-propelled particles moving in a plane. The particles are subject to proper rotation on the one hand and interactions with their neighbors through local alignment on the other hand. Two regimes have been investigated. In the small angular velocity regime, the hydrodynamic model consist of a slight modification of the previously obtained Self-Organized Hydrodynamic (SOH) model, including a source term to account for a net average angular velocity. In the large angular velocity regime, after modifying the interaction force to preserve the particle propensity to locally align with their neighbors, the resulting hydrodynamic model involves additional terms accounting for such effects as transport in the normal direction to the velocity and off-diagonal pressure tensor terms. A linearized stability analysis has been performed showing the stability of the model in some particular case. Perspectives include a deeper analytical study of the models, such as proving linearized stability in the general case and local well-posedness of smooth solutions. Numerical simulations will be performed with two purposes. The first one is to validate the hydrodynamic model by comparison to simulations of the IBM. The second one is to explore what new structures and features are exhibited by these models.

%%%%%%%%%%%%%%%%%%%%%%%%%%%%%%%%%%%%%%%%%%%%%%%%%%%%%%%%%%%%%%%%%%%%%%%%%%%%%%%%%%%%%%%%%%%%%%%%
%%%%%%%%%%%%%%%%%%%%%%%%%%%%%%%%%%%%%%%%%%%%%%%%%%%%%%%%%%%%%%%%%%%%%%%%%%%%%%%%%%%%%%%%%%%%%%%%
%%%%%%%%%%%%%%%%%%%%%%%%%%%%%%%%%%%%%%%%%%%%%%%%%%%%%%%%%%%%%%%%%%%%%%%%%%%%%%%%%%%%%%%%%%%%%%%%
%%%%%%%%%%%%%%%%%%%%%%%%%%%%%%%%%%%%%%%%%%%%%%%%%%%%%%%%%%%%%%%%%%%%%%%%%%%%%%%%%%%%%%%%%%%%%%%%

%%%%%%%%%%%%%%%%%%%%%%%%%%%%%%%%%%%%%%%%%%%%%%%%%%%%%%%%%%%%%%%%%%%%%%%%%%%%%%%%
%%%%%%%%%%%%%%%%%%%%%%%%%%%%%%%%%%%%%%%%%%%%%%%%%%%%%%%

%%%%%%%%%%%%%%%%%%%%%%%%%%%%%%%%%%%%%%%%%%%%%%%%%%%%%%%%%%%%%%%%%%%%%%%%%%%%%%%%%%%%%%%%%%%%%%%%
%%%%%%%%%%%%%%%%%%%%%%%%%%%%%%%%%%%%%%%%%%%%%%%%%%%%%%%%%%%%%%%%%%%%%%%%%%%%%%%%%%%%%%%%%%%%%%%%
%%%%%%%%%%%%%%%%%%%%%%%%%%%%%%%%%%%%%%%%%%%%%%%%%%%%%%%%%%%%%%%%%%%%%%%%%%%%%%%%%%%%%%%%%%%%%%%%
%%%%%%%%%%%%%%%%%%%%%%%%%%%%%%%%%%%%%%%%%%%%%%%%%%%%%%%%%%%%%%%%%%%%%%%%%%%%%%%%%%%%%%%%%%%%%%%%
\begin{appendices}
\setcounter{equation}{0}
\section{Small angular velocity case: proof of Theorem \ref{thm:small_hydro}}
\label{sec:small_proof}

The proof of Theorem \ref{thm:small_hydro} involves three steps which are developed in the following sections.

%%%%%%%%%%%%%%%%%%%%%%%%%%%%%%%%%%%%%%%%%%%%%%%%%%%%%%%%%%%%%%%%%%%%%%%%%%%%%%%%%%%%%%%%%%%%%%%%
\subsection{Determination of the equilibria}
\label{subsec:small_equi}

Thanks to (\ref{eq:sm_smallkin_f}), we have $Q(f^\varepsilon) = {\mathcal O}(\varepsilon)$. Taking the limit $\varepsilon \to 0$ implies $Q(f^0) = 0$. Therefore, $f^0$ is a so-called equilibrium, i.e. a solution of $Q(f) = 0$. Since $Q$ only operates on the $(v,W)$ variables, we first ignore the spatio-temporal dependence.

Let $\Omega \in {\mathbb S}^1$ be given and define the linear operator 
$$ {\mathcal Q}_\Omega(f) (v,W) = d \, \nabla_v \cdot \left[ M_{\Omega}(v) \nabla_v \left( \frac{f(v,W)}{M_{\Omega}(v)} \right) \right] . $$
Easy computations \cite{Degond_Motsch_M3AS08} show that:
$$ Q(f) = {\mathcal Q}_{\Omega_f}(f).$$ 
We now introduce the functional setting. Let $f$ and $g$ be smooth functions of $(v,W)$ with fast decay when $W \to \pm \infty$. We define the duality products: 
\begin{eqnarray*}
&&\hspace{-1cm}
\langle f,g\rangle_{0,\Omega} := \int_{(v,W) \in {\mathbb S}^1 \times {\mathbb R}} f(v,W) \, g(v,W) \, \frac{1}{M_{\Omega} (v)} \, dv \, dW , \\
&&\hspace{-1cm}
\langle f,g\rangle_{1,\Omega} = \int_{(v,W) \in {\mathbb S}^1 \times {\mathbb R}} \nabla_v \big( \frac{f(v,W)}{M_{\Omega}(v)} \big) \cdot \nabla_v \big( \frac{g(v,W)}{M_{\Omega}(v)} \big) \, M_{\Omega} (v) \, dv \, dW .
\end{eqnarray*}
Then, $\langle f,g\rangle_{0,\Omega}$ defines a duality (i.e. a continuous bilinear form) between $f \in L^1({\mathbb R}, L^2({\mathbb S}^1))$ and $f \in L^\infty({\mathbb R}, L^2({\mathbb S}^1))$. Similarly, $\langle f,g\rangle_{1,\Omega}$ defines a duality between $f \in L^1({\mathbb R}, H^1({\mathbb S}^1))$ and $f \in L^\infty({\mathbb R}, H^1({\mathbb S}^1))$. Thanks to Green's formula applied with smooth functions, we have 
\begin{eqnarray}
&&\hspace{-1cm}
- \langle {\mathcal Q}_\Omega(f),g\rangle_{0,\Omega} = d \langle f,g \rangle_{1,\Omega}. 
\label{eq:sm_smallkin_selfadjoint}
\end{eqnarray}
Therefore, for $f \in L^\infty({\mathbb R}, L^2({\mathbb S}^1))$, we define ${\mathcal Q}_\Omega(f)$ as a linear form on $L^\infty({\mathbb R}, L^2({\mathbb S}^1))$. Actually, since this linear form is defined and continuous on $C_0^0({\mathbb R}, L^2({\mathbb S}^1))$, where $C^0_0$ denotes the space of continuous functions tending to zero at infinity, ${\mathcal Q}_\Omega(f)$ is a bounded measure on ${\mathbb R}$ with values in $H^1({\mathbb S}^1)$ but we will not use this characterization. We now define the set of equilibria:

\begin{definition}
The set ${\mathcal E}$ of equilibria of $Q$ is given by 
$${\mathcal E} = \big\{ f \in L^1({\mathbb R}, H^1({\mathbb S}^1)) \, \, | \, \, f \geq 0 \mbox{ and } {\mathcal Q}_{\Omega_f} (f) = 0 \big\} .$$ 
\label{def:small_equi}
\end{definition}

The characterization of ${\mathcal E}$ is given in the following lemma. 

\begin{lemma}
The set ${\mathcal E}$ of equilibria is the set of all functions of the form  
\begin{eqnarray}
&&\hspace{-1cm}
v \mapsto \rho_W \, M_\Omega(v) , 
\label{eq:sm_smallkin_equi}
\end{eqnarray}
where the function $W \mapsto \rho_W \in {\mathbb R}_+$ and the vector $\Omega$ are arbitrary in the sets $L^1({\mathbb R})$ and ${\mathbb S}^1$ respectively.
\label{lem:smallkin_equi}
\end{lemma}

\medskip
\noindent
{\bf Proof.} First, suppose that $f \in {\mathcal E}$. Then, thanks to (\ref{eq:sm_smallkin_selfadjoint}), we have $0 = - \langle {\mathcal Q}_{\Omega_f}(f),f\rangle_{0,\Omega_f} = d \langle f,f \rangle_{1,\Omega_f}$. It follows that $\nabla_v \big( \frac{f(v,W)}{M_{\Omega_f}} \big) = 0$, i.e. there exists $\rho_W \in {\mathbb R}$, independent of $v$, such that $f(v,W) = \rho_W \, M_{\Omega_f}$. Additionally, that $f \in L^1({\mathbb R}, H^1({\mathbb S}^1))$ and $f \geq 0$ implies that $\rho_W \geq 0$ and that the function $W \in {\mathbb R} \to \rho_W \in {\mathbb R}_+$ belongs to $L^1({\mathbb R})$. Therefore, $f$ is of the form (\ref{eq:sm_smallkin_equi}). 

Conversely, suppose that $f$ is of the form (\ref{eq:sm_smallkin_equi}) with $\rho_W$ as regular as in the lemma. Then, the results follow obviously if we can show that $\Omega_f = \Omega$. But, thanks to (\ref{eq:sm_smallkin_cur_equi}), we have 
$ J_{\rho_W M_\Omega} = \int_{W \in {\mathbb R}} \rho_W \, dW \, c_1 \, \Omega$,
and since $c_1>0$ and $\rho_W >0$, we have $\Omega_{\rho_W M_\Omega} = \Omega$, which shows the result. \endproof

From this lemma, and the fact that $f^0$ is an equilibrium, we deduce that $f^0$ is given by (\ref{eq:sm_smallkin_f0}). 
Now, $\rho_W = \rho_W(x,t)$ and $\Omega=\Omega(x,t)$ are a priori arbitrary functions of $(x,t)$. Indeed, $Q$ only acts on the $(v,W)$ variables. Hence, the fact that $Q(f^0) = 0$ does not impose any condition on the dependence of $f^0$ on $(x,t)$. In order to determine how $\rho_W$ and $\Omega$ depend on $(x,t)$, we need the second step of the proof, developed in the following section.

%%%%%%%%%%%%%%%%%%%%%%%%%%%%%%%%%%%%%%%%%%%%%%%%%%%%%%%%%%%%%%%%%%%%%%%%%%%%%%%%%%%%%%%%%%%%%%%%
\subsection{Generalized Collision Invariants (GCI)}
\label{subsec:small_GCI}

We first recall the concept of a Collision Invariant.

\begin{definition}
A collision invariant (CI) is a function $\psi \in L^\infty({\mathbb R}, H^1({\mathbb S}^1))$ such that for all functions $f \in L^1({\mathbb R}, H^1({\mathbb S}^1))$, we have
\begin{equation}
- \int_{(v,W) \in {\mathbb S}^1 \times {\mathbb R}} Q(f) \, \psi \, dv \, dW := d \langle \psi M_{\Omega_f} \, , \, f \rangle_{1,\Omega_f}= 0.
\label{eq:sm_def_CI}
\end{equation}
We denote by ${\mathcal C}$ the set of CI. The set ${\mathcal C}$ is a vector space. 
\label{def:CI}
\end{definition}

We first have the obvious result:

\begin{proposition}
Any function $\phi$: $W \in {\mathbb R} \mapsto \phi(W) \in {\mathbb R}$ belonging to $L^\infty({\mathbb R})$ is a CI. 
\label{prop:determination_CI}
\end{proposition}

\medskip
\noindent
{\bf Proof.} Let $\phi \in  L^\infty({\mathbb R})$ and $f \in L^1({\mathbb R}, H^1({\mathbb S}^1))$. Then, obviously $\phi M_{\Omega_f} \in L^\infty({\mathbb R}, H^1({\mathbb S}^1))$ and since $\phi$ does not depend on $v$, it satisfies (\ref{eq:sm_def_CI}). 
\endproof

\medskip
We will see that this set of CI does not suffice to provide the spatio-temporal evolution of $\rho_W$ and $\Omega$ in the hydrodynamic limit. In the absence of other obvious CI, we introduce a weaker concept, that of 'Generalized Collision Invariant' (GCI). The rationale for introducing this concept is discussed in details in \cite{Degond_etal_Schwartz13, Degond_Motsch_M3AS08}. 

\begin{definition}
Let $\Omega \in {\mathbb S}^1$ be given. A Generalized Collision Invariant (GCI) associated to $\Omega$ is a function $\psi \in L^\infty({\mathbb R}, H^1({\mathbb S}^1))$ which satisfies the following property: for all functions $f(v, W)$ such that $f \in L^1({\mathbb R}, H^1({\mathbb S}^1))$ and that $P_{\Omega^\bot} \Omega_f  = 0$, we have
\begin{equation}
- \int_{(v,W) \in {\mathbb S}^1 \times {\mathbb R}} {\mathcal Q}_\Omega(f) \, \psi \, dv \, dW: = d \langle \psi M_\Omega \, , \, f\rangle_{1,\Omega} = 0.
\label{eq:sm_def_GCI}
\end{equation}
We denote by ${\mathcal G}_\Omega$ the set of GCI associated to $\Omega$. It is a vector space. 
\label{def:GCI}
\end{definition}

\noindent
Of course, if $\psi \in L^\infty({\mathbb R}, H^1({\mathbb S}^1))$, so does $ \psi M_\Omega $ and (\ref{eq:sm_def_GCI}) is well-defined. Before determining ${\mathcal G}_\Omega$, we introduce an appropriate functional setting for functions of $v$ only. We consider the space $V_0 = \{ \varphi \in H^1({\mathbb S}^1), \, \, \int_{v \in {\mathbb S}^1} \varphi(v) \, dv = 0 \}$. Let $\Omega \in {\mathbb S}^1$ be given. We define the following norms or semi-norms on $L^2({\mathbb S}^1)$ and $H^1({\mathbb S}^1)$ respectively, by: 
\begin{eqnarray*}
&&\hspace{-1cm}
|f|^2_{0,\Omega} := \int_{v \in {\mathbb S}^1} |f(v)|^2 \, \frac{1}{M_{\Omega} (v)} \, dv , \qquad 
|f|^2_{1,\Omega} = \int_{v \in {\mathbb S}^1} \Big|\nabla_v \big( \frac{f(v)}{M_{\Omega}(v)} \big) \Big|^2 \, M_{\Omega} (v) \, dv .
\end{eqnarray*}
Of course, these two semi-norms are respectively equivalent to the classical $L^2$ norm and $H^1$ semi-norm on $L^2({\mathbb S}^1)$ and $H^1({\mathbb S}^1)$.
We have the following Poincaré inequality: 
\begin{eqnarray}
&&\hspace{-1cm}
|\varphi|^2_{1,\Omega} \geq C |\varphi|^2_{0,\Omega}, \qquad  \forall \varphi \in  V_0,
\label{eq:sm_Poincare_0} 
\end{eqnarray}
with a positive constant $C$. We denote by $(f,g)_{0,\Omega}$ and $(f,g)_{1,\Omega}$ the associated bilinear forms. 

\begin{proposition}
We have
$$ {\mathcal G}_\Omega = \big\{ \beta \chi_\Omega(v) + \phi(W), \quad \beta  \in {\mathbb R}, \quad \phi \in L^\infty({\mathbb R}) \big\}, $$
where $\varphi_\Omega = \chi_\Omega M_\Omega$ is the unique solution in $V_0$ of the variational formulation 
\begin{eqnarray}
&&\hspace{-1.5cm}
\mbox{Find } \quad  \varphi  \in V_0 \quad  \mbox{ such that } \quad
(\varphi,f)_{1,\Omega} = ( \Omega^\bot \cdot v \, \, M_\Omega \, , f )_{0,\Omega}, \quad  \forall f \in H^1({\mathbb S}^1) . 
\label{eq:sm_GCI_variational_3}
\end{eqnarray}
\label{prop:exist_GCI}
\end{proposition}

\noindent
{\bf Proof.} The existence of a unique solution $\varphi_\Omega \in V_0$ of the variational problem (\ref{eq:sm_GCI_variational_3}) is an easy consequence of Lax-Milgram's theorem and the Poincaré inequality (\ref{eq:sm_Poincare_0}). We refer the reader to \cite{Degond_Motsch_M3AS08, Frouvelle_M3AS12}. 

Now, let $\Omega \in {\mathbb S}^1$ be given, $\psi \in {\mathcal G}_\Omega$ and $f \in  L^1({\mathbb R}, H^1({\mathbb S}^1))$. First we note that the condition $P_{\Omega^\bot} \Omega_f  = 0$ is equivalent to $P_{\Omega^\bot} J_f  = 0$ and can be written 
$$ \int_{(v,W) \in {\mathbb S}^1 \times {\mathbb R}} f \,\, \Omega^\bot \cdot v \,\, dv \, dW = 0, $$ 
or equivalently, $\langle \Omega^\bot \cdot v \, \, M_\Omega, f \rangle_{0,\Omega} = 0$. Then, by (\ref{eq:sm_def_GCI}), $\psi$ is a GCI if and only if $\psi \in L^\infty ({\mathbb R}, H^1({\mathbb S}^1))$ and the following implication holds: for all $f \in L^1({\mathbb R}, H^1({\mathbb S}^1))$, 
\begin{eqnarray*}
&&\hspace{-1cm}
\langle \Omega^\bot \cdot v \, \, M_\Omega, f \rangle_{0,\Omega} = 0  \qquad \Longrightarrow \qquad 
\langle f , \psi M_\Omega \rangle_{1,\Omega} = 0 . 
\end{eqnarray*}
By a standard functional analytic argument, this means that there exists a real number $\beta$ such that
\begin{eqnarray}
&&\hspace{-1cm}
\langle\psi M_\Omega,f \rangle_{1,\Omega} = \beta \, \langle \Omega^\bot \cdot v \, \, M_\Omega, f \rangle_{0,\Omega}, \qquad \forall f \in  L^1({\mathbb R}, H^1({\mathbb S}^1)).
\label{eq:sm_GCI_variational}
\end{eqnarray}
Therefore, $\psi$ is the solution of an elliptic variational problem. 

Now, we remark that the function $(v,W) \to \beta \chi_\Omega(v) + \phi(W)$, with $\phi \in L^\infty({\mathbb R})$ belongs to $L^\infty ({\mathbb R}, H^1({\mathbb S}^1))$ and satisfies the variational problem (\ref{eq:sm_GCI_variational}). 
These are the only ones. Indeed, by linearity, the difference $\psi$ of two such solutions is an element of $ L^\infty({\mathbb R}, H^1({\mathbb S}^1))$ and satisfies 
\begin{eqnarray*}
&&\hspace{-1cm}
\langle \psi M_\Omega,f \rangle_{1,\Omega} = 0, \qquad \forall f \in  L^1({\mathbb R}, H^1({\mathbb S}^1)).
\end{eqnarray*}
Then, introducing the indicator function $\zeta_A(W)$ of the interval $[-A,A]$, with $A>0$ and taking $f = \psi M_\Omega \zeta_A$ as a test function in $L^1({\mathbb R}, H^1({\mathbb S}^1))$, we get
$$ \int_{(v,W) \in {\mathbb S}^1 \times [-A,A]} |\nabla_v \psi|^2 \, M_\Omega\, dv \, dW = 0, $$
which implies that $\psi$ does not depend on $v$ and is therefore of the form $\psi(W)$ with $\psi \in L^\infty({\mathbb R})$. This concludes the proof. \endproof

Interpreting the variational problem (\ref{eq:sm_GCI_variational_3}) in the distributional sense, we see that $\chi_\Omega$ is a solution of the following elliptic problem: 
\begin{eqnarray}
&&\hspace{-1cm}
- \nabla_v \cdot ( M_\Omega \nabla_v \chi_\Omega) = v \cdot \Omega^\bot \,\, M_\Omega, \qquad \int_{v \in {\mathbb S}^1} \chi_\Omega (v) \, M_\Omega (v) \,d v = 0.  
\label{eq:sm_prb_chi}
\end{eqnarray}
Additionally, we can write \cite{Frouvelle_M3AS12} $\chi_\Omega (v) = g(\theta)$, where $\theta = \widehat{(\Omega, v)}$ and $g$ is the odd $2 \pi$-periodic function in $H^1_{\mbox{{\scriptsize loc}}}({\mathbb R})$ (which can be identified to $H^1_0(0,\pi)$) which uniquely solves the problem  
\begin{eqnarray}
&&\hspace{-1cm}
- \frac{d}{d \theta} \big(  e^{ \frac{\cos\theta}{d}} \, \,\frac{dg}{d \theta}(\theta) \big) = \sin \theta \, e^{ \frac{\cos\theta}{d}}.
\label{eq:sm_prb_g}
\end{eqnarray}
A closed formula for $g$ can be obtained \cite{Frouvelle_M3AS12}:
\begin{eqnarray}
&&\hspace{-1cm}
g(\theta) = d \, \theta - d \, \pi \frac{\int_0^\theta e^{- \frac{\cos \varphi}{d}} \, d \varphi}{\int_0^\pi e^{- \frac{\cos \varphi}{d}} \, d \varphi}. 
\label{eq:sm_closform_g}
\end{eqnarray}
Since the function $\frac{g(\theta)}{\sin \theta}$ is even and $2 \pi$-periodic, it can be expressed as a function of $\cos \theta$. Thus, we introduce the function $h$ defined on $[-1,1]$ such that 
\begin{eqnarray}
h(\cos \theta) = \frac{g(\theta)}{\sin \theta}. 
\label{eq:sm_express_g}
\end{eqnarray}
Then, we can write 
\begin{eqnarray}
\chi_\Omega(v) = h \big( \Omega \cdot v) \, \, \Omega^\bot \cdot v, 
\label{eq:sm_express_chi}
\end{eqnarray}
and the function $h$ is bounded. We are now well equipped to derive the hydrodynamic limit $\varepsilon \to 0$ of (\ref{eq:sm_smallkin_f}). This is done in the next section.

%%%%%%%%%%%%%%%%%%%%%%%%%%%%%%%%%%%%%%%%%%%%%%%%%%%%%%%%%%%%%%%%%%%%%%%%%%%%%%%%%%%%%%%%%%%%%%%%
\subsection{Hydrodynamic limit $\varepsilon \to 0$}
\label{subsec:small_hydro}

This section is devoted to the proof of Theorem \ref{thm:small_hydro}.

\medskip
\noindent
{\bf Proof of Theorem \ref{thm:small_hydro}.} We recall that, as a consequence of Lemma \ref{lem:smallkin_equi} and the fact that $f^0 = \lim_{\varepsilon \to 0} f^\varepsilon$ is an equilibrium, $f^0$ is given by (\ref{eq:sm_smallkin_f0}). In the remainder of the proof, we omit the superscript $0$ for the sake of clarity. 

We first prove (\ref{eq:sm_smallhydro_rhoW}). Taking an arbitrary function $\phi \in L^\infty({\mathbb R})$, multiplying (\ref{eq:sm_smallkin_f}) by $\phi$, integrating with respect to $(v,W) \in {\mathbb S}^1 \times {\mathbb R}$, using the fact that $\phi$ is a GCI thanks to Proposition \ref{prop:exist_GCI} and taking the limit $\varepsilon \to 0$, we get:
$$ \int_{W \in {\mathbb R}} \big( \partial_t \rho_W + \nabla_x \cdot (c_1 \rho_W \Omega) \big) \, \phi(W) \, dW = 0. $$
In the second term, we have used (\ref{eq:sm_smallkin_cur_equi}), as well as the definition (\ref{eq:sm_smallhydro_moments}). Since this equation is valid for any $\phi \in L^\infty({\mathbb R})$, we immediately deduce (\ref{eq:sm_smallhydro_rhoW}).

We now prove (\ref{eq:sm_smallhydro_Omega}). We multiply (\ref{eq:sm_smallkin_f}) by $\chi_{\Omega_{f^\varepsilon}}$ and integrate with respect to $v$. Since $\chi_{\Omega_{f^\varepsilon}}$ is a GCI associated to $\Omega_{f^\varepsilon}$ and since $f^\varepsilon$ has precisely mean direction $\Omega_{f^\varepsilon}$, we have 
$$ \int_{(v,W) \in {\mathbb S}^1 \times {\mathbb R}} Q(f^\varepsilon) \, \chi_{\Omega_{f^\varepsilon}} \, dv \, dW = 0. $$
Then we get 
\begin{eqnarray}
&&\hspace{-1cm}
\int_{(v,W) \in {\mathbb S}^1 \times {\mathbb R}} ( {\mathcal T}^1 f^\varepsilon + {\mathcal T}^2 f^\varepsilon ) \, \chi_{\Omega_{f^\varepsilon}}  \, dv \, dW = 0, 
\label{eq:sm_int_GCI_eps}
\end{eqnarray}
where ${\mathcal T}^k$, $k=1,2$ are the following operators: 
$$ {\mathcal T}^1 f = \partial_t f + \nabla_x \cdot (v f) , \qquad {\mathcal T}^2 f = W \nabla_v \cdot (v^\bot f) . $$
Taking the limit $\varepsilon \to 0$ in (\ref{eq:sm_int_GCI_eps}) and using the fact that $f^\varepsilon \to \rho_W M_\Omega$ we get:
\begin{eqnarray}
&&\hspace{-1cm}
\int_{(v,W) \in {\mathbb S}^1 \times {\mathbb R}} ( {\mathcal T}^1 (\rho_W M_\Omega) + {\mathcal T}^2 (\rho_W M_\Omega) ) \, \chi_\Omega  \, dv \, dW := T_1 + T_2 = 0, 
\label{eq:sm_int_GCI_eps_2}
\end{eqnarray}

The contribution of the first term of (\ref{eq:sm_int_GCI_eps_2}) has been computed in \cite{Degond_etal_MAA13, Frouvelle_M3AS12}. Using the expression (\ref{eq:sm_express_chi}) of $\chi_\Omega$, it leads to 
\begin{eqnarray}
&&\hspace{-1cm}
T_1 = \Omega^\bot \cdot \int_{W \in {\mathbb R}} \Big[  \rho_W \, \big( \frac{\alpha}{d} \, \partial_t \Omega + \gamma (\Omega \cdot \nabla_x) \Omega \big)  + \alpha P_{\Omega^\bot} \nabla_x \rho_W \Big] \, dW,
\label{eq:sm_int_GCI_eps_3}
\end{eqnarray}
with 
\begin{eqnarray*}
&&\hspace{-1cm}
\alpha = \int_{v \in {\mathbb S}^1} M_\Omega(v) \, \big( 1 - (v \cdot \Omega)^2 \big) \, h(v \cdot \Omega) \, d v, \\
&&\hspace{-1cm}
\gamma = \frac{1}{d} \int_{v \in {\mathbb S}^1} M_\Omega(v) \, \big( 1 - (v \cdot \Omega)^2 \big) \, h(v \cdot \Omega) \, \cos (v \cdot \Omega) \, d v.
\end{eqnarray*}
Since $\alpha$ and $\gamma$ do not depend on $W$, we can integrate the variable $W$ out and (\ref{eq:sm_int_GCI_eps_3}) leads to:
\begin{eqnarray}
&&\hspace{-1cm}
T_1 = \Omega^\bot \cdot \Big[ \rho \, \big( \frac{\alpha}{d} \, \partial_t \Omega + \gamma (\Omega \cdot \nabla_x) \Omega \big)  + \alpha P_{\Omega^\bot} \nabla_x \rho \Big].
\label{eq:sm_int_GCI_eps_4}
\end{eqnarray}

We now turn towards the second term. We have 
\begin{eqnarray*}
&&\hspace{-1cm}
T_2 = \int_{(v,W) \in {\mathbb S}^1 \times {\mathbb R}} W  \, \nabla_v \cdot \big( v^\bot \rho_W M_\Omega \big) \, (v \cdot \Omega^\bot) \, h(v \cdot \Omega) \, d v \, dW.
\end{eqnarray*}
Owing to the fact that $\nabla_v \cdot \big( v^\bot M_\Omega \big) = - \frac{v \cdot \Omega^\bot}{d} \, M_\Omega$, we get
\begin{eqnarray}
T_2 &=& - \frac{1}{d} \, \int_{W \in {\mathbb R}} W  \, \rho_W \, dW  \, \int_{v \in {\mathbb S}^1} M_\Omega \, (v \cdot \Omega^\bot)^2 \, h(v \cdot \Omega) \, d v \\
&=& - \frac{\alpha}{d} Y.
\label{eq:sm_int_GCI_eps_5}
\end{eqnarray}

Now, collecting (\ref{eq:sm_int_GCI_eps_4}) and (\ref{eq:sm_int_GCI_eps_5}) and multiplying by $\frac{d}{\alpha}$, we get (\ref{eq:sm_smallhydro_Omega}) with $c_2 = \frac{\gamma d}{\alpha}$, i.e.
\begin{eqnarray}
c_2 &=& \frac{\int_{v \in {\mathbb S}^1} M_\Omega(v) \, \big( 1 - (v \cdot \Omega)^2 \big) \, h(v \cdot \Omega) \, \cos (v \cdot \Omega) \, d v}{\int_{v \in {\mathbb S}^1} M_\Omega(v) \, \big( 1 - (v \cdot \Omega)^2 \big) \, h(v \cdot \Omega) \, d v}, 
\label{eq:sm_express_c2}
\\
\mbox{} \nonumber \\
&=& \frac{\int_0^\pi e^{\frac{\cos \theta}{d}} \, \sin^2 \theta \, \, h(\cos \theta) \, \cos \theta \, d \theta}{\int_0^\pi e^{\frac{\cos \theta}{d}} \, \sin^2 \theta \, \, h(\cos \theta) \, d \theta}. 
\label{eq:sm_express_c2_theta}
\\
\mbox{} \nonumber \\
&=& \frac{\int_0^\pi e^{\frac{\cos \theta}{d}} \, g(\theta) \, \sin \theta \, \cos \theta \, d \theta}{\int_0^\pi e^{\frac{\cos \theta}{d}} \, g(\theta) \, \sin \theta \,  d \theta}, 
\label{eq:sm_express_c2_g}
\end{eqnarray}
where we use (\ref{eq:sm_express_g}) in the last equality.

%%%%%%%%%%%%%%%%%%%%%%%%%%%%%%%%%%%%%%%%%%%%%%%%%%%%%%%%%%%%%%%%%%%%%%%%%%%%%%%%%%%%%%%%%%%%%%%%
%%%%%%%%%%%%%%%%%%%%%%%%%%%%%%%%%%%%%%%%%%%%%%%%%%%%%%%%%%%%%%%%%%%%%%%%%%%%%%%%%%%%%%%%%%%%%%%%
%%%%%%%%%%%%%%%%%%%%%%%%%%%%%%%%%%%%%%%%%%%%%%%%%%%%%%%%%%%%%%%%%%%%%%%%%%%%%%%%%%%%%%%%%%%%%%%%
%%%%%%%%%%%%%%%%%%%%%%%%%%%%%%%%%%%%%%%%%%%%%%%%%%%%%%%%%%%%%%%%%%%%%%%%%%%%%%%%%%%%%%%%%%%%%%%%
\setcounter{equation}{0}
\section{Large angular velocity case: proof of Theorem \ref{thm:la_hydro}}
\label{sec:large_proof}

The proof of Theorem \ref{thm:la_hydro} is divided into the same three steps as that of Theorem \ref{thm:small_hydro}. However, there are substantial differences and new difficulties which justify why we develop this proof in full detail below.

%%%%%%%%%%%%%%%%%%%%%%%%%%%%%%%%%%%%%%%%%%%%%%%%%%%%%%%%%%%%%%%%%%%%%%%%%%%%%%%%%%%%%%%%%%%%%%%%
\subsection{Determination of the equilibria}
\label{subsec:la_equi}

We first prove Lemmas \ref{lem:la_exist_Phi} to \ref{lem:la_exist_omOM}. 

\medskip
\noindent
{\bf Proof of Lemma \ref{lem:la_exist_Phi}.} We show the existence and uniqueness of $\Phi_W$. For simplicity, we omit the index $W$. Defining $G(\theta) = \frac{1}{d} (W - \sin \theta)$, (\ref{eq:la_def_Phi}) can be rewritten 
\begin{eqnarray}
&&\hspace{-1cm}
\Phi' - G \Phi = C,
\label{eq:la_Phi_deriv}
\end{eqnarray}
where $C$ is a constant. This equation can be integrated elementarily on the interval $[0,2\pi[$ and leads to 
$$ \Phi(\theta) = e^{H(\theta)} \big( C \int_0^\theta e^{-H(s)} \, ds  + D \big), \quad \theta \in [0,2 \pi[, $$
where $D$ is another constant and $H$ is the antiderivative of $G$ which vanishes at $0$: $H(\theta) = \frac{1}{d} (W \theta + \cos \theta - 1)$. The constants $C$ and $D$ are determined from the requirement that, on the one hand $\Phi$ is $2 \pi$-periodic and smooth, hence leading to $\Phi(0) = \Phi(2\pi)$ and on the other hand it is normalized to unity, i.e. $\int_0^{2 \pi} \Phi(\theta) \, d \theta = 1$. These two conditions lead to the following linear system for $C$ and $D$: 
\begin{eqnarray*}
&&\hspace{-1cm}
\left\{ \begin{array}{l}
 e^{H(2 \pi)} \int_0^{2\pi} e^{-H(s)} \, ds \, \, \, C + \big( e^{H(2 \pi)} -1 \big) \, \, D = 0, \\
\int_0^{2\pi} e^{H(\theta)}  \int_0^{\theta} e^{-H(s)} \, ds  \, d \theta \, \, \, C + \int_0^{2\pi} e^{H(\theta)} \, d \theta \, \,  \, D = 1.
\end{array} \right.
\end{eqnarray*}
The determinant $\Delta$ of this system can be written 
\begin{eqnarray*}
&&\hspace{-1cm}
\Delta =  e^{H(2 \pi)} \int_0^{2\pi} e^{H(\theta)} \, \int_\theta^{2\pi} e^{-H(s)} \, ds \, d \theta +  
\int_0^{2\pi} e^{H(\theta)}  \int_0^{\theta} e^{-H(s)} \, ds  \, d \theta,
\end{eqnarray*}
and is clearly strictly positive. Therefore, there exists a unique pair of constants $(C,D)$ which satisfies the required conditions. These constants can be computed readily and are given by: 
\begin{eqnarray*}
&&\hspace{-1cm}
C = - \frac{1}{\Delta} \, \big( e^{H(2 \pi)} -1 \big), \qquad D = \frac{1}{\Delta} \,\, e^{H(2 \pi)} \int_0^{2\pi} e^{-H(s)} \, ds. 
\end{eqnarray*}
Then, the solution can finally be written:
\begin{eqnarray}
&&\hspace{-1cm}
\Phi(\theta) = \frac{e^{H(\theta)}}{\Delta} \,   \big( e^{H(2 \pi)} \, \int_\theta^{2\pi} e^{-H(s)} \, ds  + \int_0^{\theta} e^{-H(s)} \, ds \big), \quad \theta \in [0,2\pi[,
\label{eq:formula_Phi}
\end{eqnarray}
and is again, clearly positive. Finally, (\ref{eq:formula_Phi}) shows that the function $\Phi$ is smooth, except may be at the cut point $\theta = 0$. However, by using the equation recursively, it is easy to see that $\Phi^{(k)}(2 \pi) = \Phi^{(k)}(0)$, showing that $\Phi$ defines a function of $C^\infty({\mathbb S}^1)$. This concludes the proof. \endproof

\medskip
\noindent
{\bf Proof of Lemma \ref{lem:la_equilibria_Qom}.} Let $f(v,W)$ be such that $\tilde {\mathcal Q}_\omega f = 0$. Using the angular coordinate $\theta = \widehat{(\omega,v)}$, and writing $f(v,W) = \rho_W \psi_W(\theta)$, with $\rho_W = \int_{v\in {\mathbb S}^1} f(v,W) \, dv$, we find that $\psi_W$ satisfies (\ref{eq:la_def_Phi}). Hence, by the uniqueness of the solution of (\ref{eq:la_def_Phi}), $\psi_W$ must be equal to $\Phi_W$, leading to the expression (\ref{eq:la_def_equ_tildeQom}). The converse is obvious. \endproof

\medskip
\noindent
{\bf Proof of Lemma \ref{lem:la_c1_indep_om}.} Let $\theta = \widehat{(\omega(W),v)}$. Then, we have:
\begin{eqnarray}
&&\hspace{-1cm}
\tilde c_{1 \, \omega} (W) = \Big| \int_0^{2 \pi} \Phi_W(\theta) (\cos \theta, \sin \theta)^T \, d \theta \Big|, 
\label{eq_la_express_tilde_C1}
\end{eqnarray}
and is clearly independent of $\omega(W)$. \endproof

\medskip
\noindent
{\bf Proof of Lemma \ref{lem:la_exist_omOM}.} We compute the components of $\Psi_\omega(W)$ in the basis $(\omega,\omega^\bot)$. We get:
\begin{eqnarray}
&&\hspace{-1cm}
\Psi_\omega(W) = \frac{1}{\tilde c_1(W)} \int_0^{2 \pi} \Phi_W(\theta) (\cos \theta, \sin \theta)^T \, d \theta,
\label{eq:la_Psi(W)_deriv}
\end{eqnarray}
where the exponent 'T' denotes the transpose of a vector or matrix. This expression shows that the angle 
\begin{eqnarray}
&&\hspace{-1cm}
\psi(W) = \widehat{(\omega(W),\Psi_\omega(W))}, 
\label{eq:la_psi(W)_def}
\end{eqnarray}
does not depend on $\omega$ and can be computed a priori from the knowledge of $\Phi_W$. Thus, given $\Omega$, if we choose $\omega$ such that $\widehat{(\omega(W),\Omega)}= \psi(W)$, $\forall W \in {\mathbb R}$, we get that $\Psi_\omega(W) = \Omega$ and that this is the unique choice of $\omega$ which realizes this equality.   \endproof

\medskip
Now, we recall that $\tilde Q(f)$ is defined by (\ref{eq:la_kin_coll_op}). We turn to the definition and determination of the equilibria of $\tilde Q$.

\begin{definition} 
The set $\tilde {\mathcal E}$ of equilibria of $\tilde Q$ is defined by 
$$\tilde {\mathcal E} = \big\{ f \in L^1({\mathbb R}, C^2({\mathbb S}^1)) \, \, | \, \, f \geq 0 \, \,  \mbox{ and } \, \, \tilde Q (f) = 0 \big\} .$$ 
\label{def:la_equi_tildeQ}
\end{definition}

\noindent
The following proposition characterizes the elements of $\tilde {\mathcal E}$:

\medskip
\begin{proposition}
The set $\tilde {\mathcal E}$ is the set of all functions of the form 
\begin{eqnarray}
&&\hspace{-1cm}
(v,W) \mapsto \rho_W \tilde M_{\omega_\Omega}(v,W), 
\label{eq:la_def_equ_tildeQ}
\end{eqnarray}
where the function $W \mapsto \rho_W \in {\mathbb R}_+$  and the vector $\Omega$ are arbitrary in $L^1({\mathbb R})$ and ${\mathbb S}^1$ respectively. 
\label{prop:la_equilibria_Q}
\end{proposition}

\noindent
{\bf Proof.} We first show that all equilibria are necessarily of the form (\ref{eq:la_def_equ_tildeQ}). Indeed, let $f(v,W)$ be such that $\tilde Q(f) = 0$. Then, it satisfies $\tilde {\mathcal Q}_{\omega_f}(f) = 0$ and is therefore an element of $\tilde {\mathcal E}_{\omega_f}$. From Lemma \ref{lem:la_equilibria_Qom}, there exists $\rho_W \geq 0$ such that $f = \rho_W \, \tilde M_{\omega_f}$. But, by Definition \ref{def:la_def_omf}, $\omega_f = \omega_{\Omega_f}$. Therefore, there exist $\Omega$ (namely $\Omega_f$) such that $f$ is of the form (\ref{eq:la_def_equ_tildeQ}). 

Conversely, suppose that $f$ is of the form (\ref{eq:la_def_equ_tildeQ}). By Lemma \ref{lem:la_equilibria_Qom}, $f \in \tilde {\mathcal E}_{\omega_\Omega}$. By (\ref{eq:la_kin_coll_op}), Definition \ref{def:la_equi} and Definition \ref{def:la_equi_tildeQ}, we have the equivalence:
$$ f \in \tilde {\mathcal E} \quad \Longleftrightarrow \quad f \in \tilde {\mathcal E}_{\omega_f}. $$
Therefore, to prove that $f \in \tilde {\mathcal E}$, it is sufficient to prove that $\omega_f = \omega_\Omega$. 
But from (\ref{eq:la_def_equ_tildeQ}), we have 
$$J_f = \int_{w \in {\mathbb R}} \rho_W \, \tilde c_1(W) \, \Psi_{\omega_\Omega} (W) \, dW.$$
But, with (\ref{eq:la_tildeom=Om}), we deduce that 
$$J_f = \int_{w \in {\mathbb R}} \rho_W \, \tilde c_1(W) \, dW \, \, \Omega,$$
and that 
$$\Omega_f = \frac{J_f}{|J_f|} = \Omega.$$
Therefore, by (\ref{eq:la_def_omf}), we have $\omega_f = \omega_{\Omega_f} = \omega_\Omega$. This concludes the proof. \endproof

%%%%%%%%%%%%%%%%%%%%%%%%%%%%%%%%%%%%%%%%%%%%%%%%%%%%%%%%%%%%%%%%%%%%%%%%%%%%%%%%%%%%%%%%%%%%%%%%
\subsection{Generalized collision invariants}
\label{subsec:la_GCI}

We define the notion of a GCI for the collision operator $\tilde Q$:

\begin{definition}
Let $\Omega \in {\mathbb S}^1$ be given. A Generalized Collision Invariant (GCI) associated to $\Omega$ is a function $\psi \in L^\infty_{\mbox{\scriptsize{loc}}} ({\mathbb R}, H^1({\mathbb S}^1))$ which satisfies the following property: 
\begin{equation}
\int_{(v,W) \in {\mathbb S}^1 \times {\mathbb R}^2} \tilde {\mathcal Q}_{\omega_\Omega} (f) \, \psi \, dv \, dW = 0, \quad \forall f \, \,  \mbox{ such that } \, \, P_{\Omega^\bot} \Omega_f  = 0, 
\label{eq:la_def_GCI}
\end{equation}
where the integral is understood in the distributional sense. We denote by $\tilde {\mathcal G}_\Omega$ the set of GCI associated to $\Omega$. It is a vector space. 
\label{def:la_GCI}
\end{definition}

The determination of $\tilde {\mathcal G}_\Omega$ is performed in the next proposition. We introduce $H^1_0({\mathbb S}^1) = \{ \phi \in H^1({\mathbb S}^1) \, \, | \, \, \int_{v \in {\mathbb S}^1} \phi(v) \, dv = 0 \}$.

\begin{proposition}
We have
$$ \tilde {\mathcal G}_\Omega = \big\{ \beta \tilde \chi_\Omega(v,W) + \phi(W), \quad \beta  \in {\mathbb R}, \quad \phi \in L^\infty_{\mbox{\scriptsize{loc}}}({\mathbb R}) \big\}, $$
where for each $W \in {\mathbb R}$, the function $v \in {\mathbb S}^1 \mapsto \tilde \chi_\Omega(v,W) $ is the unique solution in $H^1_0({\mathbb S}^1)$ of the problem 
\begin{eqnarray}
&&\hspace{-1cm}
- d \Delta_v \chi - (P_{v^\bot} \omega_\Omega(W) + W v^\bot) \cdot \nabla_v \chi = \Omega^\bot \cdot v. 
\label{eq:la_GCI_1}
\end{eqnarray}
\label{prop:la_exist_GCI}
\end{proposition}

\medskip
\noindent
{\bf Proof.} The proof starts like that of Prop. \ref{prop:exist_GCI}. Let $\Omega \in {\mathbb S}^1$ be given. The constraint $P_{\Omega^\bot} \Omega_f  = 0$ is a linear constraint on $f$, which can be written 
$\int_{(v,W) \in {\mathbb S}^1 \times {\mathbb R}} f \, \, \Omega^\bot \cdot v \,  \, dv \, dW = 0 . $
By Definition \ref{def:la_GCI}, $\psi$ is a GCI if and only if the following implication holds: 
$$ \int_{(v,W) \in {\mathbb S}^1 \times {\mathbb R}} f \, \, \Omega^\bot \cdot v \,  \, dv \, dW = 0 \quad \Longrightarrow \quad 
- \int_{(v,W) \in {\mathbb S}^1 \times {\mathbb R}} \tilde {\mathcal Q}_{\omega_\Omega} (f) \, \psi \, dv \, dW = 0,  $$
which is equivalent to the existence of a real number $\beta$ such that 
$$ - \int_{(v,W) \in {\mathbb S}^1 \times {\mathbb R}} \tilde {\mathcal Q}_{\omega_\Omega} (f) \, \psi \, dv \, dW = \beta \int_{(v,W) \in {\mathbb S}^1 \times {\mathbb R}} f \, \, \Omega^\bot \cdot v \,  \, dv \, dW , $$
for all functions $f$. 
By introducing the formal $L^2$ adjoint $\tilde {\mathcal Q}^*_{\omega_\Omega}$ of $\tilde {\mathcal Q}_{\omega_\Omega}$, this is again equivalent to the problem:
\begin{eqnarray}
&&\hspace{-1cm}
- \tilde {\mathcal Q}^*_{\omega_\Omega} \psi = \beta \, \Omega^\bot \cdot v, 
\label{eq:la_Q*psi}
\end{eqnarray}
which is nothing but the elliptic problem (\ref{eq:la_GCI_1}). We note that the different values of $W$ are decoupled in problem (\ref{eq:la_GCI_1}) and that, for any given $W \in {\mathbb R}$, it can be solved as a function of $v$ only. Therefore, from now on, we omit the dependence of $\omega_\Omega$ in $W$ and simply write it $\omega$. 

We solve this equation in the space $H^1({\mathbb S}^1)$ by using a variational formulation. For $\psi, \varphi \in H^1({\mathbb S}^1)$, we denote by $\ell(\psi, \varphi)$ the bilinear form associated to (\ref{eq:la_GCI_1}), i.e. 
$$ \ell (\psi, \varphi) = d \int_{v \in {\mathbb S}^1} \nabla_v \psi \cdot \nabla_v \varphi \, dv - \int_{v \in {\mathbb S}^1}  
\big( (\omega + W v^\bot) \cdot \nabla_v \psi \big) \, \varphi \, dv. $$
The bilinear form $\ell$ in continuous on $H^1({\mathbb S}^1)$. By Young's inequality applied to the second term, we have 
$$ \ell(\varphi, \varphi) \geq \frac{d}{2} \int_{v \in {\mathbb S}^1} |\nabla_v \varphi|^2 \, dv - C \, \int_{v \in {\mathbb S}^1} |\varphi|^2 \, dv , $$
for all $\varphi \in H^1({\mathbb S}^1)$. Therefore, there exists $\lambda$ large enough such that the bilinear form 
$$ a (\psi, \varphi) = \ell (\psi, \varphi) + \lambda \int_{v \in {\mathbb S}^1} \psi \, \varphi \, dv, $$
is coercive on $H^1({\mathbb S}^1)$. Then, by Lax-Milgram theorem, for all $\zeta \in L^2({\mathbb S}^1)$ there exists a unique solution $\psi \in H^1({\mathbb S}^1)$ such that 
\begin{eqnarray}
&&\hspace{-1cm}
a (\psi, \varphi) = \int_{v \in {\mathbb S}^1} \zeta \, \varphi \, dv, \quad \forall \varphi \in H^1({\mathbb S}^1) , 
\label{eq:la_var_psi}
\end{eqnarray}
and the mapping $T^\lambda$ which to each $\zeta \in L^2({\mathbb S}^1)$ associates this solution $\psi \in H^1({\mathbb S}^1)$ is a bounded linear operator. By the compact embedding of $H^1({\mathbb S}^1)$ into $L^2({\mathbb S}^1)$, the mapping $T^\lambda$ is a compact operator of $L^2({\mathbb S}^1)$. 

Now, we specify $\zeta = \zeta_0:= \beta v \cdot \Omega^\bot$. $\zeta_0$ is a function of $L^2({\mathbb S}^1)$. The variational solution $\psi$ of (\ref{eq:la_Q*psi}) can be written: 
$$ a(\psi, \varphi) = \int_{v \in {\mathbb S}^1} (\zeta_0 + \lambda \psi) \, \varphi \, dv, \quad \forall \varphi \in H^1({\mathbb S}^1) , $$
or equivalently
$$ \psi = T^\lambda (\zeta_0 + \lambda \psi). $$
This is a fixed point equation. Changing unknown to $\xi = \zeta_0 + \lambda \psi$, the equation is transformed into 
\begin{eqnarray}
&&\hspace{-1cm}
(\mbox{Id} - \lambda T^\lambda) \xi = \zeta_0.
\label{eq:la_fixpoint}
\end{eqnarray}
We denote by $\mbox{Im}(\mbox{Id} - \lambda T^\lambda)$ and $\mbox{Ker}(\mbox{Id} - \lambda T^{\lambda *})$ the image of $\mbox{Id} - \lambda T^\lambda$ and the kernel of its adjoint respectively (where $T^{\lambda *}$ denotes the adjoint of $T^\lambda$). Eq (\ref{eq:la_fixpoint}) has a solution if and only if $\zeta_0 \in \mbox{Im}(\mbox{Id} - \lambda T^\lambda)$. Since $T^\lambda$ is compact, we can apply the Fredhlom alternative and this condition is equivalent to $\zeta_0 \in \big( \mbox{Ker}(\mbox{Id} - \lambda T^{\lambda *}) \big)^\bot$. 

We show that $\mbox{Ker}(\mbox{Id} - \lambda T^{\lambda *}) = \mbox{Span} \{ \tilde M_\omega \}$, where, by abuse of notation, we denote by $\tilde M_\omega$ the function $v \to \tilde M_{\omega_\Omega} (v,W)$, for the considered particular value of $W$. First, $T^{\lambda *}$ is defined as follows: let $\zeta \in L^2({\mathbb R})$. Then, $\psi = T^{\lambda *} \zeta$ if and only if $\psi$ is the solution of the variational formulation: 
$$ a(\varphi, \psi) = \int_{v \in {\mathbb S}^1} \zeta \, \varphi \, dv, \quad \forall \varphi \in H^1({\mathbb S}^1) , $$
or equivalently, using Green's formula: 
\begin{eqnarray}
&&\hspace{-1cm}
d \int_{v \in {\mathbb S}^1} \nabla_v \psi \cdot \nabla_v \varphi \, dv + \int_{v \in {\mathbb S}^1}  
\nabla_v \cdot \big( (\omega + W v^\bot) \psi \big) \, \, \varphi \, dv + \lambda \int_{v \in {\mathbb S}^1} \psi \, \varphi \, dv = \nonumber \\
&&\hspace{6cm}
= \int_{v \in {\mathbb S}^1} \zeta \, \varphi \, dv, \quad \forall \varphi \in H^1({\mathbb S}^1) . 
\label{eq:la_var_GCI}
\end{eqnarray}
When $\zeta = \tilde M_\omega$, we see that this variational formulation is solved with $\psi = \frac{1}{\lambda} \tilde M_\omega$. This is due to the fact that, by construction, $\tilde M_\omega$ cancels the first two terms of (\ref{eq:la_var_GCI}). Therefore, $T^{\lambda *} \tilde M_\omega = \frac{1}{\lambda} \tilde M_\omega$, or $(\mbox{Id} - \lambda T^{\lambda *}) \tilde M_\omega = 0$. Thus $\mbox{Span} \{ \tilde M_\omega \} \subset \mbox{Ker}(\mbox{Id} - \lambda T^{\lambda *})$. Reciprocally, let $\mu \in \mbox{Ker}(\mbox{Id} - \lambda T^{\lambda *})$. Then $T^{\lambda *} \mu = \frac{1}{\lambda} \mu$. Inserting $\psi = \frac{1}{\lambda} \mu$ and $\zeta = \mu$ in (\ref{eq:la_var_GCI}), we see that $\mu$ satisfies 
\begin{eqnarray*}
&&\hspace{-1cm}
d \int_{v \in {\mathbb S}^1} \nabla_v \mu \cdot \nabla_v \varphi \, dv + \int_{v \in {\mathbb S}^1}  
\nabla_v \cdot \big( (\omega + W v^\bot) \mu \big) \, \, \varphi \, dv  = 0,  \quad \forall \varphi \in H^1({\mathbb S}^1) , 
\end{eqnarray*}
which is the weak formulation of: 
\begin{eqnarray*}
&&\hspace{-1cm}
d \Delta_v \mu - \nabla_v \cdot \big( (P_{v^\bot} \omega_\Omega(W) + W v^\bot) \mu \big) = 0. 
\end{eqnarray*}
By Lemma \ref{lem:la_equilibria_Qom}, we know that the only solutions to this equation are proportional to $\tilde M_\omega$. This shows that $\mbox{Ker}(\mbox{Id} - \lambda T^{\lambda *}) \subset \mbox{Span} \{ \tilde M_\omega \}$ and finally proves the identity of these two spaces. 

Now, (\ref{eq:la_fixpoint}) has a solution if and only if $\zeta_0 \in \big( \mbox{Span} \{ \tilde M_\omega \} \big)^\bot$. We compute: 
\begin{eqnarray*}
\int_{v \in {\mathbb S}^1} \zeta_0(v) \, \tilde M_\omega (v) \, dv &=& \beta \, \Omega^\bot \cdot \int_{v \in {\mathbb S}^1} v M_{\omega_\Omega} (v,W) \, dv \\
&=& \beta \, \tilde c_1(W) \, \Omega^\bot \, \Psi_{\omega_\Omega} (W) \\
&=& 0, 
\end{eqnarray*}
by virtue of (\ref{eq:la_tildeom=Om}). Consequently, there exists a solution in $H^1({\mathbb S}^1)$ to (\ref{eq:la_fixpoint}). 

Now, the Fredholm theory also tells that  $\mbox{dim}(\mbox{Ker}(\mbox{Id} - \lambda T^{\lambda})) =\mbox{dim}(\mbox{Ker}(\mbox{Id} - \lambda T^{\lambda *})) = 1$, where dim stands for the dimension of a space. But, we easily see that the constants belong to $\mbox{Ker}(\mbox{Id} - \lambda T^{\lambda})$. Indeed, $\psi = \frac{1}{\lambda}$ solves the variational formulation (\ref{eq:la_var_psi}) for $\zeta = 1$. Therefore, $T^\lambda 1 = \frac{1}{\lambda}$ and $(\mbox{Id} - \lambda T^{\lambda}) 1 = 0$. It follows that $\mbox{Ker}(\mbox{Id} - \lambda T^{\lambda}) = \mbox{Span} \{ 1 \}$. Therefore, the general solution of (\ref{eq:la_fixpoint}) is obtained from any particular solution by adding an arbitrary constant. We can select a unique solution, denoted by $\psi_\beta$ by imposing the extra constraint that $\int_{v \in {\mathbb S}^1} \psi_\beta \, dv = 0$. We realize that $\psi_\beta = \beta \psi_1$ (which follows easily from the uniqueness).

Now, we construct the function $\tilde \chi_\Omega(v,W)$ such that for all $W \in {\mathbb R}$, the function $v \mapsto \tilde \chi_\Omega(v,W)$ coincides with the function $\psi_1$ obtained by the construction above for the considered value of $W$. This function is a solution of (\ref{eq:la_Q*psi}) with $\beta = 1$. We obtain a solution of (\ref{eq:la_Q*psi}) for an arbitrary $\beta$ by taking $\beta \tilde \chi_\Omega(v,W)$. Now, suppose that there are two solutions of (\ref{eq:la_Q*psi}) for the same value of $\beta$. The difference is a solution of (\ref{eq:la_Q*psi}) for $\beta = 0$. We obtain such solutions by following the same steps above, except that the right-hand side $\zeta_0$ is now equal to $0$. The corresponding changed unknown $\xi$ solves the homogeneous version of  (\ref{eq:la_fixpoint}), i.e. is an element of $\mbox{Ker}(\mbox{Id} - \lambda T^{\lambda})$. Therefore, $\xi$ is a constant in $v$, and so is $\psi$. When restoring the dependence in $W$, this means that the solutions of (\ref{eq:la_Q*psi}) for $\beta = 0$ consist of the functions of $W$ only. Therefore, any solution of $(\ref{eq:la_Q*psi})$ is written $\beta \tilde \chi_\Omega(v,W) + \phi(W)$, with an arbitrary function $\phi(W)$. Since $\beta$ is any real number, the set of GCI is spanned by such elements when $\beta \in {\mathbb R}$ and the function $\phi(W)$ are arbitrary. This is what is stated in Proposition \ref{prop:la_exist_GCI}, and ends the proof. \endproof

%%%%%%%%%%%%%%%%%%%%%%%%%%%%%%%%%%%%%%%%%%%%%%%%%%%%%%%%%%%%%%%%%%%%%%%%%%%%%%%%%%%%%%%%%%%%%%%%
\subsection{Hydrodynamic limit $\varepsilon \to 0$}
\label{subsec:la_hydro}

This section is devoted to the proof of Theorem \ref{thm:la_hydro}.

\medskip
\noindent
{\bf Proof of Theorem \ref{thm:la_hydro}.} The beginning of the proof is analogous to that of Theorem \ref{thm:small_hydro}. Let $f^\varepsilon$ be a solution of (\ref{eq:la_kin_f_0}) with $\omega_{f^\varepsilon}$ given by (\ref{eq:la_def_omf}). Thanks to Proposition \ref{prop:la_equilibria_Q}, there exist two functions $\rho_W(x,t)$ and $\Omega(x,t)$ where, for fixed $(x,t)$, the function $W \to \rho_W(x,t)$ and the vector $\Omega(x,t)$ belong to $L^1({\mathbb R})$ and ${\mathbb S}^1$ respectively, such that (\ref{eq:la_smallkin_f0}) holds. The derivation of (\ref{eq:la_smallhydro_rhoW}) is also similar as in the proof of Theorem \ref{thm:small_hydro}.

We concentrate on the proof of (\ref{eq:la_smallhydro_Omega}). We omit the superscript $0$ on $f^0$ for the sake of clarity. Again, the beginning of the proof is similar and we end up getting 
\begin{eqnarray}
&&\hspace{-1cm}
\int_{(v,W) \in {\mathbb S}^1 \times {\mathbb R}} {\mathcal T} (\rho_W \tilde M_{\omega_\Omega}) \, \, \tilde \chi_\Omega  \, \, dv \, dW =  0, 
\label{eq:la_int_GCI_eps_1}
\end{eqnarray}
with ${\mathcal T} = \partial_t + v \cdot \nabla_x$. We compute:
\begin{eqnarray}
&&\hspace{-1cm}
{\mathcal T} (\rho_W \tilde M_{\omega_\Omega}) =  \tilde M_{\omega_\Omega} \{ A_\rho + \rho_W A_\Omega \}, 
\label{eq:la_pat+vnabxrhoMom}
\end{eqnarray}
where, using (\ref{eq:la_smallhydro_rhoW}),  
\begin{eqnarray}
A_\rho &=&  {\mathcal T} \rho_W = (\partial_t + \tilde c_1 \Omega \cdot \nabla_x) \rho_W + (v - \tilde c_1 \Omega) \cdot \nabla_x \rho_W \nonumber \\
&=& - \tilde c_1 \rho_W (\nabla_x \cdot  \Omega) + (v - \tilde c_1 \Omega) \cdot \nabla_x \rho_W, 
\label{eq:la_Arho}
\end{eqnarray}
and $A_\Omega =  {\mathcal T} \ln \tilde M_{\omega_\Omega}$ i.e.
\begin{eqnarray}
A_\Omega (x,t) &=&  \frac{\partial \ln \tilde M_{\omega_\Omega}}{\partial \Omega} \Big|_{\Omega(x,t)} \, \, {\mathcal T} \Omega (x,t)  . 
\label{eq:la_AOmega}
\end{eqnarray}

The quantity $\frac{\partial \ln \tilde M_{\omega_\Omega}}{\partial \Omega} |_\Omega$ is a linear form acting on the tangent line to ${\mathbb S}^1$ at $\Omega$. By the chain rule:
\begin{eqnarray}
\frac{\partial \ln \tilde M_{\omega_\Omega}}{\partial \Omega}\Big|_\Omega &=& \frac{\partial \ln \tilde M_{\omega}}{\partial \omega}\Big|_{\omega_\Omega} \, \, \frac{\partial \omega_\Omega}{\partial \Omega}\Big|_\Omega. 
\label{eq:la_ident_aux_6}
\end{eqnarray}
where $\frac{\partial \ln \tilde M_{\omega}}{\partial \omega}|_{\omega_\Omega}$ is a linear form acting on the tangent line to ${\mathbb S}^1$ at $\omega_\Omega$ and $\frac{\partial \omega_\Omega}{\partial \Omega}|_\Omega$ is a linear application from the tangent line to ${\mathbb S}^1$ at $\Omega$ into the tangent line to ${\mathbb S}^1$ at $\omega_\Omega$. We compute the first factor. Since $\ln \tilde M_{\omega} = \ln \Phi_W(\theta)$ with $\theta = \widehat{(\omega,v)}$ and $\Phi_W$ given at Lemma~\ref{lem:la_exist_Phi}, we can write, thanks to (\ref{eq:la_Phi_deriv}):
\begin{eqnarray}
\frac{\partial \ln \tilde M_{\omega}}{\partial \omega}\Big|_{\omega} \, \, \tau &=& \frac{\partial \ln \Phi_W}{\partial \theta} \Big|_{\widehat{(\omega,v)}} \, \, \frac{\partial \widehat{(\omega,v)}}{\partial \omega}\Big|_{\omega} \, \, \tau \nonumber \\
&=& \big( \, \,  \frac{1}{d} \, (\omega^\bot \cdot v - W) \, - \frac{C}{\tilde M_\omega}  \, \big) \, \,  \omega^\bot \cdot \tau , 
\label{eq:la_parlnMpartom}
\end{eqnarray}
for all tangent vectors $\tau$ to ${\mathbb S}^1$ at $\omega$. We now compute the second factor. We differentiate relation (\ref{eq:la_tildeom=Om}) with respect to $\Omega$ and we get that 
\begin{eqnarray}
\frac{\partial \omega_\Omega}{\partial \Omega}\Big|_\Omega &= &  \big( \, \frac{\partial \Psi_\omega}{\partial \omega}\Big|_{\omega_\Omega} \, \big)^{-1}. 
\label{eq:la_ident_aux_3}
\end{eqnarray}
Let $\tau$ be a tangent vector to ${\mathbb S}^1$ at $\omega$. We have, using Lemma \ref{lem:la_c1_indep_om} and Eqs. (\ref{eq:la_def_u_tildeom_1}), (\ref{eq:la_parlnMpartom}), 
\begin{eqnarray}
\frac{\partial \Psi_\omega}{\partial \omega}\Big|_{\omega} \, \tau  &=& \frac{1}{\tilde c_1} \,  \frac{\partial u_\omega}{\partial \omega}\Big|_{\omega} \, \tau \nonumber \\
&=& \frac{1}{\tilde c_1} \, \int_{v \in {\mathbb S}^1} \big( \frac{\partial \tilde M_\omega}{\partial \omega}\Big|_{\omega} \tau \big) \, v \, dv \nonumber \\
&=& \frac{1}{\tilde c_1} \, \int_{v \in {\mathbb S}^1} \big( \, \,  \frac{1}{d} \, (\omega^\bot \cdot v - W) \, \tilde M_\omega \, - C  \, \big)  \, v \, dv \, \,  \omega^\bot \cdot \tau \nonumber \\
&=& \frac{1}{d \, \tilde c_1} \, \int_{v \in {\mathbb S}^1}   (\omega^\bot \cdot v - W) \, \tilde M_\omega  \, v \, dv \, \,  \omega^\bot \cdot \tau ,
\label{eq:la_ident_aux_2}
\end{eqnarray}
where the term in factor of $C$ vanishes by oddness considerations. Now, we note that  
\begin{eqnarray}
\int_{v \in {\mathbb S}^1}   (\omega^\bot \cdot v - W) \, \tilde M_\omega  \, \, v \cdot \Psi_\omega \, \, dv = 0 . 
\label{eq:la_ident_aux}
\end{eqnarray}
Indeed, differentiating the equation $|u_\omega| = \tilde c_1$ with respect to $\omega$, we get
\begin{eqnarray*}
0 = \frac{\partial |u_\omega|}{\partial \omega}\Big|_{\omega} \, \tau  &=& \Psi_\omega \cdot \frac{\partial u_\omega}{\partial \omega}\Big|_{\omega} \, \tau \\
&=& \Psi_\omega \cdot \int_{v \in {\mathbb S}^1} \big( \frac{\partial \tilde M_\omega}{\partial \omega}\Big|_{\omega} \tau \big) \, v \, dv  \\ 
&=& \frac{1}{d} \, \int_{v \in {\mathbb S}^1}   (\omega^\bot \cdot v - W) \, \tilde M_\omega \,  (v \cdot \Psi_\omega)  \, dv \, \, \,  \omega^\bot \cdot \tau , 
\end{eqnarray*}
which implies (\ref{eq:la_ident_aux}). Then, decomposing $v = (v \cdot \Psi_\omega) \Psi_\omega + (v \cdot \Psi_\omega^\bot) \Psi_\omega^\bot$, (\ref{eq:la_ident_aux_2}) leads to
\begin{eqnarray}
\frac{\partial \Psi_\omega}{\partial \omega}\Big|_{\omega} \, \tau  &=& \frac{1}{d \, \tilde c_1} \, \int_{v \in {\mathbb S}^1}   (\omega^\bot \cdot v - W) \, \tilde M_\omega  \, (v \cdot \Psi_\omega^\bot) \,  \, dv \, \, \, (\omega^\bot \cdot \tau) \, \Psi_\omega^\bot \label{eq:la_ident_aux_9} \\
&=& \lambda \, (\omega^\bot \cdot \tau) \, \Psi_\omega^\bot, 
\label{eq:la_ident_aux_8}
\end{eqnarray}
with 
\begin{eqnarray*}
\lambda  &=& \frac{1}{d \, \tilde c_1} \, \int_{v \in {\mathbb S}^1}   (\omega^\bot \cdot v)  \, (\Psi_\omega^\bot \cdot v) \, \tilde M_\omega  \,  \,  \, dv , 
\end{eqnarray*}
using that the second term in (\ref{eq:la_ident_aux_9}) vanishes, thanks to the definition of $\Psi_\omega$. 
Now, using (\ref{eq:la_ident_aux_3}) and (\ref{eq:la_ident_aux_8}), we get, for all tangent vector $\bar \tau$ to ${\mathbb S}^1$ at $\Omega$ 
\begin{eqnarray}
\frac{\partial \omega_\Omega}{\partial \Omega}\Big|_\Omega \, \bar \tau = \frac{1}{\lambda} \, (\Omega^\bot \cdot \bar \tau) \, \omega^\bot.  
\label{eq:la_ident_aux_4}
\end{eqnarray}
Then, inserting (\ref{eq:la_parlnMpartom}) and (\ref{eq:la_ident_aux_4}) into (\ref{eq:la_ident_aux_6}), we get, for all tangent vector $\bar \tau$ to ${\mathbb S}^1$ at $\Omega$
\begin{eqnarray}
\frac{\partial \ln \tilde M_{\omega_\Omega}}{\partial \Omega}\Big|_\Omega \bar \tau &=& \frac{1}{\lambda} \, \big( \, \,  \frac{1}{d} \, (\omega_\Omega^\bot \cdot v - W) \, - \frac{C}{\tilde M_{\omega_\Omega}}  \, \big) \, \Omega^\bot \cdot \bar \tau,
\label{eq:la_ident_aux_7}
\end{eqnarray}
with 
\begin{eqnarray*}
\lambda  &=& \frac{1}{d \, \tilde c_1} \, \int_{v \in {\mathbb S}^1}   (\omega_\Omega^\bot \cdot v)  \, (\Omega^\bot \cdot v) \, \tilde M_{\omega_\Omega}  \,  \,  \, dv . 
\end{eqnarray*}
We note that $\lambda = \lambda(W)$ only depends on $W$. Indeed, introducing $\theta = \widehat{(\omega_\Omega,v)}$ and $\psi(W)= \widehat{(\omega_\Omega,\Omega)}$, we can write
\begin{eqnarray}
\lambda(W)  &=& \frac{1}{d \, \tilde c_1(W)} \, \int_0^{2 \pi} \sin \theta  \, \sin (\theta - \psi(W)) \, \Phi_W(\theta) \, d \theta, 
\label{eq:la_def_lambda}
\end{eqnarray}
which clearly defines a function of $W$ only.

\medskip
Inserting (\ref{eq:la_ident_aux_7}) into (\ref{eq:la_AOmega}) and collecting it with (\ref{eq:la_Arho}) to insert it into (\ref{eq:la_pat+vnabxrhoMom}), we get
\begin{eqnarray*}
&&\hspace{-1cm}
{\mathcal T} (\rho_W \tilde M_\Omega) 
= \tilde M_{\omega_\Omega} \big\{ - \tilde c_1(W) \, \rho_W \, (\nabla_x \cdot  \Omega) + (v - \tilde c_1(W) \, \Omega) \cdot \nabla_x \rho_W \nonumber \\
&&\hspace{2cm}
+   \frac{\rho_W}{\lambda(W)} \, \big( \,   \frac{1}{d} \, ( \omega_\Omega^\bot \cdot v - W ) \,  - \frac{C(W)}{\tilde M_{\omega_\Omega}}  \, \big) \, \, \,  \Omega^\bot \cdot \, (\partial_t + v \cdot \nabla_x) \Omega  \, \, \big\} 
\end{eqnarray*}
which can be rewritten, by decomposing $v = (v \cdot \Omega) \Omega + (v \cdot \Omega^\bot) \Omega^\bot$:
\begin{eqnarray}
&&\hspace{-1cm}
{\mathcal T} (\rho_W \tilde M_\Omega) =  - \tilde c_1(W) \, \rho_W \, \tilde M_{\omega_\Omega} \, \, \, \nabla_x \cdot  \Omega + (v \cdot \Omega - \tilde c_1(W)) \, \tilde M_{\omega_\Omega} \, \, \, \Omega \cdot \nabla_x \rho_W  \nonumber \\
&&\hspace{0.5cm}
+ (v \cdot \Omega^\bot ) \, \tilde M_{\omega_\Omega} \, \, \, \Omega^\bot \cdot \nabla_x \rho_W +   \frac{\rho_W}{\lambda(W)} \, \big( \,   \frac{1}{d} \, ( \omega_\Omega^\bot \cdot v - W ) \, \tilde M_{\omega_\Omega}  - C(W)\, \big) \, \, \,   \partial_t  \Omega \cdot \Omega^\bot \nonumber \\
&&\hspace{0.5cm}
+   \frac{\rho_W}{\lambda(W)} \, \big( \,   \frac{1}{d} \, ( \omega_\Omega^\bot \cdot v - W ) \, \tilde M_{\omega_\Omega}  - C(W)\, \big) \,  (v \cdot \Omega) \, \, \,  (\Omega \cdot \nabla_x) \Omega \cdot \Omega^\bot \nonumber \\
&&\hspace{0.5cm}
+   \frac{\rho_W}{\lambda(W)} \, \big( \,   \frac{1}{d} \, ( \omega_\Omega^\bot \cdot v - W ) \, \tilde M_{\omega_\Omega}  - C(W)\, \big) \,  (v \cdot \Omega^\bot) \, \, \,   (\Omega^\bot \cdot \nabla_x) \Omega \cdot \Omega^\bot
\label{eq:la_pat+vnabxrhoMom_2}
\end{eqnarray}

Now, we define the following quantities:
\begin{eqnarray*}
a_1 &=& \frac{1}{\lambda(W) \, d} \, \int_{v \in {\mathbb S}^1}  ( \omega_\Omega^\bot \cdot v - W ) \, \tilde M_{\omega_\Omega} 
\, \tilde \chi_\Omega (v,W) \, dv, 
\\
a_2 &=& \frac{1}{\lambda(W) \, d} \, \int_{v \in {\mathbb S}^1}  ( \omega_\Omega^\bot \cdot v - W ) \, (v \cdot \Omega) \, \tilde M_{\omega_\Omega} \, \tilde \chi_\Omega (v,W) \, dv\nonumber \\
&&\hspace{3cm}
- \frac{C(W)}{\lambda(W)} \, \int_{v \in {\mathbb S}^1} (v \cdot \Omega) \,  \tilde \chi_\Omega (v,W) \, dv, 
\\
a_3 &=& \frac{1}{\lambda(W) \, d} \, \int_{v \in {\mathbb S}^1}  ( \omega_\Omega^\bot \cdot v - W ) \, (v \cdot \Omega^\bot) \, \tilde M_{\omega_\Omega} \, \tilde \chi_\Omega (v,W) \, dv \nonumber \\
&&\hspace{3cm}
- \frac{C(W)}{\lambda(W)} \, \int_{v \in {\mathbb S}^1} (v \cdot \Omega^\bot) \,  \tilde \chi_\Omega (v,W) \, dv, 
\\
a_4 &=& - \tilde c_1(W) \, \int_{v \in {\mathbb S}^1} \tilde M_{\omega_\Omega}(v,W) \, \tilde \chi_\Omega (v,W) \, dv, 
\\
a_5 &=& \int_{v \in {\mathbb S}^1} (v \cdot \Omega^\bot) \, \tilde M_{\omega_\Omega} \, \tilde \chi_\Omega (v,W) \, dv, 
\\
a_6 &=& \int_{v \in {\mathbb S}^1} (v \cdot \Omega - \tilde c_1(W)) \, \tilde M_{\omega_\Omega} \, \tilde \chi_\Omega (v,W) \, dv. 
\end{eqnarray*}
From (\ref{eq:la_GCI_1}), the function $\tilde \chi_\Omega(v,W)$ can be written $\tilde \chi_\Omega(v,W) = X_W(\theta)$, with $\theta = \widehat{(\omega_\Omega,v)}$ and $X_W$ the unique $2 \pi$-periodic solution of 
\begin{eqnarray}
&&\hspace{-1cm}
- X_W''+(\sin \theta - W) X_W' = \sin(\theta - \psi(W)), \qquad \int_0^{2 \pi} X_W (\theta) \, d \theta = 0, 
\label{eq:la_def_X}
\end{eqnarray}
with $\psi(W) = \widehat{(\omega_\Omega,\Omega)}$. Therefore, the quantities $a_1$ through $a_6$ can be written:
\begin{eqnarray}
a_1 &=& \frac{1}{\lambda(W) \, d} \, \int_0^{2 \pi}  ( \,\sin \theta - W \,) \, \Phi_W(\theta) \, X_W(\theta) \, d \theta, 
\label{eq:def_a1} \\
a_2 &=& \frac{1}{\lambda(W) \, d} \, \int_0^{2 \pi}  ( \, \sin \theta - W \, ) \, \cos (\,\theta - \psi(W)\,)  \, \Phi_W(\theta) \, X_W(\theta) \, d \theta, \nonumber \\
&&\hspace{3cm}
- \frac{C(W)}{\lambda(W)} \, \int_0^{2 \pi} \cos (\, \theta - \psi(W)\, ) \,  X_W(\theta) \, d \theta, 
\label{eq:def_a2} \\
a_3 &=& \frac{1}{\lambda(W) \, d} \, \int_0^{2 \pi}  ( \, \sin  \theta - W \, ) \, \sin (\, \theta - \psi(W)\, ) \, \Phi_W(\theta) \, X_W(\theta) \, d \theta, \nonumber \\
&&\hspace{3cm}
- \frac{C(W)}{\lambda(W)} \, \int_0^{2 \pi} \sin (\, \theta - \psi(W)\, ) \,  X_W(\theta) \, d \theta, 
\label{eq:def_a3} \\
a_4 &=& - \tilde c_1(W) \, \int_0^{2 \pi} \Phi_W(\theta) \, X_W(\theta) \, d \theta, 
\label{eq:def_a4} \\
a_5 &=& \int_0^{2 \pi} \sin (\,\theta - \psi(W) \,) \, \Phi_W(\theta) \, X_W(\theta) \, d \theta, 
\label{eq:def_a5} \\
a_6 &=& \int_0^{2 \pi} \big( \cos ( \,\theta - \psi(W) \, ) - \tilde c_1(W) \Big) \Phi_W(\theta) \, X_W(\theta) \, d \theta. 
\label{eq:def_a6}
\end{eqnarray}
We notice that they depend only on $W$ and we shall denote them by $a_k(W)$, $k=1, \ldots , 6$. We now define the following moments of $\rho_W$: 
\begin{eqnarray}
m_k[\rho_W] &=& \int_{w \in {\mathbb R}}  a_k(W) \, \rho_W \, dW, \quad k=1, \ldots , 6 . 
\label{eq:la_mom_rhoW}
\end{eqnarray}
With these definitions, we can multiply (\ref{eq:la_pat+vnabxrhoMom_2}) by $\chi_\Omega \, \Omega^\bot$ and integrate the resulting expression on $(v,W) \in {\mathbb S}^1 \times {\mathbb R}$. Thanks to (\ref{eq:la_int_GCI_eps_1}), we get (\ref{eq:la_smallhydro_Omega}), which ends the proof of Theorem \ref{thm:la_hydro}. \endproof

%%%%%%%%%%%%%%%%%%%%%%%%%%%%%%%%%%%%%%%%%%%%%%%%%%%%%%%%%%%%%%%%%%%%%%%%%%%%%%%%%%%%%%%%%%%%%%%%
%%%%%%%%%%%%%%%%%%%%%%%%%%%%%%%%%%%%%%%%%%%%%%%%%%%%%%%%%%%%%%%%%%%%%%%%%%%%%%%%%%%%%%%%%%%%%%%%
%%%%%%%%%%%%%%%%%%%%%%%%%%%%%%%%%%%%%%%%%%%%%%%%%%%%%%%%%%%%%%%%%%%%%%%%%%%%%%%%%%%%%%%%%%%%%%%%
%%%%%%%%%%%%%%%%%%%%%%%%%%%%%%%%%%%%%%%%%%%%%%%%%%%%%%%%%%%%%%%%%%%%%%%%%%%%%%%%%%%%%%%%%%%%%%%%
\setcounter{equation}{0}
\section{Small angular velocity limit of the SOHR-L model. Proofs}
\label{sec:la_small_zeta_proof}

\subsection{Proof of Proposition \ref{prop:la_limit_SOHRL_small_vel}.} 
\label{subsec:app_small_ang_vel}

We first need to let $\zeta \to 0$ in the coefficients (\ref{eq:la_mom_rhoW_zeta}) of the SOHR-L model. For this, we need the following lemma:

\begin{lemma}
(i) For fixed $W$, the functions $\Phi_W$ and $X_W$ respectively given by (\ref{eq:la_def_Phi}) and (\ref{eq:la_def_X}) are such that
\begin{eqnarray}
&&\hspace{-1cm}
\Phi_{\zeta W} (\theta) = \Phi_0(\theta) + \zeta W \Phi_1(\theta) + {\mathcal O} (\zeta^2), \quad X_{\zeta W} (\theta) = X_0(\theta) + \zeta W X_1(\theta) + {\mathcal O} (\zeta^2), 
\label{eq:la_phi_X_expan_zeta}
\end{eqnarray}
where $\Phi_0$, $X_1$ are even and $X_0$, $\Phi_1$ are odd functions of $\theta$. Furthermore, we have 
\begin{eqnarray}
&&\hspace{-1cm}
\Phi_0(\theta) = M_\Omega(v) = \frac{1}{Z_d} e^{\frac{\cos \theta}{d}}, \qquad X_0 (\theta)  =  \chi_\Omega(v) = g (\theta), 
\label{eq:la_phi_X_zeta=0}
\end{eqnarray}
where $\theta = \widehat{(\Omega,v)}$, $M_\Omega(v)$ and $\chi_\Omega(v)$ are the VMF distribution (\ref{eq:sm_smallkin_VMF}) and the GCI (\ref{eq:sm_prb_chi}) associated to the small angular velocity case, $g$ is given by (\ref{eq:sm_prb_g}) or (\ref{eq:sm_closform_g}) and $Z_d$ is the normalization factor (\ref{eq:sm_smallkin_VMF}) . 

\medskip
\noindent
(ii) We have 
\begin{eqnarray}
&&\hspace{-1cm}
\tilde c_1(\zeta W) = \tilde c_1 (0) + {\mathcal O}(\zeta^2), \quad \tilde c_1(0) = \int_0^{2\pi} \Phi_0(\theta) \, \cos \theta \, d \theta = c_1, 
\label{eq:la_c1_zeta=0} \\
&&\hspace{-1cm}
\lambda(\zeta W) = \lambda (0) + {\mathcal O}(\zeta^2), \quad \lambda(0) = \frac{1}{d \, c_1} \int_0^{2\pi} \Phi_0(\theta) \, \sin^2 \theta \, d \theta, 
\label{eq:la_lambda_zeta=0} \\
&&\hspace{-1cm}
a_1(\zeta W) = a_1 (0) + {\mathcal O}(\zeta^2), \quad a_1(0) = \frac{1}{d \, \lambda(0)} \int_0^{2\pi} \Phi_0(\theta) \, X_0(\theta) \, \sin \theta \, d \theta, 
\label{eq:la_a1_zeta=0} \\
&&\hspace{-1cm}
a_2(\zeta W) = a_2 (0) + {\mathcal O}(\zeta^2), \quad a_2(0) = \frac{1}{d \, \lambda(0)} \int_0^{2\pi} \Phi_0(\theta) \, X_0(\theta) \, \cos \theta \, \sin \theta \, d \theta, 
\label{eq:la_a2_zeta=0} \\
&&\hspace{-1cm}
a_5(\zeta W) = a_5 (0) + {\mathcal O}(\zeta^2), \quad a_5(0) = d \, \lambda (0) a_1(0),
\label{eq:la_a5_zeta=0} 
\end{eqnarray}
where $c_1$ is the order parameter of the VMF distribution in the small angular case, given by (\ref{eq:sm_smallkin_cur_equi}). 
\label{lem:la_lim_zeta_small}
\end{lemma}

\medskip
\noindent
{\bf Proof of Lemma \ref{lem:la_lim_zeta_small}.} Changing $W$ into $\zeta W$ into (\ref{eq:la_parity_phi_W}) and inserting expansions (\ref{eq:la_phi_X_expan_zeta}), we immediately get that $\Phi_0$, $X_1$ are even and $X_0$, $\Phi_1$ are odd functions of $\theta$. 

Now, changing $W$ into $\zeta W$ into (\ref{eq:la_def_Phi}) and again inserting the expansion (\ref{eq:la_phi_X_expan_zeta}), we get that $\Phi_0$ is a smooth periodic solution of 
\begin{eqnarray*}
&&\hspace{-1cm}
\Phi_0'' + \frac{1}{d} (\sin \theta \, \Phi_0)' = 0, \qquad \int_0^{2\pi} \Phi_0(\theta) \, d\theta = 1. 
\end{eqnarray*}
Such a solution is unique and given by the first eq. (\ref{eq:la_phi_X_zeta=0}). Inserting expansion (\ref{eq:la_phi_X_expan_zeta}) into  (\ref{eq_la_express_tilde_C1}) gives (\ref{eq:la_c1_zeta=0}). 

Before expanding $X_W(\theta)$, we need to expand $\psi(\zeta W) = \widehat{(\omega^\zeta_\Omega(W),\Omega)}$. We have, by (\ref{eq:la_om_zeta}) and (\ref{eq:la_tildeom=Om}), 
$$\psi(\zeta W) = \widehat{(\omega_\Omega(\zeta W),\Omega)} = \widehat{(\omega_\Omega(\zeta W),\Psi_{\omega_\Omega}(\zeta \, W))} = \widehat{(\omega,\Psi_{\omega}(\zeta \, W))}.$$ 
The last equality comes from the fact that $\psi(W)$ does not depend on the particular choice of $\omega(W)$. Then, inserting expansion (\ref{eq:la_phi_X_expan_zeta}) into (\ref{eq:la_Psi(W)_deriv}) and using the evenness of $\Phi_0$ and the oddness of $\Phi_1$, we get
$$ \Psi_{\omega}(\zeta \, W) = \omega + \frac{\beta}{c_1}  \, \zeta \, W \, \omega^\bot + {\mathcal O}(\zeta^2), \qquad \beta =   \int_0^{2\pi} \Phi_1(\theta) \, \sin \theta \, d \theta. $$
It follows that 
\begin{eqnarray}
&&\hspace{-1cm}
\psi(\zeta W) = \frac{\beta}{c_1}  \, \zeta \, W + {\mathcal O}(\zeta^2). 
\label{eq:la_psi_expan}
\end{eqnarray}
We deduce that the right-hand side of (\ref{eq:la_def_X}) (with $W$ changed into $\zeta W$) can be expanded into:
\begin{eqnarray}
&&\hspace{-1cm}
\sin (\theta - \psi(\zeta W))  = \sin \theta - \frac{\beta}{c_1}  \, \zeta \, W \, \cos \theta + {\mathcal O}(\zeta^2). 
\label{eq:la_sin_theta-psi_expan}
\end{eqnarray}
Now, inserting (\ref{eq:la_phi_X_expan_zeta}) into (\ref{eq:la_def_X}) (with $W$ changed into $\zeta W$), we find that $X_0(\theta)$ is a smooth periodic solution of
\begin{eqnarray*}
&&\hspace{-1cm}
- X_0'' + \frac{1}{d} \sin \theta \, X_0' = \sin \theta, \qquad \int_0^{2\pi} X_0(\theta) \, d\theta = 0. 
\end{eqnarray*}
Now, by comparing with (\ref{eq:sm_prb_chi}), we realize that the second relation (\ref{eq:la_phi_X_zeta=0}) holds. 

Now, inserting the expansions (\ref{eq:la_phi_X_expan_zeta}), (\ref{eq:la_c1_zeta=0}) and (\ref{eq:la_psi_expan}) successively into 
(\ref{eq:la_def_lambda}) and (\ref{eq:def_a1}), (\ref{eq:def_a2}), (\ref{eq:def_a5}), we get (\ref{eq:la_lambda_zeta=0}), (\ref{eq:la_a1_zeta=0}), (\ref{eq:la_a2_zeta=0}), (\ref{eq:la_a5_zeta=0}), which ends the proof of the Lemma. \endproof 

\medskip
\noindent
{\bf End of proof of Proposition \ref{prop:la_limit_SOHRL_small_vel}.} Since $a_3$, $a_4$ and $a_6$ are even functions of $W$, the expansion $a_k(\zeta W) = {\mathcal O}(\zeta)$ for $k = 3, \, 4, \, 6$, when $\zeta \to 0$ holds. Therefore, in this limit, $m_k^\zeta[\rho_W] \to 0$ for $k = 3, \, 4, \, 6$. Now, using (\ref{eq:la_a1_zeta=0}), (\ref{eq:la_a2_zeta=0}), (\ref{eq:la_a5_zeta=0}), we have $m_k^\zeta[\rho_W] \to a_k(0) \, \rho$, with $\rho$ given by (\ref{eq:sm_smallhydro_moments}). This leads~to:
\begin{eqnarray*}
&&\hspace{-1cm}
\rho \, a_1(0) \partial_t \Omega + \rho \, a_2(0) \, (\Omega \cdot \nabla_x)  \Omega + a_5(0) \, P_{\Omega^\bot} \nabla_x \rho  = 0. 
\end{eqnarray*}
Dividing by $a_1(0)$, we get (\ref{eq:la_limSOHRL_Omega}) with the coefficients $c_2$ and $c_5$ given by:
\begin{eqnarray}
&&\hspace{-1cm}
c_2 = \frac{a_2(0)}{a_1(0)}, \qquad c_5 = \frac{a_5(0)}{a_1(0)}.
\label{eq:la_c2_Theta}
\end{eqnarray}
Now, using (\ref{eq:la_a1_zeta=0}), (\ref{eq:la_a2_zeta=0}), (\ref{eq:la_a5_zeta=0}) together with (\ref{eq:la_phi_X_zeta=0}), we notice that the first eq. (\ref{eq:la_c2_Theta}) is nothing but (\ref{eq:sm_express_c2}), while the second eq. (\ref{eq:la_c2_Theta}) can be recast into (\ref{eq:la_limSOHRL_Theta}). Finally, Eq. (\ref{eq:la_limSOHRL_rhoW}) directly follows from (\ref{eq:la_smallhydro_rhoW}) and (\ref{eq:la_c1_zeta=0}). This ends the proof of Proposition \ref{prop:la_limit_SOHRL_small_vel}. \endproof

\subsection{Proof of Proposition \ref{prop:la_expan_forder}.} 
\label{subsec:app_small_angvel_forder}

To compute the order ${\mathcal O}(\zeta)$ terms in the expansion of the SOHR-L model when $\zeta \to 0$, we need to complement Lemma \ref{lem:la_lim_zeta_small} by information about the first-order corrections to the terms $a_3$, $a_4$ and $a_6$ (see (\ref{eq:def_a3}), (\ref{eq:def_a4}), (\ref{eq:def_a6})). This is the purpose of the following lemma: 

\begin{lemma}
(i) The perturbations $\Phi_1$ and $X_1$ are the unique smooth $2 \pi$ periodic solutions to the problems
\begin{eqnarray}
&&\hspace{-1cm}
\Phi_1'' + \frac{1}{d} (\sin \theta \, \Phi_1)' = \frac{1}{d} \Phi_0', \qquad \int_0^{2\pi} \Phi_1(\theta) \, d\theta = 0, 
\label{eq:la_phi_expan_forder}\\
&&\hspace{-1cm}
X_1'' - \frac{1}{d} \,\sin \theta \, X_1' = - \frac{X_0'}{d} - \frac{\beta}{c_1} \, \cos \theta, \qquad \int_0^{2\pi} \Phi_1(\theta) \, d\theta = 0.
\label{eq:la_X_expan_forder}
\end{eqnarray}

\medskip
\noindent
We have the expansions: 
\begin{eqnarray}
&&\hspace{-1cm}
a_3(\zeta W) = a_3^1 \, \zeta \, W + {\mathcal O}(\zeta^3), \nonumber \\
&&\hspace{0cm}
a_3^1 = \frac{1}{d \, \lambda(0)} \int_0^{2\pi} \big[ - \sin \theta \, \Phi_0 \, X_0 \big(1 + \frac{\beta}{c_1} \big) + \sin^2 \theta \, ( \Phi_0 X_1 + \Phi_1 X_0 ) \big] \, d \theta
\label{eq:la_a3_zeta=0} \\
&&\hspace{-1cm}
a_4(\zeta W) = a_4^1 \, \zeta \, W + {\mathcal O}(\zeta^3), \quad  a_4^1 = c_1 \int_0^{2\pi} ( \Phi_0 X_1 + \Phi_1 X_0 ) \, d \theta, 
\label{eq:la_a4_zeta=0} \\
&&\hspace{-1cm}
a_6(\zeta W) = a_6^1 \, \zeta \, W + {\mathcal O}(\zeta^3), \quad  a_6^1 = \int_0^{2\pi} (\cos \theta - c_1 ) \, ( \Phi_0 X_1 + \Phi_1 X_0 ) \, d \theta,
\label{eq:la_a6_zeta=0} 
\end{eqnarray}

\label{lem:la_lim_zeta_forder}
\end{lemma}

\medskip
\noindent
{\bf Proof of Lemma \ref{lem:la_lim_zeta_forder}.} Eqs. (\ref{eq:la_phi_expan_forder}) and (\ref{eq:la_X_expan_forder}) follow easily from (\ref{eq:la_def_Phi}) and (\ref{eq:la_def_X}) (changing $W$ into $\zeta W$ and expanding up to second order in $\zeta$). Then, from (\ref{eq:la_Phi_deriv}) and (\ref{eq:la_parity_phi_W}), we find that the constant $C(W)$ is odd with respect to $W$. Therefore, $C(\zeta W)$ is expanded in $\zeta$ according to $ C(W) = C_1 \, \zeta \, W$, where the expression of the constant $C_1$ can be obtained from $\Phi_0$, $\Phi_1$ but will not be needed. Indeed, in the expansion of $a_3(\zeta W)$, the term containing $C$ has non contribution by oddness with respect to $\theta$. The other term can be expanding using the auxiliary computations already done in the proof of Lemma \ref{lem:la_lim_zeta_small}. They lead to the expressions (\ref{eq:la_a3_zeta=0}), (\ref{eq:la_a4_zeta=0}), (\ref{eq:la_a6_zeta=0}). \endproof

\medskip
Once Lemma \ref{lem:la_lim_zeta_forder} is proved, the proof of Proposition \ref{prop:la_expan_forder} is straighforward and left to the reader. \endproof

%%%%%%%%%%%%%%%%%%%%%%%%%%%%%%%%%%%%%%%%%%%%%%%%%%%%%%%%%%%%%%%%%%%%%%%%%%%%%%%%%%%%%%%%%%%%%%%%
%%%%%%%%%%%%%%%%%%%%%%%%%%%%%%%%%%%%%%%%%%%%%%%%%%%%%%%%%%%%%%%%%%%%%%%%%%%%%%%%%%%%%%%%%%%%%%%%
%%%%%%%%%%%%%%%%%%%%%%%%%%%%%%%%%%%%%%%%%%%%%%%%%%%%%%%%%%%%%%%%%%%%%%%%%%%%%%%%%%%%%%%%%%%%%%%%
%%%%%%%%%%%%%%%%%%%%%%%%%%%%%%%%%%%%%%%%%%%%%%%%%%%%%%%%%%%%%%%%%%%%%%%%%%%%%%%%%%%%%%%%%%%%%%%%
\setcounter{equation}{0}
\section{Graphical representations}
\label{sec:la_graphics}

\begin{figure}
        \centering
        \begin{subfigure}[b]{0.5\textwidth}
                \centering
                \includegraphics[width=\textwidth]{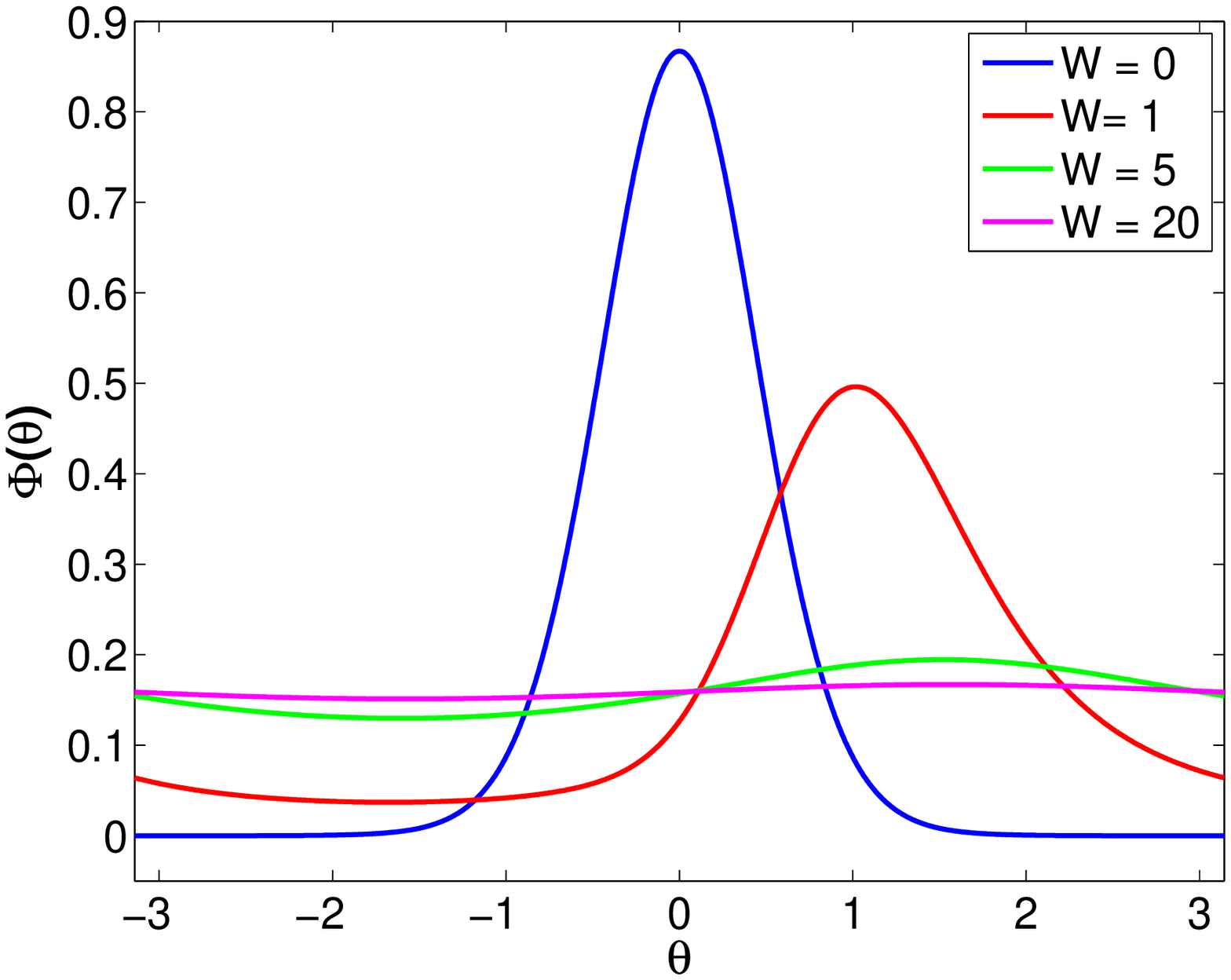}
                \caption{d = 0.2}
                \label{GVM02}
        \end{subfigure}%
        ~ %add desired spacing between images, e. g. ~, \quad, \qquad etc.
        \begin{subfigure}[b]{0.5\textwidth}
                \centering
                \includegraphics[width=\textwidth]{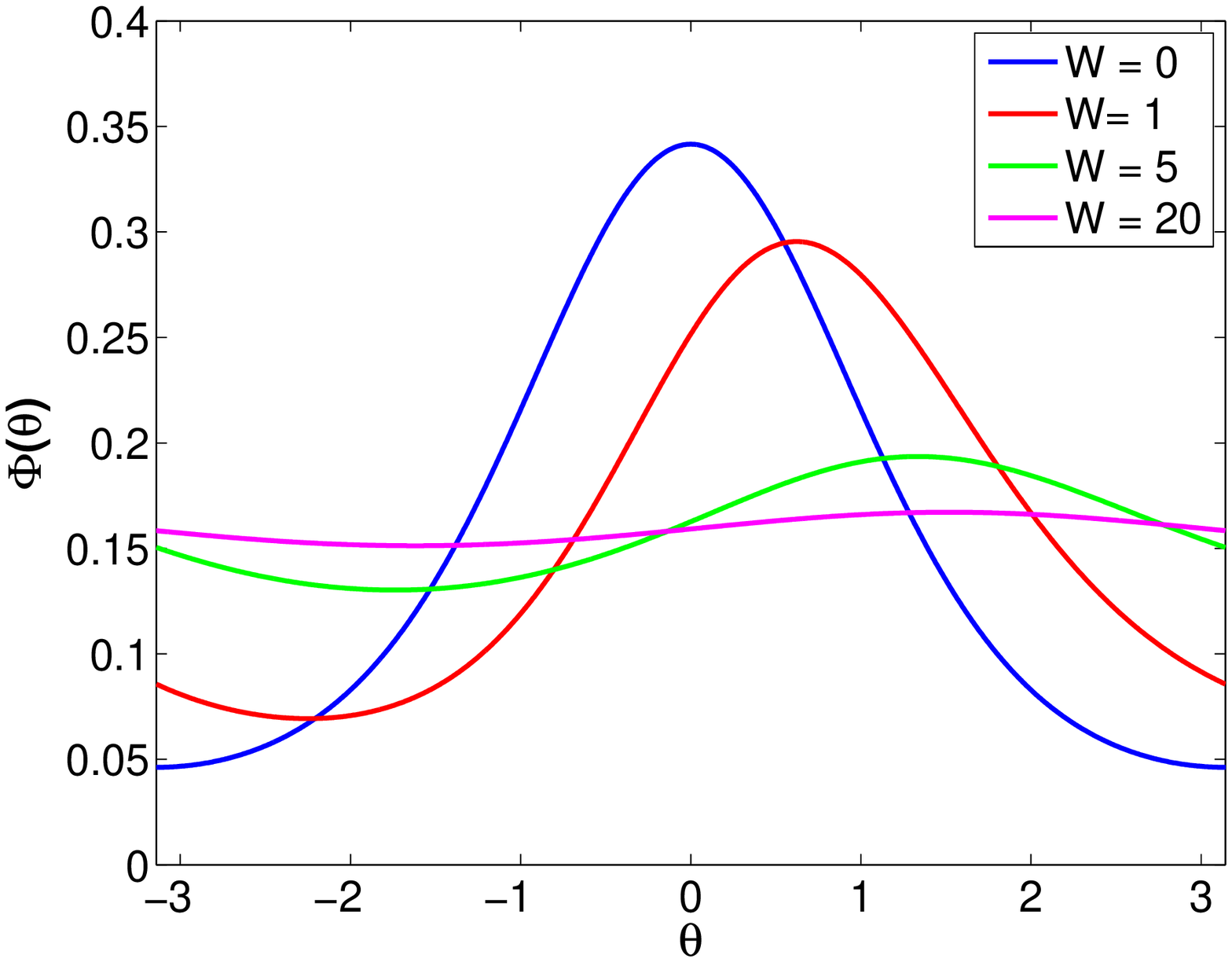}
                \caption{d = 1}
                \label{GVM1}
        \end{subfigure}
        ~ %add desired spacing between images, e. g. ~, \quad, \qquad etc.
        \begin{subfigure}[b]{0.5\textwidth}
                \centering
                \includegraphics[width=\textwidth]{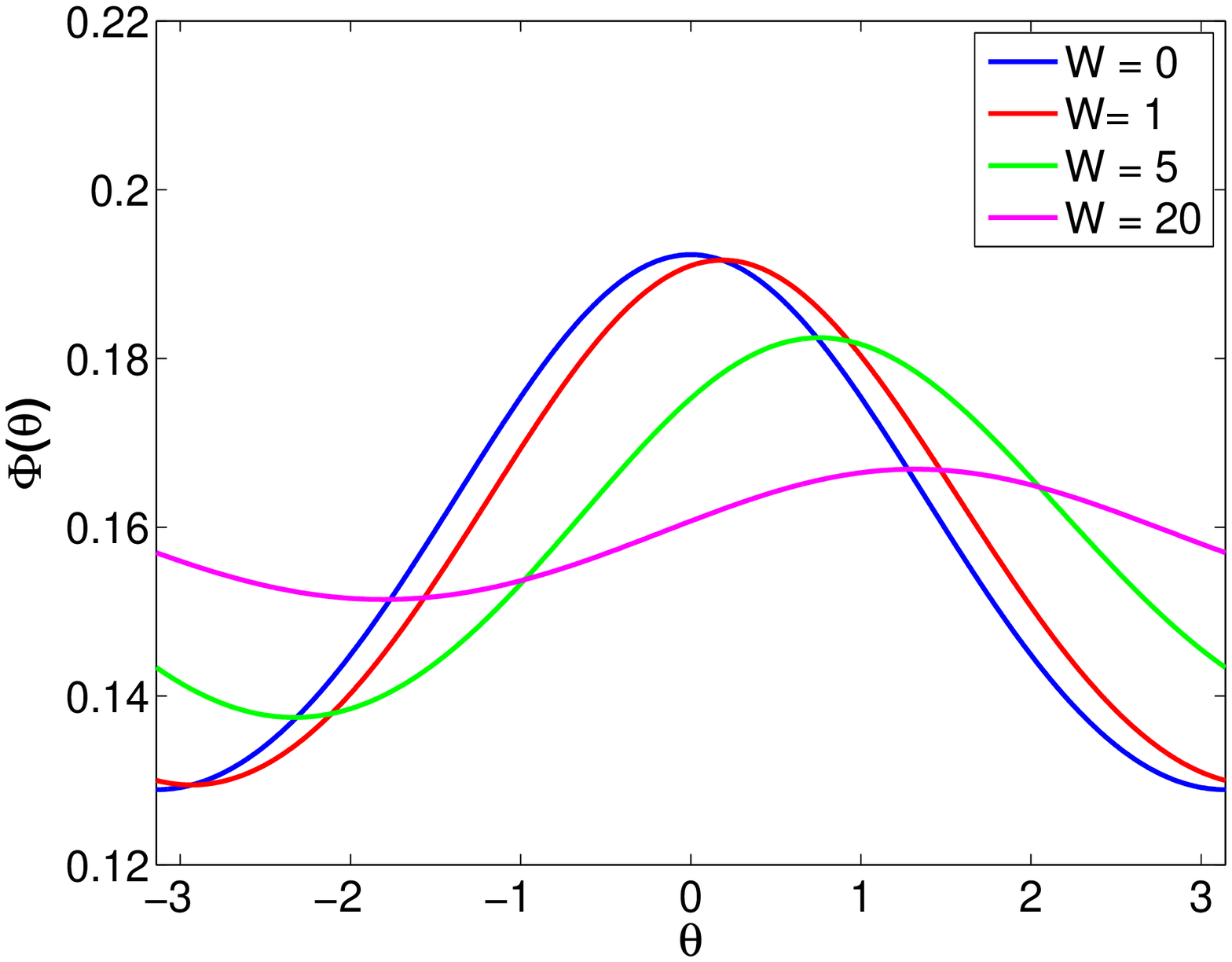}
                \caption{d = 5}
                \label{GVM5}
        \end{subfigure}
        \caption{(Color online) The Generalized von Mises-Fisher (GVM) $\Phi_W(\theta)$ as a function of $\theta$ for three values of the noise parameter:  $d=0.2$ (Fig. \ref{GVM02}), $d=1$ (Fig. \ref{GVM1}) and $d=5$ (Fig. \ref{GVM5}). In each figure, four values of the angular velocity $W$ are represented: $W=0$ (blue curve),  $W=1$ (red curve), $W=5$ (green curve) and $W=20$ (magenta curve).}
\label{fig:GVM}
\end{figure}

\begin{figure}
        \centering
        \begin{subfigure}[b]{0.5\textwidth}
                \centering
                \includegraphics[width=\textwidth]{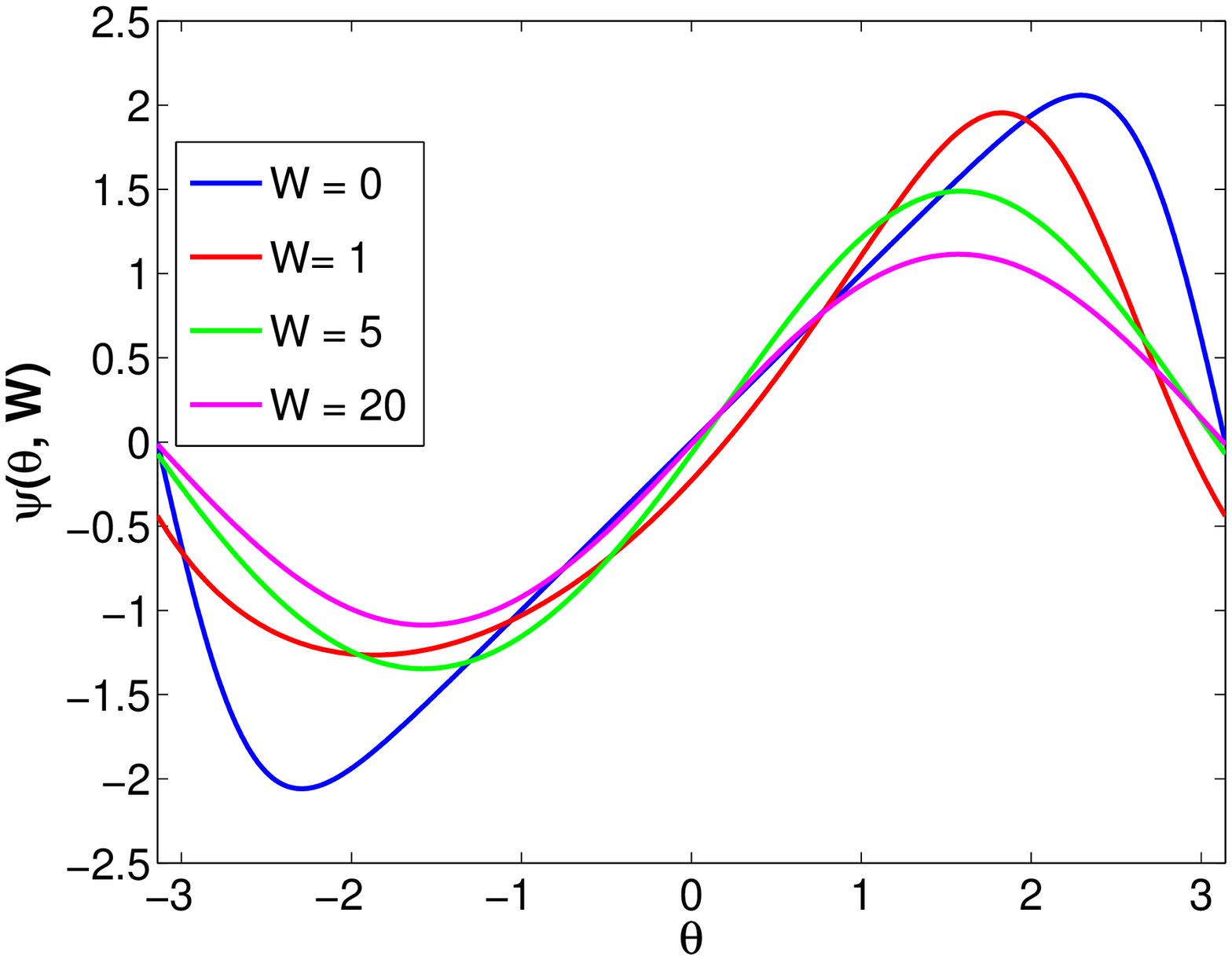}
                \caption{d = 0.2}
                \label{GCI02}
        \end{subfigure}%
        ~ %add desired spacing between images, e. g. ~, \quad, \qquad etc.
        \begin{subfigure}[b]{0.5\textwidth}
                \centering
                \includegraphics[width=\textwidth]{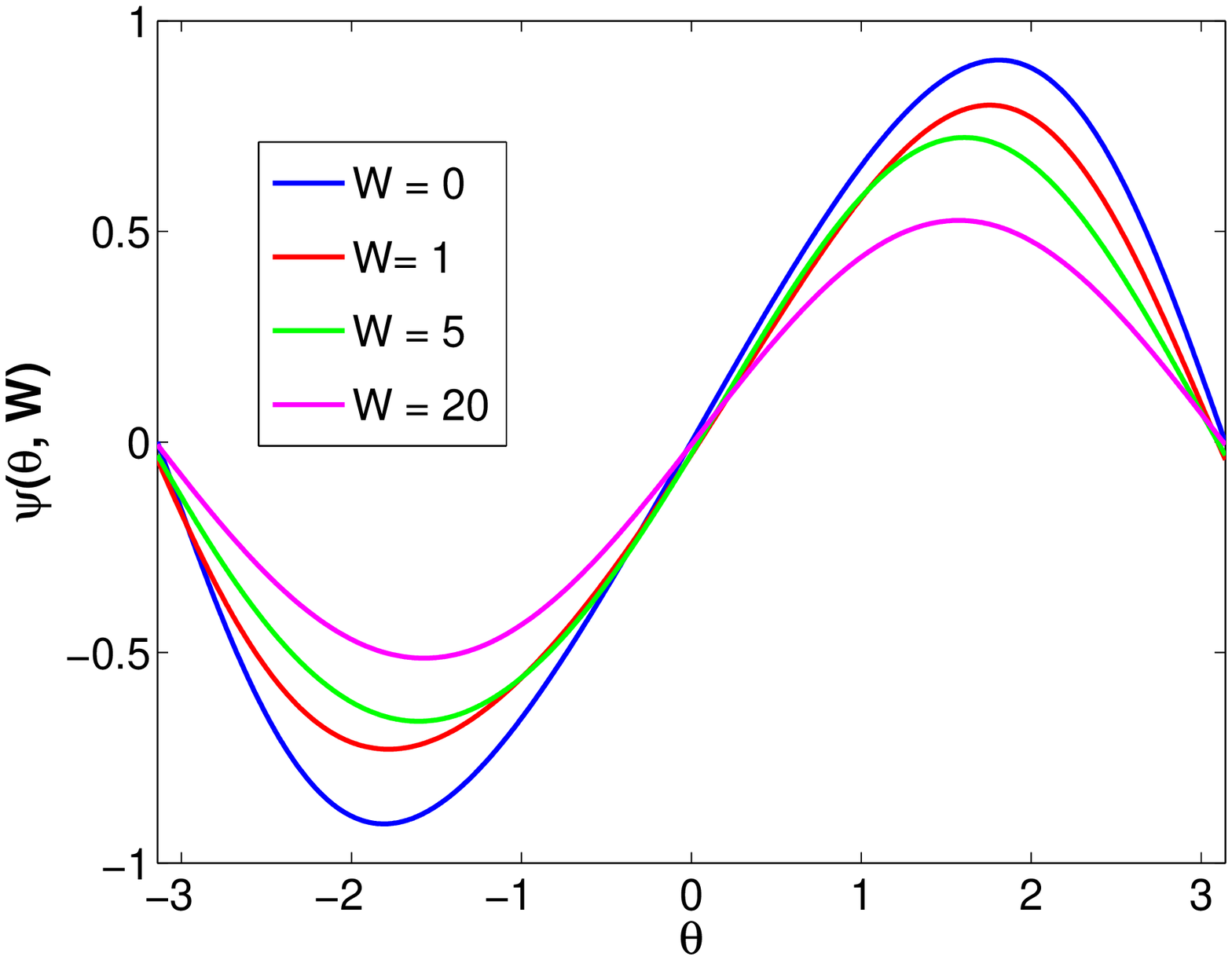}
                \caption{d = 1}
                \label{GCI1}
        \end{subfigure}
        ~ %add desired spacing between images, e. g. ~, \quad, \qquad etc.
        \begin{subfigure}[b]{0.5\textwidth}
                \centering
                \includegraphics[width=\textwidth]{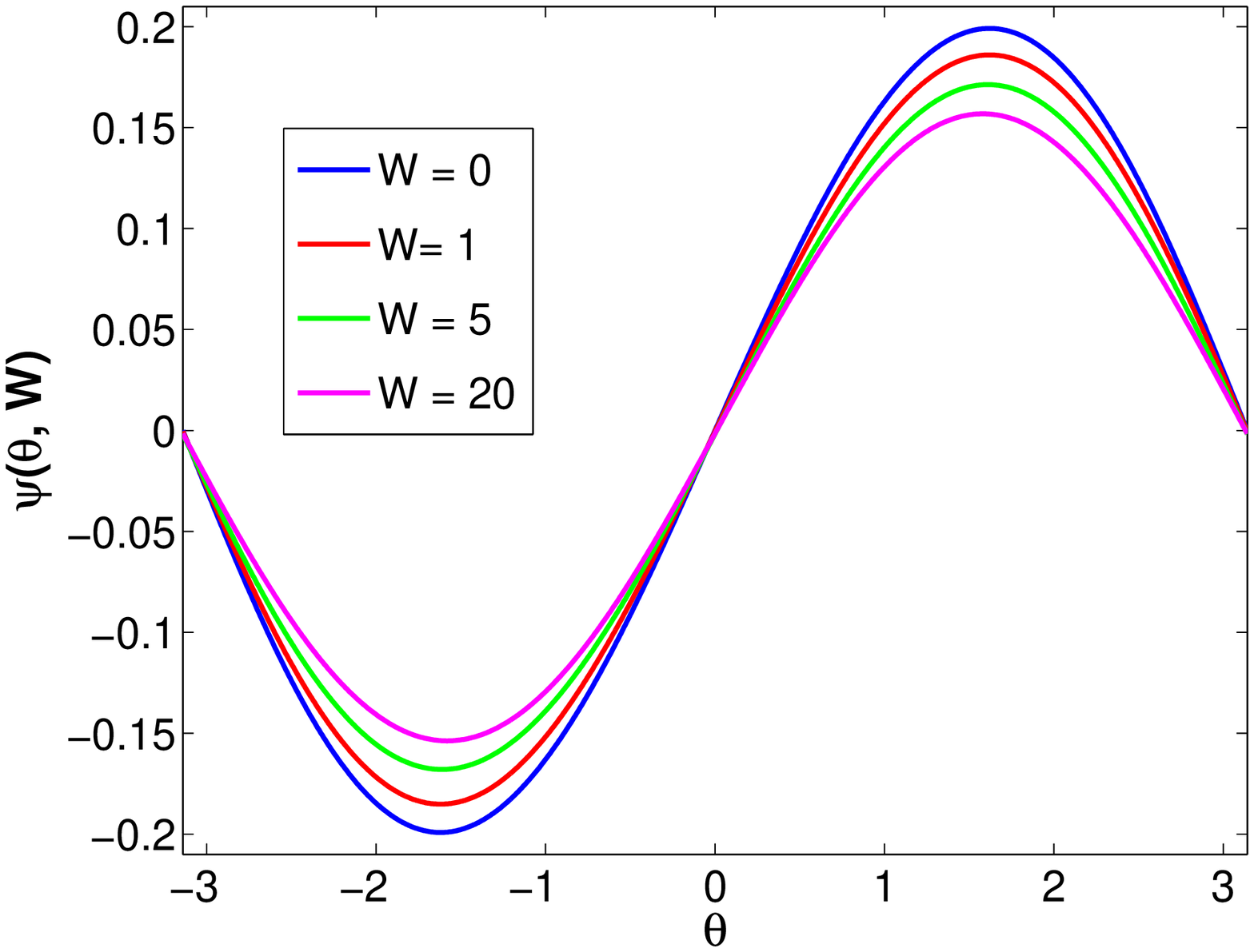}
                \caption{d = 5}
                \label{GCI5}
        \end{subfigure}
        \caption{(Color online) The generalized collision invariant $\beta X_W(\theta)$ as a function of $\theta$ for three values of the noise parameter:  $d=0.2$ (Fig. \ref{GCI02}), $d=1$ (Fig. \ref{GCI1}) and $d=5$ (Fig. \ref{GCI5}). In each figure, four values of the angular velocity $W$ are represented: $W=0$ (blue curve),  $W=1$ (red curve), $W=5$ (green curve) and $W=20$ (magenta curve). The scaling parameter $\beta$ is adjusted in such a way that the maxima of the various curves have similar orders of magnitude, for an easier comparison. }
\label{fig:GCI}
\end{figure}

In this appendix, we provide some graphical representations of the equilibrium GVM distribution, of the GCI and of the coefficients $a_1, \ldots, a_6$ of the large angular rotation case. Fig. \ref{fig:GVM} provides the Generalized von Mises-Fisher (GVM) distribution $\tilde M_\omega(v,W)$ (\ref{eq:la_def_Mom}) as a function of the angle $\theta = \widehat{\omega_\Omega,v)}$, i.e. the function $\Phi_W(\theta)$ defined at Def. \ref{lem:la_exist_Phi}. Fig. \ref{fig:GCI} provides the Generalized Collision Invariant (GCI) $\chi_\Omega(v,W)$ defined at Prop. \ref{prop:la_exist_GCI} as a function of the angle $\theta = = \widehat{\omega_\Omega,v)}$, i.e. the function $X_W(\theta)$ defined by (\ref{eq:la_def_X}). The GCI have been scaled to present similar maxima and be more easily compared (in other words, the function represented is $\beta X_W(\theta)$ for some value of the scaling parameter $\beta$ ; we notice that the final SOHR-L model is independent of the use of $\beta X_W(\theta)$ instead of $X_W(\theta)$, as the GCI form a vector space). The GVM and GCI are represented for three values of the noise parameter:  $d=0.2$ (Fig. \ref{GVM02} and \ref{GCI02}), $d=1$ (Fig. \ref{GVM1}  and \ref{GCI1}) and $d=5$ (Fig. \ref{GVM5} and \ref{GCI5}). In each figure, four values of the angular velocity $W$ are represented: $W=0$ (blue curve),  $W=1$ (red curve), $W=5$ (green curve) and $W=20$ (magenta curve). 

On Fig. \ref{fig:GVM}, we observe that the GVM have Gaussian shapes which become more uneven with maxima drifting towards the right when the angular velocity $W$ increases. As $W$ becomes large (see the magenta curves corresponding to $W=20$), the GVM becomes close to a uniform distribution, and the difference to the uniform distribution seems close to an odd function. The influence of $W$ is stronger when the noise parameter $d$ is small. Indeed, comparing the blue and red curves respectively corresponding to $W=0$ and $W=1$, we observe a fairly large difference in the case $d=0.2$ (Fig. \ref{GVM02}) while the difference is tiny in the case $d=5$ (Fig. \ref{GVM5}). In particular, we observe that the position of the peak is strongly drifted towards the right in the case $d=0.2$ (Fig. \ref{GVM02}) and to a lesser extent, in the case $d=1$ (Fig. \ref{GVM1}). The drift of the peak towards the right shows that the angle $\psi(W) = \widehat{(\omega_\Omega(W), \Omega)}$ can be significant. For instance, here, in the case $d=0.2$ (Fig. \ref{GVM02}), we see that this angle is about $1$ radian (if we estimate it as the position of the peak). As expected, the width of the peak increases with the noise parameter $d$. 

On Fig. \ref{fig:GCI}, we notice that the GCI are close to odd functions of $\theta$ and are rigorously odd functions in the case $W=0$. The influence of increasing values of $W$ is similar as for the GVM, with a deformation of the GCI towards the right (compare the cases $W=0$ (blue curve) and $W=1$ (red curve) for the noise parameter $d=0.2$ (Fig. \ref{GCI02})). The influence of $W$ is less pronounced for increasing values of $d$, with almost no difference between the cases $W=0$ (blue curve) and $W=1$ (red curve) for the noise parameter $d=5$ (Fig. \ref{GCI5}). When both $W$ and $d$ are small, the GCI have sharp variations around $\theta = \pm \pi$ and smoother variation around $\theta = 0$ (see the cases $W=0$ (blue curve) for $d=0.2$ (Fig. \ref{GCI02})). When either $d$ or $W$ increases, the GCI becomes closer and closer to the sine function.

\begin{figure}
        \centering
        \begin{subfigure}[b]{0.5\textwidth}
                \centering
                \includegraphics[width=\textwidth]{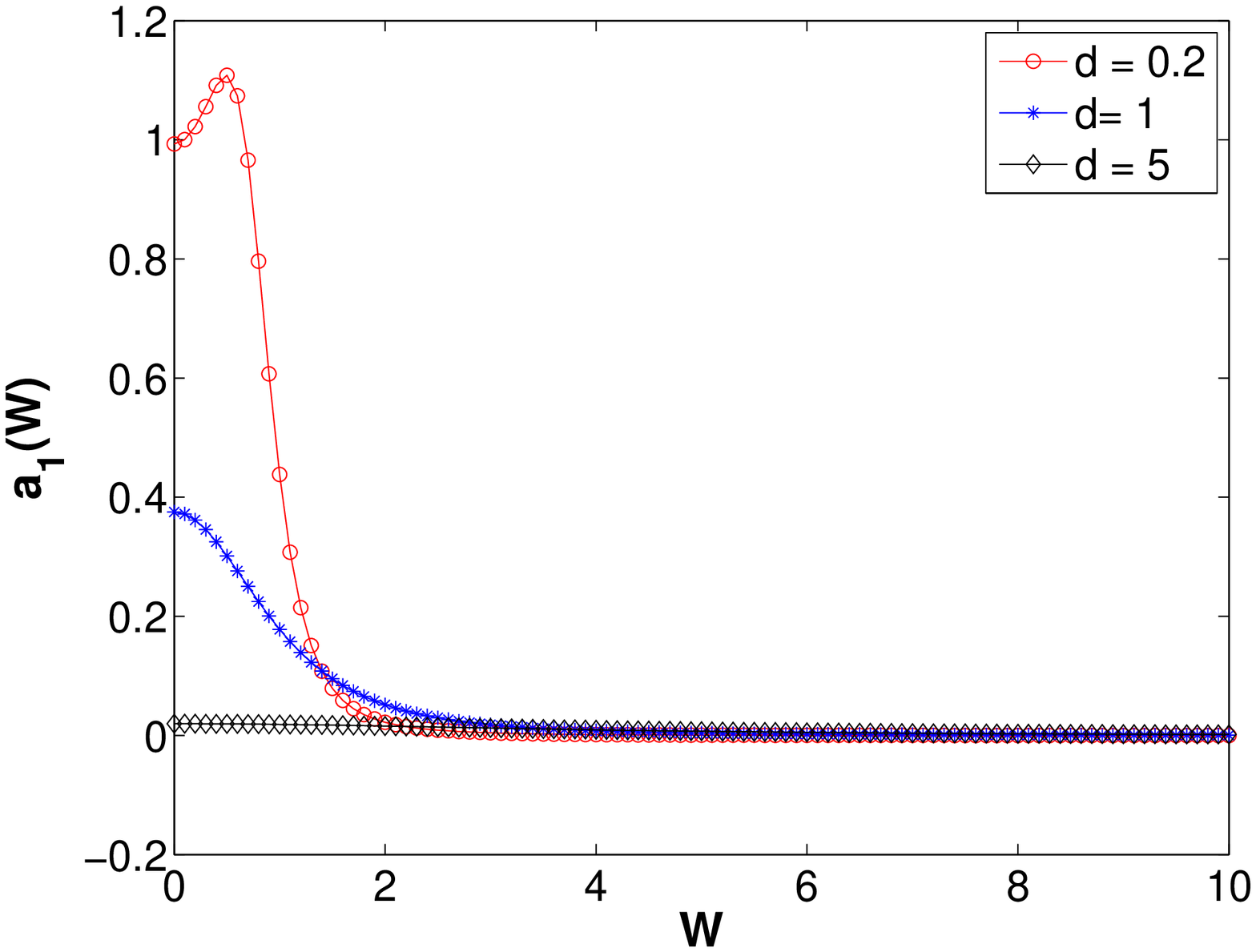}
                \caption{$a_1(W)$}
                \label{a1}
        \end{subfigure}%
        ~ %add desired spacing between images, e. g. ~, \quad, \qquad etc.
        \begin{subfigure}[b]{0.5\textwidth}
                \centering
                \includegraphics[width=\textwidth]{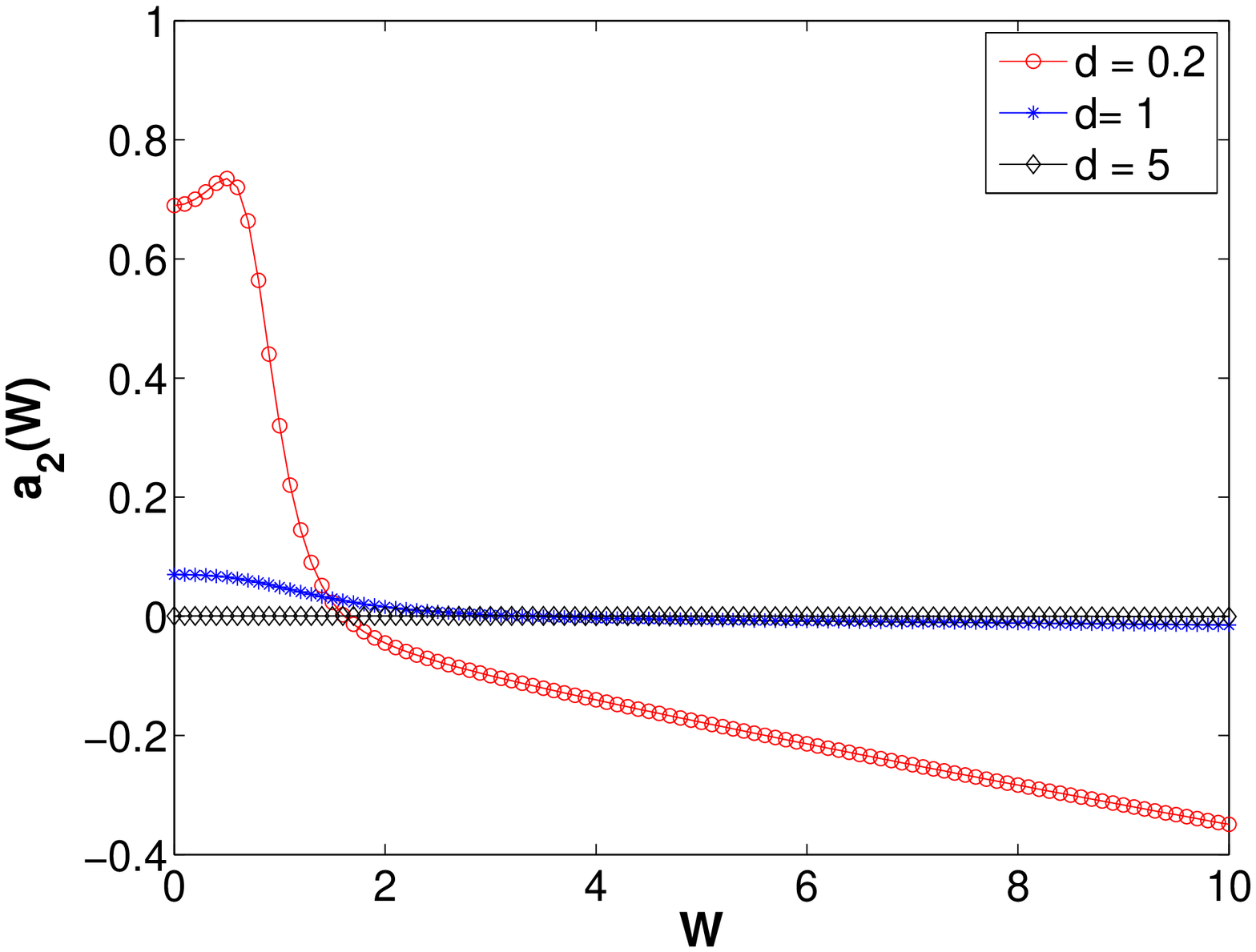}
                \caption{$a_2(W)$}
                \label{a2}
        \end{subfigure}

        \begin{subfigure}[c]{0.5\textwidth}
                \centering
                \includegraphics[width=\textwidth]{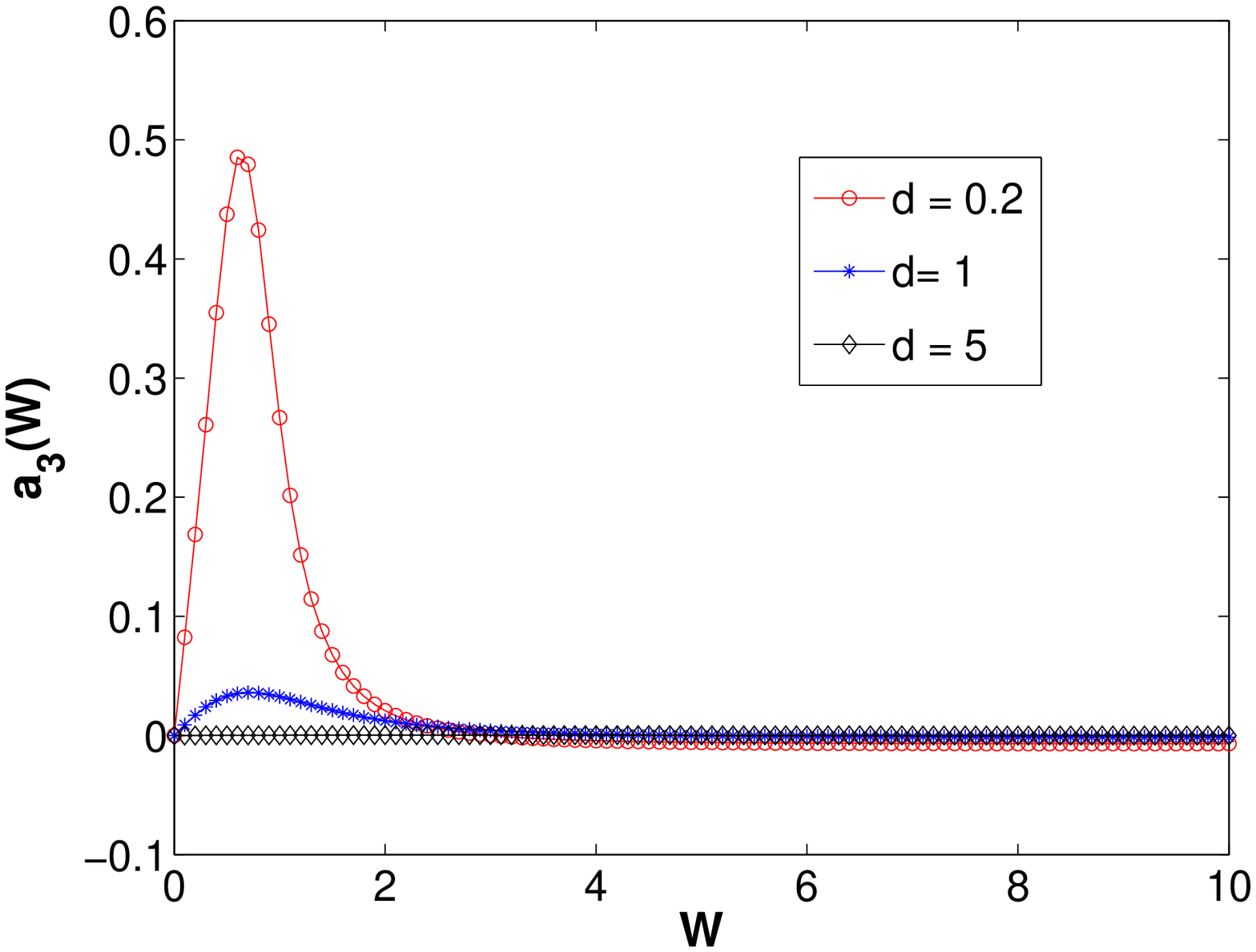}
                \caption{$a_3(W)$}
                \label{a3}
        \end{subfigure}%
        ~ %add desired spacing between images, e. g. ~, \quad, \qquad etc.
        \begin{subfigure}[d]{0.5\textwidth}
                \centering
                \includegraphics[width=\textwidth]{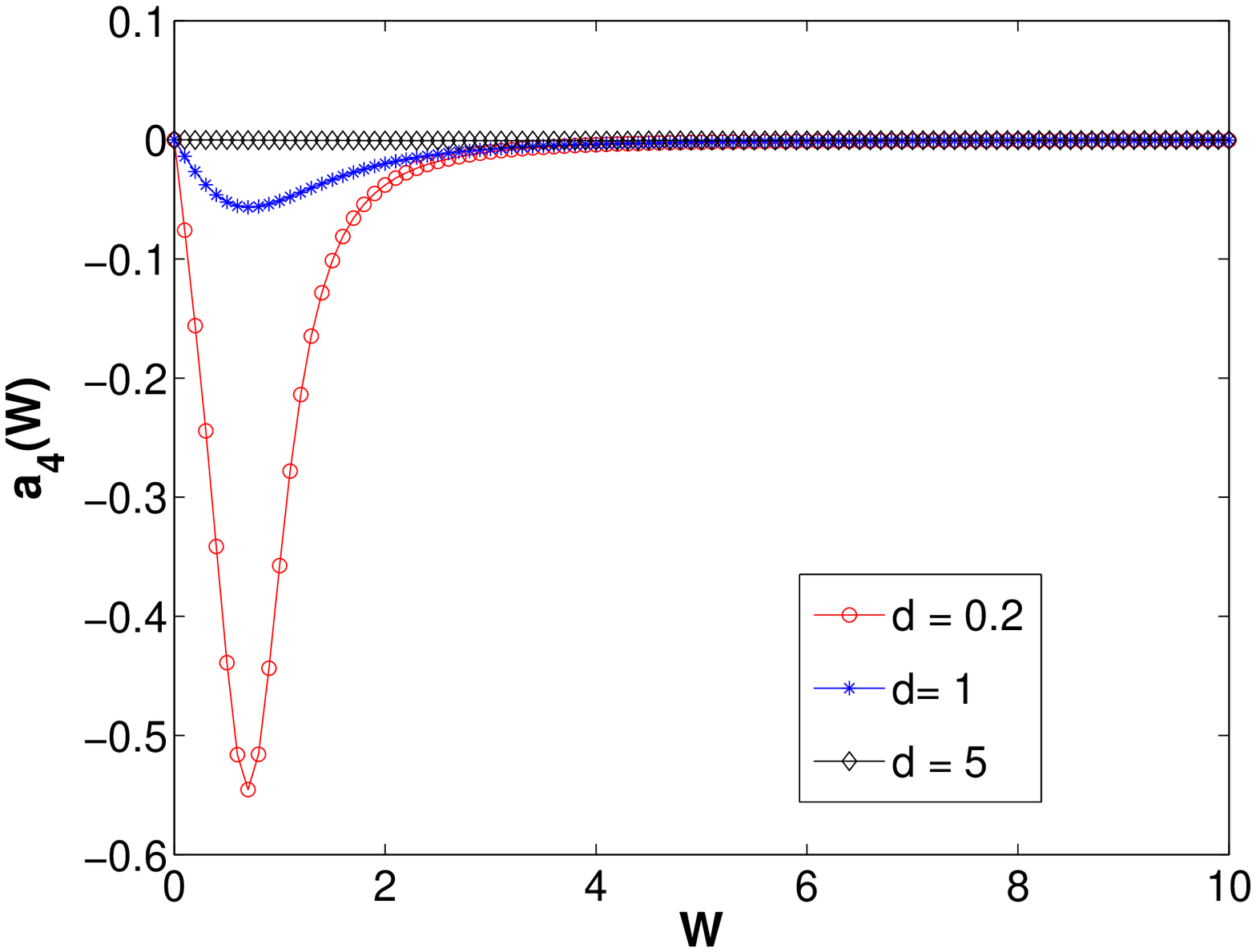}
                \caption{$a_4(W)$}
                \label{a4}
        \end{subfigure}

        \begin{subfigure}[e]{0.5\textwidth}
                \centering
                \includegraphics[width=\textwidth]{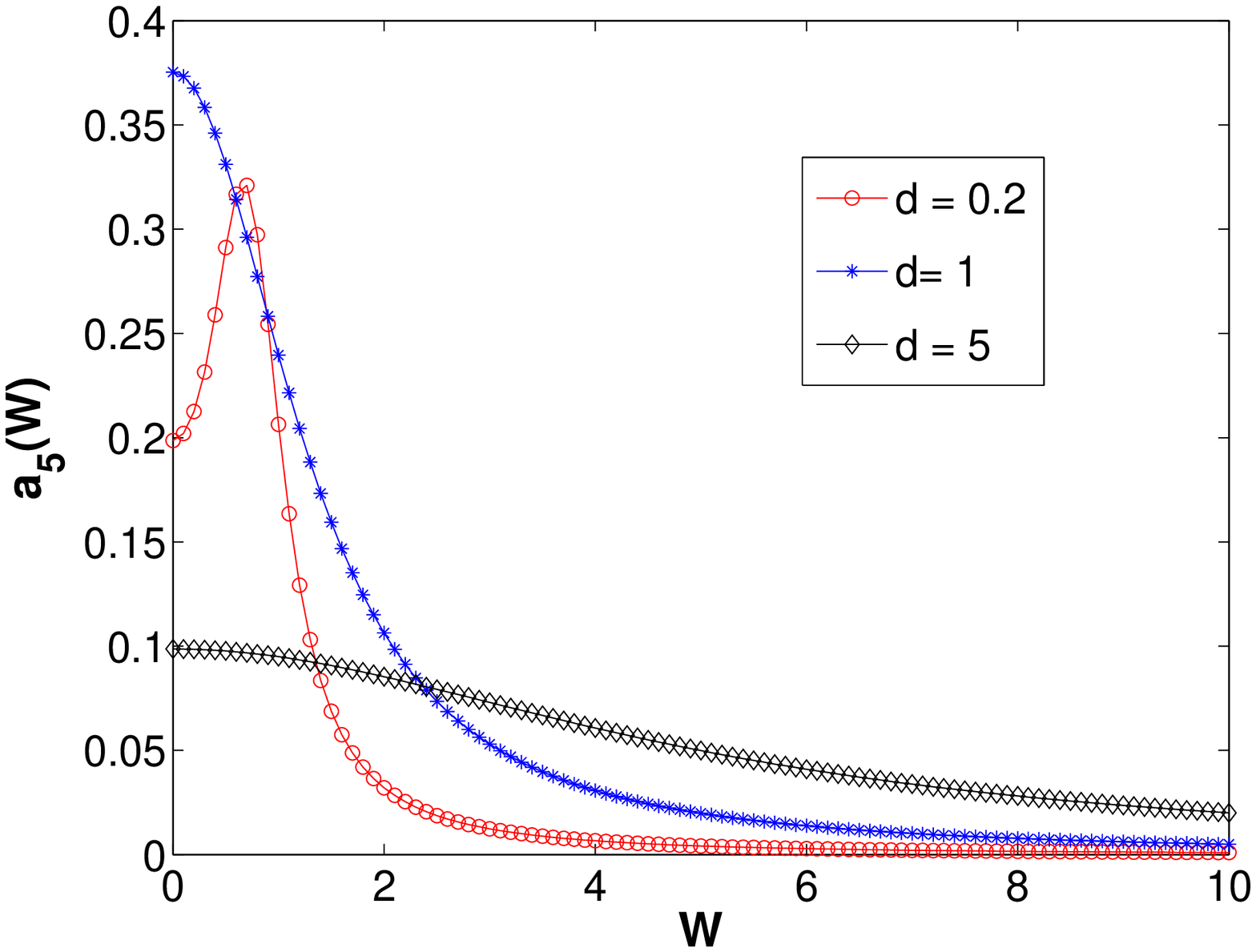}
                \caption{$a_5(W)$}
                \label{a5}
        \end{subfigure}%       
        ~ %add desired spacing between images, e. g. ~, \quad, \qquad etc.
        \begin{subfigure}[f]{0.5\textwidth}
                \centering
                \includegraphics[width=\textwidth]{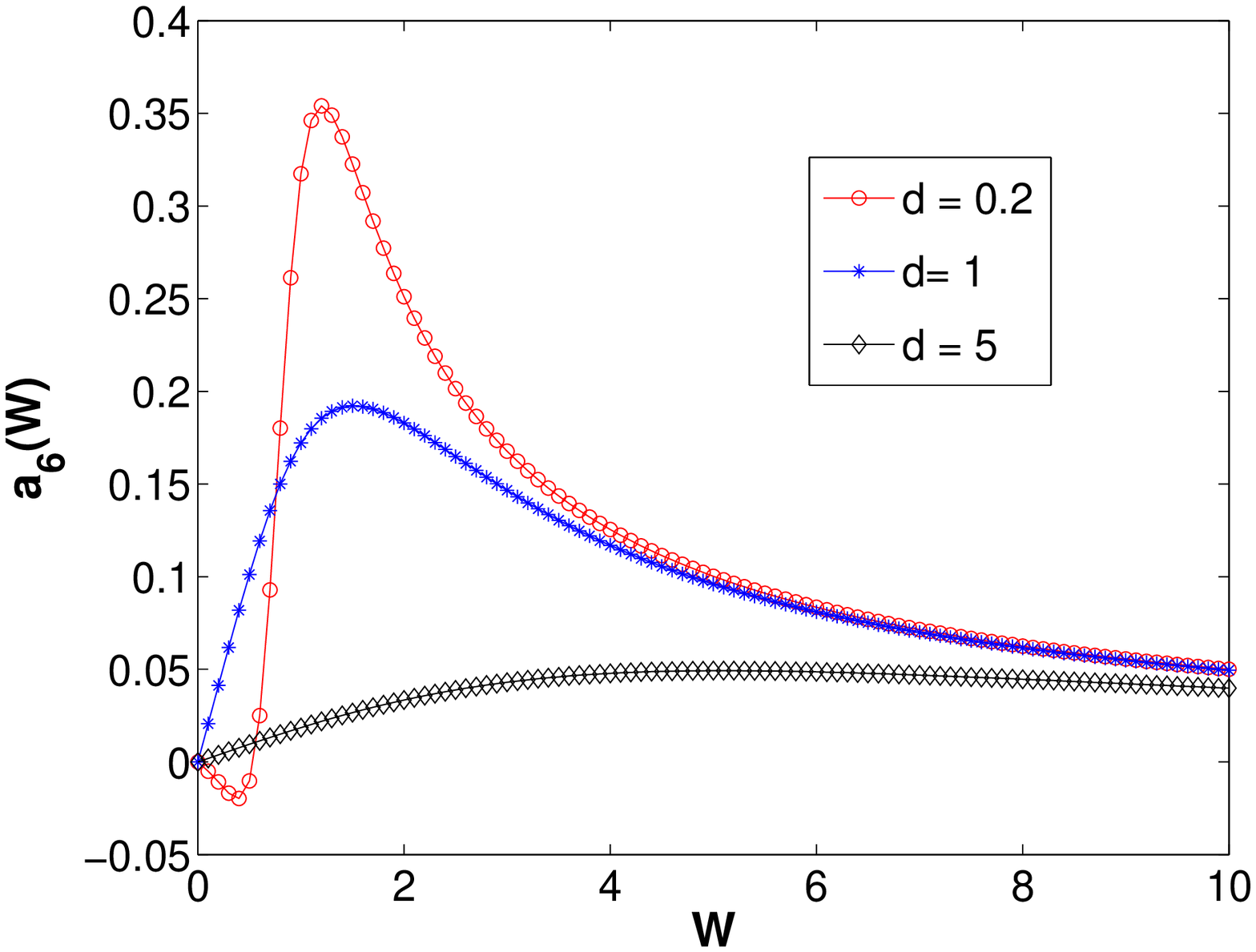}
                \caption{$a_6(W)$}
                \label{a6}
        \end{subfigure}
        \caption{(Color online) The coefficients $a_1$ (Fig. \ref{a1}), $a_2$ (Fig. \ref{a2}), $a_3$ (Fig. \ref{a3}), $a_4$ (Fig. \ref{a4}), $a_5$ (Fig. \ref{a5}), $a_6$ (Fig. \ref{a6}) as functions of the angular velocity $W$. Three values of the noise parameter $d$ are shown: $d=0.2$ (red dots), $d=1$ (blue stars), $d=5$ (black diamonds).  }
\label{fig:coefs}
\end{figure}

Finally, on Fig. \ref{fig:coefs}, the coefficients $a_1$ through $a_6$ as functions of $W$ in the range $W \in [0,10]$ are represented. Again, three values of the noise parameter $d$ are investigated: $d=0.2$ (red dots), $d=1$ (blue stars), $d=5$ (black diamonds). As announced in Prop. \ref{prop:la_lin_stab}, we realize that $a_1$ and $a_5$ are positive. We also observe that $a_1$ through $a_4$ are quite small for large values of $d$ (see the case $d=5$) and that $a_1$, $a_3$ and $a_5$ seem to converge to~$0$ as $W \to \infty$. By contrast, $a_2$ and $a_4$ seem to have a linear behavior as $W \to \infty$, while $a_6$ seems to converge to a non-zero value. Finally, as expected, the range of variation of the parameters as a function of $W$ is narrower in the low noise case ($d=0.2$) than in the large noise case ($d=5$). All these observations need to be confirmed by theoretical investigations, which will be developed in future work.

%%%%%%%%%%%%%%%%%%%%%%%%%%%%%%%%%%%%%%%%%%%%%%%%%%%%%%%%%%%%%%%%%%%%%%%%%%%%%%%%%%%%%%%%%%%%%%%%
%%%%%%%%%%%%%%%%%%%%%%%%%%%%%%%%%%%%%%%%%%%%%%%%%%%%%%%%%%%%%%%%%%%%%%%%%%%%%%%%%%%%%%%%%%%%%%%%
%%%%%%%%%%%%%%%%%%%%%%%%%%%%%%%%%%%%%%%%%%%%%%%%%%%%%%%%%%%%%%%%%%%%%%%%%%%%%%%%%%%%%%%%%%%%%%%%
%%%%%%%%%%%%%%%%%%%%%%%%%%%%%%%%%%%%%%%%%%%%%%%%%%%%%%%%%%%%%%%%%%%%%%%%%%%%%%%%%%%%%%%%%%%%%%%%
\end{appendices}

\end{document}